\documentclass[lettersize,journal]{IEEEtran}

\usepackage[algo2e]{algorithm2e} 
\usepackage{amsmath,amssymb,amsmath,amsfonts}
\usepackage{bm}
\usepackage{bbm}
\usepackage{graphicx}
\usepackage{cite}
\usepackage{epstopdf}
\usepackage{gensymb}
\usepackage{color}
\usepackage{lipsum}
\usepackage{float}
\usepackage{epstopdf}
\usepackage[mathcal]{eucal}
\usepackage{subfiles}
\usepackage[T1]{fontenc}
\usepackage{siunitx}
\usepackage[breaklinks=true,hidelinks]{hyperref}
\usepackage{tabulary}
\usepackage{epsfig,graphics,graphicx,subfig,amssymb,amstext,amsmath,algorithm,algorithmic,wrapfig,multirow}
\usepackage{array}
\usepackage{adjustbox}
\usepackage[dvipsnames]{xcolor}
\usepackage{cite}
\usepackage{times}
\usepackage{placeins}
\usepackage{textcomp}
\usepackage[font=small]{caption}
\usepackage{flushend}
\usepackage{color, colortbl}
\usepackage{tabularx}
\usepackage{bm}
\usepackage{enumerate}
\usepackage{tikz}
\usepackage{pgfplots}
\usepackage{epstopdf}
\usetikzlibrary{shapes,arrows}
\usepackage[utf8]{inputenc}
\usepackage{mathtools}
\usepackage[utf8]{inputenc}
\usepackage{xcolor}
\usepackage{tikz}
\usetikzlibrary{chains,arrows,calc,positioning}
\usepackage{amssymb}
\usepackage{pifont}
\usepackage{soul}
\usepackage{breqn} 
\usepackage{booktabs} 
\usepackage{multirow}
\usepackage{enumitem}
\usepackage{tabulary}
\usepackage[normalem]{ulem}
\usepackage{dblfloatfix}
\usepackage{makecell}
\usepackage{lipsum}
\usepackage{amsthm}
\usepackage{filecontents}
\usepackage[acronyms,nonumberlist,nopostdot,nomain,nogroupskip,acronymlists={hidden}]{glossaries}
\newglossary[algh]{hidden}{acrh}{acnh}{Hidden Acronyms}

\makeatletter
\AtBeginDocument{%
  \renewcommand{\@glslink}[4]{#4}%
}
\makeatother

\usepackage{multicol}
\usepackage{longtable}
\usepackage{blindtext}
\usepackage{tcolorbox}

\usepackage{comment}
\usepackage{ifthen}

\newif\ifshowfigtables
\showfigtablesfalse
\ifshowfigtables
  \long\def\commentfigtable#1\endcommentfigtable{#1}
\else
  \long\def\commentfigtable#1\endcommentfigtable{}
\fi

\newif\ifshowsections
\showsectionsfalse
\ifshowsections
  \long\def\commentsection#1\endcommentsection{#1}
\else
  \long\def\commentsection#1\endcommentsection{}
\fi

\makeatletter
\let\oldlt\longtable
\let\endoldlt\endlongtable
\def\longtable{\@ifnextchar[\longtable@i \longtable@ii}
\def\longtable@i[#1]{\begin{figure}[t]
\onecolumn
\begin{minipage}{0.5\textwidth}
\oldlt[#1]
}
\def\longtable@ii{\begin{figure}[t]
\onecolumn
\begin{minipage}{0.5\textwidth}
\oldlt
}
\def\endlongtable{\endoldlt
\end{minipage}
\twocolumn
\end{figure}}
\makeatother

\newacronym{dof}{DoF}{degrees of freedom}
\newacronym{snr}{SNR}{signal-to-noise ratio}
\newacronym{sinr}{SINR}{signal-to-interference-plus-noise ratio}
\newacronym[plural=\gls{cnn}s,firstplural=convolutional neural networks (CNNs)]{cnn}{CNN}{convolutional neural network}
\newacronym{ml}{ML}{machine learning}
\newacronym{phy}{PHY}{Physical layer}
\newacronym{csi}{CSI}{channel state information}
\newacronym{fcc}{FCC}{Federal Communication Commission}
\newacronym{mimo}{MIMO}{multiple-input, multiple-output}
\newacronym{iui}{IUI}{inter-user interference}
\newacronym{isi}{ISI}{inter-symbol interference}
\newacronym{ap}{AP}{access point}
\newacronym{sta}{STA}{station}
\newacronym{dl}{DL}{downlink}
\newacronym{ul}{UL}{uplink}
\newacronym{mcs}{MCS}{modulation and coding scheme}
\newacronym[plural=\gls{cfr}s,firstplural=channel frequency responses (CFRs)]{cfr}{CFR}{channel frequency response}
\newacronym{cir}{CIR}{channel impulse response}
\newacronym{cqi}{CQI}{channel quality indicator}
\newacronym{ndp}{NDP}{Null Data Packet}
\newacronym[plural=\gls{ltf}s,firstplural=long training fields (LTFs)]{ltf}{LTF}{Long Training field}
\newacronym[plural=\gls{stf}s,firstplural=short training fields (STFs)]{stf}{STF}{Short Training field}
\newacronym{vht}{VHT}{Very High Throughput}
\newacronym{ht}{HT}{High Throughput}
\newacronym{pci}{PCI}{protocol control information}
\newacronym{ofdm}{OFDM}{orthogonal frequency-division multiplexing}
\newacronym{ofdma}{OFDMA}{orthogonal frequency-division multiple access}
\newacronym{mse}{MSE}{mean-square-error}
\newacronym{mmse}{MMSE}{minimum mean-square-error}
\newacronym{zf}{ZF}{zero-forcing}
\newacronym{los}{LOS}{line-of-sight}
\newacronym{nlos}{NLOS}{non-line-of-sight}
\newacronym{mac}{MAC}{Medium Access Control}
\newacronym{fcs}{FCS}{Frame Check Sequence}
\newacronym{ifcs}{i-FCS}{intermediate Frame Check Sequence}
\newacronym{ack}{ACK}{Acknowledgment}
\newacronym{ppdu}{PPDU}{PHY protocol data unit}
\newacronym{psdu}{PSDU}{PHY service data unit}
\newacronym{mpdu}{MPDU}{MAC protocol data unit}
\newacronym{msdu}{MSDU}{MAC service data unit}
\newacronym{fft}{FFT}{fast Fourier transform}
\newacronym{ifft}{IFFT}{inverse fast Fourier transform}
\newacronym{rf}{RF}{radio frequency}
\newacronym{cp}{CP}{cyclic prefix}
\newacronym{cfo}{CFO}{carrier frequency offset}
\newacronym{ru}{RU}{resource unit}
\newacronym{svd}{SVD}{singular value decomposition}
\newacronym{ldpc}{LDPC}{low-density parity check code}
\newacronym{bcc}{BCC}{binary convolutional code}
\newacronym{csd}{CSD}{cyclic shift diversity}
\newacronym{fec}{FEC}{forward error correction}
\newacronym{gi}{GI}{guard interval}
\newacronym{qam}{QAM}{quadrature amplitude modulation}
\newacronym{psk}{PSK}{phase-shift keying}
\newacronym{evm}{EVM}{error vector magnitude}
\newacronym{ber}{BER}{bit error rate}
\newacronym{per}{PER}{packet error rate}
\newacronym{dac}{DAC}{digital-to-analog converter}
\newacronym{dhcp}{DHCP}{dynamic host configuration protocol}
\newacronym{adc}{ADC}{analog-to-digital converter}
\newacronym{sig}{SIG}{Signal field}
\newacronym{tb}{TB}{trigger-based}
\newacronym{agc}{AGC}{automatic gain control}
\newacronym{elr}{ELR}{enhanced long range}
\newacronym{qos}{QoS}{quality of service}
\newacronym{qoe}{QoE}{quality of experience}
\newacronym{txop}{TXOP}{transmission opportunity}
\newacronym{wfa}{WFA}{Wi-Fi Alliance}
\newacronym{wg}{WG}{working group}
\newacronym{sg}{SG}{study group}
\newacronym{tg}{TG}{task group}
\newacronym{sc}{SC}{Standing Committee}
\newacronym{scs}{SCS}{stream classification service}
\newacronym{tig}{TIG}{topic interest group}
\newacronym{ahc}{AHC}{ad-hoc committee}
\newacronym{wglb}{WGLB}{working group letter ballot}
\newacronym{par}{PAR}{project authorization request}
\newacronym{weca}{WECA}{Wireless Ethernet Compatibility Alliance}
\newacronym{ieee-sa}{IEEE-SA}{IEEE Standards Association}
\newacronym{pda}{PDA}{personal digital assistant}
\newacronym{hd}{HD}{high definition}
\newacronym{mlo}{MLO}{multi-link operation}
\newacronym{urllc}{URLLC}{ultra-reliable low latency}
\newacronym{bss}{BSS}{basic service set}
\newacronym{obss}{OBSS}{overlapping basic service set}
\newacronym{iiot}{IIoT}{industrial Internet of things}
\newacronym{dru}{DRU}{distributed resource unit}
\newacronym{psd}{PSD}{power spectral density}
\newacronym{ueqm}{UEQM}{unequal modulation}
\newacronym{ar}{AR}{augmented reality}
\newacronym{vr}{VR}{virtual reality}
\newacronym{iot}{IoT}{Internet of things}
\newacronym{npca}{NPCA}{non-primary channel access}
\newacronym{nat}{NAT}{network address translation}
\newacronym{wlan}{WLAN}{wireless local area network}
\newacronym{dcf}{DCF}{distributed coordination function}
\newacronym{ai}{AI}{artificial intelligence}
\newacronym{ieee}{IEEE}{Institute of Electrical and Electronics Engineers}
\newacronym{csat}{CSAT}{carrier sense adaptive transmission}
\newacronym{lbt}{LBT}{listen-before-talk}
\newacronym{lte}{LTE}{long-term evolution}
\newacronym{twt}{TWT}{target wake time}
\newacronym{tcp}{TCP}{transmission control protocol}
\newacronym{udp}{UDP}{user datagram protocol}
\newacronym{eht}{EHT}{Extremely-High Throughput}
\newacronym{tsn}{TSN}{time-sensitive networking}
\newacronym{harq}{HARQ}{hybrid automatic repeat request}
\newacronym{laa}{LAA}{licensed assisted access}
\newacronym{ism}{ISM}{industrial, scientific, and medical}
\newacronym{dsss}{DSSS}{direct sequence spread spectrum}
\newacronym{fhss}{FHSS}{frequency-hopping spread spectrum}
\newacronym{cck}{CCK}{complementary code keying}
\newacronym{sr}{SR}{spatial reuse}
\newacronym{uhr}{UHR}{Ultra-High Reliability}
\newacronym{mapc}{MAPC}{multi-access point coordination}
\newacronym{edp}{EDP}{enhanced data privacy}
\newacronym{3gpp}{3GPP}{3rd Generation Partnership Project}
\newacronym{ctdma}{Co-TDMA}{coordinated TDMA}
\newacronym{cofdma}{Co-OFDMA}{coordinated OFDMA}
\newacronym{crtwt}{Co-RTWT}{coordinated R-TWT}
\newacronym{csr}{Co-SR}{coordinated SR}
\newacronym{cbf}{Co-BF}{coordinated beamforming}
\newacronym{wba}{WBA}{Wireless Broadband Alliance}
\newacronym{rssi}{RSSI}{received signal strength indicator}
\newacronym{dps}{DPS}{dynamic power save}
\newacronym{ndpa}{NDPA}{Null Data Packet Announcement}
\newacronym{dcm}{DCM}{dual carrier modulation}
\newacronym{dup}{DUP}{duplication}
\newacronym{bsrp}{BSRP}{Buffer Status Report Poll}
\newacronym{bar}{BAR}{Block Acknowledgment Request}
\newacronym{mru}{MRU}{multiple resource unit}
\newacronym{rts}{RTS}{request to send}
\newacronym{cts}{CTS}{clear to send}
\newacronym{sifs}{SIFS}{short interframe space}
\newacronym{eifs}{EIFS}{extended interframe space}
\newacronym{eirp}{EIRP}{equivalent isotropic radiated power}
\newacronym{bfrp}{BFRP}{Beamforming Feedback Report Poll}
\newacronym{brp}{BRP}{beam refinement protocol}
\newacronym{im}{IM}{interference mitigation}
\newacronym{qc}{QC}{quasi-cyclic}
\newacronym{xr}{XR}{extended reality}
\newacronym{txbf}{TxBF}{transmit beamforming}
\newacronym{he}{HE}{High-Efficiency}
\newacronym{plcp}{PLCP}{physical layer convergence protocol}
\newacronym{nav}{NAV}{network allocation vector}
\newacronym{cca}{CCA}{clear channel assessment}
\newacronym{psr}{PSR}{parametrized spatial reuse}
\newacronym{pd}{PD}{packet detect}
\newacronym{dso}{DSO}{dynamic subband operation}
\newacronym{pcf}{PCF}{point coordination function}
\newacronym{edca}{EDCA}{enhanced distributed channel access}
\newacronym{pedca}{P-EDCA}{prioritized enhanced distributed channel access}
\newacronym{hcca}{HCCA}{hybrid coordination function controlled channel access}
\newacronym{p2p}{P2P}{peer-to-peer}
\newacronym{arq}{ARQ}{automatic repeat request protocol}
\newacronym{csma}{CSMA}{carrier sense multiple access}
\newacronym{ca}{CA}{collision avoidance}
\newacronym{difs}{DIFS}{distributed interframe space}
\newacronym{beb}{BEB}{binary exponential backoff}
\newacronym{cw}{CW}{contention window}
\newacronym[longplural=access categories]{ac}{AC}{access category}
\newacronym{aifs}{AIFS}{arbitrary inter-frame space}
\newacronym{mu-rts}{MU-RTS}{multi-user RTS}
\newacronym{su}{SU}{single user}
\newacronym{mu}{MU}{multi user}
\newacronym{psrr}{PSRR}{parameterized spatial reuse reception}
\newacronym{psrt}{PSRT}{parameterized spatial reuse transmission}
\newacronym{psm}{PSM}{power save mode}
\newacronym{rtwt}{R-TWT}{restricted TWT}
\newacronym{tim}{TIM}{traffic indication map}
\newacronym{dtim}{DTIM}{delivery traffic indication map}
\newacronym{aid}{AID}{association identifier}
\newacronym{bu}{BU}{bufferable unit}
\newacronym{uapsd}{U-APSD}{unscheduled automatic power save delivery}
\newacronym{wnm}{WNM}{wireless network management}
\newacronym{sp}{SP}{service period}
\newacronym{mld}{MLD}{multi-link device}
\newacronym{ss}{SS}{spatial stream}
\newacronym{tid}{TID}{Traffic Identifier}
\newacronym{hc}{HC}{high capability}
\newacronym{lc}{LC}{low capability}
\newacronym{icf}{ICF}{Initial Control frame}
\newacronym{drl}{DRL}{deep reinforcement learning}
\newacronym{ran}{RAN}{radio access network}
\newacronym{fl}{FL}{federated learning}
\newacronym{tgbn}{TGbn}{Task Group bn}
\newacronym{mlme}{MLME}{MAC layer management entity}
\newacronym{pasn}{PASN}{pre-association security negotiation}
\newacronym{usig}{U-SIG}{Universal Signal}
\newacronym{afc}{AFC}{automatic frequency coordination}
\newacronym{dfs}{DFS}{dynamic frequency selection}
\newacronym{amsdu}{A-MSDU}{Aggregate MAC Service Data Unit}
\newacronym{ampdu}{A-MPDU}{Aggregate MAC Protocol Data Unit}
\newacronym{vo}{VO}{voice}
\newacronym{vi}{VI}{video}
\newacronym{be}{BE}{best effort}
\newacronym{bk}{BK}{background}
\newacronym{dscp}{DSCP}{Differentiated Services Code Point}
\newacronym{muedca}{MU-EDCA}{multi-user EDCA}
\newacronym{pch}{PCH}{primary channel}
\newacronym{sch}{SCH}{secondary channel}
\newacronym{sca}{SCA}{Secondary Channel Access}
\newacronym{ed}{ED}{energy detection}
\newacronym{rnr}{RNR}{Reduced Neighbor Report}
\newacronym{mlsr}{MLSR}{multi-link single radio}
\newacronym{emlsr}{EMLSR}{enhanced multi-link single radio}
\newacronym{nstrmlmr}{NSTR MLMR}{non-simultaneous transmit and receive multi-link multi-radio}
\newacronym{strmlmr}{STR MLMR}{simultaneous transmit and receive multi-link multi-radio}
\newacronym{emlmr}{EMLMR}{enhanced multi-link multi-radio}
\newacronym{ba}{BA}{Block Acknowledgment}
\newacronym{str}{STR}{simultaneous transmit and receive}
\newacronym{nstr}{NSTR}{non-simultaneous transmit and receive}
\newacronym{voip}{VoIP}{voice over IP}
\newacronym{pll}{PLL}{phase-locked loop}
\newacronym{ssid}{SSID}{service set identifier}
\newacronym{bssid}{BSSID}{basic service set identifier}
\newacronym{unii}{U-NII}{Unlicensed National Information Infrastructure}
\newacronym{itu}{ITU}{International Telecommunication Union}
\newacronym{ietf}{IETF}{Internet Engineering Task Force}
\newacronym{pifs}{PIFS}{PCF interframe space}
\newacronym{dns}{DNS}{Domain Name System}
\newacronym{ip}{IP}{Internet Protocol}
\newacronym{bas}{BAS}{Broadcast Auxiliary Services}
\newacronym{lpi}{LPI}{low power indoor}
\newacronym{vlp}{VLP}{very low power}
\newacronym{revcom}{RevCom}{Standards Review Committee}
\newacronym{immw}{IMMW}{integrated mmWave}
\newacronym{irm}{IRM}{identifiable random MAC address}
\newacronym{id}{ID}{identifier}
\newacronym{ftm}{FTM}{Fine Timing Measurement}
\newacronym{rtt}{RTT}{round trip time}
\newacronym{tspec}{TSPEC}{traffic specification}

\newlength\maxlength
\newlength\thislength

\newglossarystyle{mystyle}
{%
  {
    \setlength{\maxlength}{0pt}%
    \forglsentries[\currentglossary]{\thislabel}%
    {%
       \settowidth{\thislength}{\glsentryshort{\thislabel}}%
       \ifdim\thislength>\maxlength
         \setlength{\maxlength}{\thislength}%
       \fi
    }%
    \setlength{\glsdescwidth}{\linewidth-\maxlength-\thislength-2\tabcolsep-1em}%
    \rowcolors{2}{BackgroundLightBlue}{BackgroundGray}
    \begin{tabular}{l@{\hspace{.5em} | \hspace{.5em}}p{\glsdescwidth}}
    \rowcolor{BackgroundLightBlue}
    \toprule
    \textbf{Acronym} & \textbf{Definition}\\%
    \midrule
  }%
  {
     \bottomrule
     \end{tabular}%
  }%
  \renewcommand*{\glsgroupheading}[1]{}%
}

\makeglossaries

\newtheoremstyle{mystyle}
  {}
  {}
  {\itshape}
  {}
  {\bfseries}
  {.}
  { }
  {}
\theoremstyle{mystyle}

\newlength \figwidth
\definecolor{bittersweet}{rgb}{1.0, 0.44, 0.37}
\definecolor{glaucous}{rgb}{0.38, 0.51, 0.71}
\definecolor{gainsboro}{rgb}{0.86, 0.86, 0.86}
\definecolor{babyblueeyes}{rgb}{0.63, 0.79, 0.95}
\definecolor{silver}{rgb}{0.75, 0.75, 0.75}
\definecolor{neoncarrot}{rgb}{1.0, 0.64, 0.26}
\definecolor{Gray}{gray}{0.9}
\definecolor{LightCyan}{rgb}{0.88,1,1}
\definecolor{BackgroundLightBlue}{rgb}{0.97,0.97,1}
\definecolor{BackgroundGray}{gray}{0.98}

\newcommand{\red}[1]{{\textcolor[rgb]{1,0,0}{#1}}}

\makeatletter

 \let\oldforeign@language\foreign@language
 \DeclareRobustCommand{\foreign@language}[1]{%
   \lowercase{\oldforeign@language{#1}}}

\newcommand{\blue}[1]{\textcolor{black}{#1}}

\makeatother

\usepackage{xr}

\makeatletter
\newcommand*{\addFileDependency}[1]{
\typeout{(#1)}
\@addtofilelist{#1}
\IfFileExists{#1}{}{\typeout{No file #1.}}
}\makeatother

\newcommand*{\myexternaldocument}[1]{%
\externaldocument{#1}%
\addFileDependency{#1.tex}%
\addFileDependency{#1.aux}%
}
\myexternaldocument{supplementary}

\newwrite\usedacronyms
\immediate\openout\usedacronyms=used-acronyms.tex

\let\oldgls\gls
\renewcommand{\gls}[1]{%
  \immediate\write\usedacronyms{\string\glsadd{#1}}%
  \oldgls{#1}%
}

\begin{document}

\bstctlcite{IEEEexample:BSTcontrol}

\title{Wi-Fi: Twenty-Five Years and Counting}

\author{
\IEEEauthorblockN{
\makebox[\textwidth][c]{Giovanni Geraci, Francesca Meneghello, Francesc Wilhelmi, David Lopez-Perez, Iñaki Val,}\\
\makebox[\textwidth][c]{Lorenzo Galati Giordano, Carlos Cordeiro, Monisha Ghosh, Edward Knightly, and Boris Bellalta}\vspace{-1cm}
}
\thanks{G. Geraci is with Nokia Standards and Universitat Pompeu Fabra, Spain. 
F. Meneghello is with Northeastern University, USA, and the University of Padova, Italy.
F. Wilhelmi and B. Bellalta are with Universitat Pompeu Fabra, Spain. 
D. Lopez-Perez is with Universitat Politècnica de València, Spain. 
I. Val is with MaxLinear, Spain. 
L. Galati Giordano is with Nokia Bell Labs, Germany. 
C. Cordeiro is with Intel Corporation, USA. 
M. Ghosh is with the University of Notre Dame, USA. 
E. Knightly is with Rice University, USA.
} 
\thanks{
This work was in part supported by the Spanish Research Agency through grants PID2021-123999OB-I00, PID2021-123995NB-I00, PID2024-156488OB-I00, CEX2021-001195-M, CNS2023-145384, and CNS2023-144333; by SGR 00955-2021 AGAUR; by ICREA Academia, by the European Union - Next Generation EU under the Italian National Recovery and Resilience Plan (NRRP), Mission 4, Component 2, Investment 1.2, CUP C93C24004880002, project ``CAMELIA''; by the Generalitat Valenciana, Spain, through grant CIDEXG/2022/17; by Horizon Europe SNS-JU through grant UNITY-6G 101192650; by Cisco; by Intel; and by NSF grants 2433923, 2402783, and 2211618.
}
\thanks{Last revised: December 20, 2025.}
}

\maketitle


\begin{abstract}
Today, Wi-Fi is over 25 years old. Yet, despite sharing the same branding name, today's Wi-Fi boasts entirely new capabilities that were not even on the roadmap 25 years ago.  This article aims to provide a holistic and comprehensive technical and historical tutorial on Wi-Fi, beginning with IEEE 802.11b (Wi-Fi 1) and looking forward to IEEE 802.11bn (\mbox{Wi-Fi} 8). This is the first tutorial article to span these eight generations. Rather than a generation-by-generation exposition, we describe the key mechanisms that have advanced Wi-Fi. We begin by discussing spectrum allocation and coexistence, and detailing the IEEE 802.11 standardization cycle. Second, we provide an overview of the physical layer and describe key elements that have enabled data rates to increase by over 1,000$\times$. Third, we describe how \mbox{Wi-Fi} Medium Access Control has been enhanced from the original Distributed Coordination Function to now include capabilities spanning from frame aggregation to wideband spectrum access. \blue{Fourth,} we describe how Wi-Fi 5 first broke the one-user-at-a-time paradigm and introduced multi-user access. \blue{Fifth,} given the increasing use of mobile, battery-powered devices, we describe Wi-Fi's energy-saving mechanisms over the generations. Sixth, we discuss how Wi-Fi was enhanced to seamlessly aggregate spectrum across 2.4\,GHz, 5\,GHz, and 6\,GHz bands to improve throughput, reliability, and latency. Finally, we describe how Wi-Fi enables nearby Access Points to coordinate in order to improve performance and efficiency. 
In the Appendix, we further discuss Wi-Fi developments beyond 802.11bn, including integrated mmWave operations, sensing, security and privacy extensions, and the adoption of AI/ML.\!
\end{abstract}


\setcounter{tocdepth}{2}

\begin{IEEEkeywords}
Wi-Fi, WLAN, IEEE 802.11, Wi-Fi Alliance, 802.11bn, unlicensed spectrum, PHY, MAC, multi-user technologies, energy savings, multi-link operation, multi-AP coordination.\vspace{-0.2cm}
\end{IEEEkeywords}

\setlength{\textfloatsep}{0.1cm}

\section{Introduction}
\label{sec:Introduction}

\blue{You have probably downloaded this article over Wi-Fi.} On September 15, 1999, 
the \gls{weca}---the future Wi-Fi Alliance~\cite{WiFiAlliance}---unveiled the term \mbox{Wi-Fi} as the consumer-facing brand for the \gls{ieee} 802.11 standard, ensuring that consumers never needed to learn to say ``I triple E eight oh two dot eleven.'' 
Last year marked the 25th anniversary of this transformative technology~\cite{Kasslin2024}, a milestone that underscores how deeply Wi-Fi has embedded itself into modern society. \blue{For Generation Z, the idea of a world without Wi-Fi is nearly unimaginable, and even millennials struggle to remember a time when a wired Ethernet cable was needed to get online \cite{berlinskyschine_genz_2020}.} Today, Wi-Fi is the most popular way to access the Internet across the globe. It is embedded in everything, from smartphones and laptops to televisions, smart appliances, and industrial sensors. With more than 21 billion Wi-Fi-enabled devices in use, and over 63\% of global Internet traffic flowing through Wi-Fi, its ubiquity is both a technological achievement and a societal cornerstone~\cite{WiFiEconomicValue2025}.

Few technologies have had such far-reaching and sustained impact. Wi-Fi is not just a convenience. It is an essential~utility that underpins the fabric of our digital lives.
It has revolutionized how we work, learn, communicate, and entertain ourselves. It powers online classrooms in rural areas, enables telemedicine in urban hospitals, supports logistics in global supply chains, and fuels social interaction in every corner of the world. During the COVID-19 pandemic, Wi-Fi was nothing short of indispensable~\cite{WiFiCoviWiFiCovid2020}: It enabled a rapid shift to remote work, kept students engaged in virtual classrooms, and allowed businesses to continue operating amid global disruption.\!

But the story of Wi-Fi is not only one of societal diffusion; it is also a narrative of relentless technological evolution. Since its inception in the late 1990s, when early versions of \gls{ieee} 802.11 offered data rates of just 1–2\,\blue{Mb/s} and basic wireless connectivity for laptops and \glspl{pda}, Wi-Fi has undergone continuous reinvention to keep pace with escalating user demands and emerging applications.


\blue{\subsection{Use Case Evolution: From Browsing to Real-Time Critical Applications}}

Wi-Fi’s early use cases centered on best-effort broadband access \cite{pahlavan2021evolution}, 
supporting basic activities such as web browsing, music streaming, and downloading files. 
As computing and communication increasingly shifted to the cloud and user expectations evolved, 
Wi-Fi was drawn into more demanding roles~\cite{IEEE_Wi-Fi_Evolution}. Today, home and enterprise networks require multi-gigabit transmission rates and low-latency performance to support 4K/8K video, real-time collaborative tools, and secure remote access. The rise of hybrid work and cloud-native productivity platforms has further elevated Wi-Fi to the status of a mission-critical utility.

Beyond these mainstream productivity and media demands,
Wi-Fi is now being stretched to meet the stringent performance requirements of a growing set of specialized, latency-sensitive, and mission-critical applications:
\begin{itemize}
    \item \Gls{ar}/\gls{vr} and immersive media applications require high throughput and tight latency bounds (<10\,ms) to deliver realistic, lag-free experiences. 
    \item Online gaming, particularly competitive and cloud-based platforms, demands sub-30\,ms latency and high reliability, making jitter and packet loss as detrimental as slow speeds.
    \item \Gls{iiot} and smart manufacturing are increasingly looking to Wi-Fi as a viable alternative to wired Ethernet and 5G \gls{urllc}. Real-time control loops, time-sensitive networking, and safety-critical functions require bounded latency, high availability, and deterministic behavior.
    \item 
    Healthcare spaces adopt Wi-Fi for everything from real-time patient monitoring to robotic-assisted surgery preparation. These applications cannot tolerate performance lapses.
    \item 
    Public infrastructure, including airports, smart cities, and connected public transportation systems, relies on Wi-Fi to deliver not only connectivity to users but also telemetry, surveillance, and machine coordination.
\end{itemize}

As these applications mature, the bar continues to rise. Beyond throughput and latency, predictability, reliability, and coordination are becoming new pillars of Wi-Fi performance.


\blue{\subsection{From Speed to Efficiency: The Generational March}}

\begin{table*}[t]
\centering
\caption{Evolution of IEEE standard amendments and associated Wi-Fi generations, use cases, key features, and peak data rates.}
\label{tab:wifi_evolution}
\colorbox{BackgroundGray}{
\begin{tabular}{p{2.62cm}|p{1.35cm}|p{0.65cm}|p{4cm}|p{5.25cm}|p{1.25cm}}
\toprule
\rowcolor{BackgroundLightBlue} \textbf{IEEE Standard} & \textbf{Generation} & \textbf{Year} & \textbf{Representative Use Cases} & \textbf{Key Features} & \textbf{Peak Rate} \\
\midrule
\textbf{802.11b} & --- & 1999 & Basic broadband: email, browsing, file sharing (home/small office) 
& 2.4\,GHz, DSSS & 11\,\blue{Mb/s} \\
\midrule

\textbf{802.11a} & --- & 1999 & Faster enterprise connections, early media streaming 
& 5\,GHz, OFDM & 54\,\blue{Mb/s} \\
\midrule

\textbf{802.11g} & --- & 2003 & Home media, consumer electronics integration 
& 2.4\,GHz, OFDM, backward compatible with 802.11b & 54\,\blue{Mb/s} \\
\midrule

\textbf{802.11n} \gls{ht} & Wi-Fi 4  & 2009 
& HD video, cloud access, multi-device homes 
& Up to 40\,MHz channels, packet aggregation, MIMO (up to 4×4) & 600\,\blue{Mb/s} \\
\midrule

\textbf{802.11ac} \gls{vht} & Wi-Fi 5 & 2013 
& 4K video, online gaming, mobile offload, enterprise WLANs 
& Up to 160\,MHz channels, DL MU-MIMO (up to 4×4), 256-QAM & 6.9\,\blue{Gb/s} \\
\midrule

\textbf{802.11ax} \blue{\gls{he}} & Wi-Fi 6/6E & 2021 
& Dense deployments (stadiums, campuses), IoT, teleconferencing 
& DL/UL OFDMA and MU-MIMO (up to 8×8), 1024-QAM, spatial reuse, TWT, 6\,GHz band & 9.6\,\blue{Gb/s}    \\
\midrule

\textbf{802.11be} \blue{\gls{eht}} & Wi-Fi 7 & 2024 
& 8K video, real-time collaboration, cloud gaming, AR/VR, IIoT pilot 
& Up to 320\,MHz channels, 4096-QAM, MLO, multi-RU, enhanced QoS & 23\,\blue{Gb/s} \\
\midrule

\textbf{802.11bn} \blue{\gls{uhr}} & Wi-Fi 8 & 2028+ 
& Robotic surgery prep, industrial automation, holography, ultra-reliable closed control loops 
& ELR PPDU, distributed-RU, LDPC enhancements, unequal modulation, seamless roaming, dynamic power save, NPCA, DSO, MAPC & \blue{23\,Gb/s}\\

\bottomrule
\end{tabular}
}\vspace{-10pt}
\end{table*}

Wi-Fi is based on the \gls{ieee} 802.11 standard, which sets the foundations of the \gls{phy} and \gls{mac} layer of this technology to be implemented by \gls{ap} and \gls{sta} devices (in the 802.11 specification, referred to as \gls{ap} and non-\gls{ap} devices, respectively). The 802.11 standard has significantly evolved in the last two and a half decades by following an iterative and backward-compatible approach, where each generation of Wi-Fi, marked by a new 802.11 amendment, has been introduced to address emerging technological and application needs. Early amendments (1990s-2000s) laid the groundwork for distributed channel access but employed relatively simple radio technologies for data transmissions. Over the years, newer 802.11 amendments gradually enriched the standard with features intended not only for higher capacity, but also for higher efficiency, more flexibility, and enhanced security. 
With this, Wi-Fi ensured support for new applications and devices with increasing communication requirements and, at the same time, fostered their proliferation.

Among some of the major breakthroughs in the history of Wi-Fi, we highlight: 802.11n (2009), 
which introduced packet aggregation and \gls{mimo}; 
802.11ac (2013) with its multi-user \gls{mimo} (MU-MIMO) capabilities and wider channels to reach higher peak data rates; 
802.11ax (2021), featuring \gls{ofdma} for scheduled uplink access, \gls{twt} for energy savings, and \gls{sr} for spectral efficiency; 802.11be (2024), which introduced \gls{mlo}. Today, with 802.11bn \blue{(expected in 2028)} on the horizon, Wi-Fi aims to add increased reliability to its wide portfolio of features.

This generational evolution is more than a series of speed upgrades. It represents a deepening of capabilities to support an increasingly complex and demanding ecosystem of applications and devices. More details on Wi-Fi generations and the standardization process are included in Section~\ref{sec:StandardizationCertificationSpectrum}.
To aid the reader, Table~\ref{tab:wifi_evolution} presents a summary of the key technological developments in each amendment and the use cases they target.


\subsection{The Road to Wi-Fi~8: Low Latency and High Reliability}

To support the next decade of wireless innovation, set by the previously highlighted use cases, Wi-Fi must expand into new domains including time-sensitive applications, highly-reliable control systems, and synchronized multi-device environments. This challenge is being taken up by the 802.11bn amendment \cite{reshef2022future}, which is developing the next-generation standard expected to underpin Wi-Fi~8.

\blue{The primary objective of 802.11be was to increase capacity and link throughput and also improve worst-case latency and jitter with at least one mode of operation. While the latter was a novel endeavor compared to previous Wi-Fi generations, target latency and jitter were not quantified, making this only an initial step towards reliability.} 
802.11bn builds upon 802.11be foundational features, while making low-latency and high reliability first-class citizens. This shift is more than incremental, it represents a paradigm change and an architectural evolution in how Wi-Fi is designed, scheduled, and coordinated~\cite{GalGerCar2024} \cite{liu2024wi}. To achieve such targets,
key innovations agreed in the 802.11bn draft (D1.1) include~\cite{80211bnD1.1}:
\begin{itemize} 
    \item \Gls{ldpc} enhancements: 
    Upgrading the error correction capabilities to support higher reliability under poor channel conditions or for critical data streams (see Section~\ref{subsec:transmitter_block}).
    \item \Gls{ueqm}: 
    Allowing different modulation orders per spatial stream, optimizing spectral efficiency in multi-antenna systems with heterogeneous link qualities (see Section~\ref{sec:mcs}).
    \item \Gls{elr} \glspl{ppdu}: 
    Extending coverage and improving link robustness for long-distance transmissions, especially in sparse or high-interference environments (see Section~\ref{subsec:ppdus}).
    \item \Gls{npca}: Allowing stations and \glspl{ap} to initiate transmissions over secondary or tertiary channels, increasing flexibility in channel utilization and reducing contention (see Section~\ref{sec:wideband_op}).
    \item \Gls{dso}: Providing very high flexibility to narrow-band \glspl{sta} (e.g., using 20\,MHz channels) by allowing them to be allocated to any subchannels among the \gls{ap}’s available bandwidth (see Section~\ref{sec:wideband_op}).
    \item \Glspl{dru}: 
    Enhancing power distribution flexibility in \gls{psd}-limited regulatory regimes, enabling improved energy concentration and transmission robustness (see Section~\ref{sec:OFDMA}).
    \item \Gls{dps}: 
    Enabling finer-grained power management strategies that adapt to application needs, helping extend the battery life of constrained \gls{iot} and mobile devices (see Section~\ref{sec:dynamic_power_save}).
    \item Seamless roaming: 
    Introducing mechanisms for fast and non-disruptive handovers across \glspl{ap} to support mobility-sensitive applications such as \gls{ar}/\gls{vr} and real-time industrial control (see Section~\ref{subsec:seamless_roaming}).
    \item \Gls{mapc}: Moving beyond traditional mesh or roaming architectures, to support coordinated scheduling across \glspl{ap}. This will enable more deterministic performance and spatial reuse in dense deployments and may form the cornerstone for network-wide latency and reliability guarantees (see Section~\ref{sec:MAPC}).
\end{itemize}

\subsection{Other Related Surveys and Tutorials}


\begin{table}
\vspace{1mm}
\caption{Other relevant articles on Wi-Fi.} 
\label{tab:otherTutorials}
\centering
\colorbox{BackgroundGray}{%
\begin{tabular}{ |m{4cm}|m{4.1cm}|} 
\toprule
\rowcolor{BackgroundLightBlue}
 \textbf{802.11 for WLAN (core)} & \textbf{Selected references}
 \\ \midrule
 {Surveys and tutorials} &  
{\cite{karmakar2017impact},
\cite{chen2017coex},
\cite{bellalta2016next},
\cite{khorov2018tutorial},
\cite{mozaffariahrar2022survey},
\cite{khorov2020current},
\cite{deng2020ieee},
\cite{verma2024multiap},
\cite{szott2022wi}}   
\\ \midrule
 {802.11bn (Wi-Fi 8)} &  
{\cite{reshef2022future}, 
\cite{GalGerCar2024}, 
\cite{liu2024wi}, 
\cite{oughton2024future}}   
\\ \midrule
 {802.11be (Wi-Fi 7)} &  
{\cite{lopez2019ieee}, 
\cite{khorov2020current},
\cite{deng2020ieee}, 
\cite{yang2020survey}, 
\cite{garcia2021ieee}, 
\cite{liu2023ieee}, 
\cite{liu2023first}, 
\cite{verma2024multiap},
\cite{murad2024introduction}, 
\cite{oughton2024future}}   
\\ \midrule
 {802.11ax (Wi-Fi 6)} &  
{\cite{afaqui2016ieee},
\cite{bellalta2016next},
\cite{khorov2018tutorial}, 
\cite{qu2019survey},
\cite{yang2021mac},
\cite{mozaffariahrar2022survey}}   
\\ \midrule
 {Older 802.11 for WLAN standards} &  
{\cite{karmakar2017impact}}   
\\ \bottomrule 
\toprule
\rowcolor{BackgroundLightBlue}
 \textbf{802.11 for other use cases} & \textbf{Selected references}   \\ \midrule
 {Wi-Fi coexistence (Coex SC)} &  
{\cite{chen2017coex},
\cite{wang2017survey},
\cite{zinno2018fair},
\cite{naik2020next}}  
\\ \midrule
 {802.11ah (Wi-Fi HaLow)} &  
{\cite{khorov2015survey},
\cite{meera2017survey}}   
\\ \midrule
 {802.11s (Wi-Fi multihop)} &  
{\cite{carrano2010ieee}}   
\\ \midrule
 {802.11ad/ay (WiGig/mmWave)} &  
{\cite{zhou2018ieee}, 
\cite{tarafder2022mac}, 
\cite{koda2024survey}}
\\ \bottomrule
\toprule
\rowcolor{BackgroundLightBlue}
 \textbf{Future of 802.11} & \textbf{Selected references}   \\ \midrule
 {Wi-Fi sensing (TGbf) and positioning (TGbk)} &  
{\cite{khalili2020wi},
\cite{singh2021machine}, 
\cite{chen2022wi}, 
\cite{meneghello2023toward},
\cite{dai2023survey}, 
\cite{du2024overview},
\cite{wang2025survey}}  
\\ \midrule
{Integrated mmWave (TGbq)} &  
{\cite{liu2024wi}}  
\\ \midrule
 {Artificial intelligence and machine learning (AIML SC)} &  
{\cite{szott2022wi},
\cite{wilhelmi2024machine}}  
\\ \bottomrule
\end{tabular}
}
\end{table}


Wi-Fi has evolved significantly over the past two decades, with each generation introducing key innovations to address growing performance demands and increasingly diverse use cases. In parallel, a number of surveys and tutorials have emerged, each offering different perspectives on the evolution of the \gls{ieee} 802.11 standard. In Table~\ref{tab:otherTutorials}, we report a selection of the most relevant and representative review articles from the literature. Despite these valuable articles, the literature currently lacks a comprehensive, tutorial-style account that traces Wi-Fi's full evolution from 802.11b to the emerging 802.11bn, Wi-Fi~8. Existing works tend to focus on specific generations or technical layers, often omitting broader system-level perspectives and tutorial-level explanations. Crucially, no survey to date covers the final set of 802.11be features and 802.11bn overall. This article fills the gap by providing an integrated, accessible guide to the historical and technical development of Wi-Fi, offering both a foundation for new researchers and a forward-looking reference for ongoing work.

\commentsection
\subsection{Other Related Surveys and Tutorials\blue{[David] [FrancescW]}}

Wi-Fi has evolved significantly over the past two decades, with each generation introducing key innovations to address growing performance demands and increasingly diverse use cases, as shown earlier. In parallel, a number of surveys and tutorials have emerged,
each offering different perspectives on the evolution of the \gls{ieee} 802.11 standard. In the following, we review a selection of the most relevant and representative survey papers from the literature, organized chronologically by the Wi-Fi generation they primarily address (see Table~\ref{tab:survey-summary} for a summary). While these form the core of our analysis, other notable works have also been identified and are classified alongside them in Table~\ref{tab:otherTutorials}. 
We highlight the scope and contributions of each paper, as well as their limitations---particularly the lack of a unified, tutorial-style reference that spans the complete evolution of Wi-Fi up to the emerging 802.11bn standard amendment.

\commentfigtable
\begin{table*}[ht]
\centering
\caption{Summary of Key Survey and Tutorial Papers on IEEE 802.11}
\label{tab:survey-summary}
\begin{tabular}{p{4.2cm} p{1.5cm} p{3.6cm} p{3.8cm} p{1.5cm}}
\toprule
\textbf{Reference} & \textbf{Gen(s). Covered} & \textbf{Main Focus} & \textbf{Final Feature Coverage} & \textbf{Wi-Fi 8} \\
\midrule
Karmakar et al. (2017) \cite{karmakar2017impact} 
& 11n/ac 
& PHY/MAC impact on transport/application layers 
& Yes — Covers finalized 802.11n/ac features and their cross-layer effects
& No 
\\
\addlinespace[0.3em]
Chen et al. (2017) \cite{chen2017coex} 
& 11ac 
& Wi-Fi/\gls{lte}-\gls{laa} coexistence in 5\,GHz 
& No — Focused on coexistence protocols, not core Wi-Fi standard features 
& No 
\\
\addlinespace[0.3em]
Bellalta et al. (2016) \cite{bellalta2016next}
& 11ac–11ax 
& Trend and use-case analysis across amendments, covering early 802.11ax intentions  
& No — Conceptual overview across generations; does not focus on finalized 802.11ax features
& No
\\
\addlinespace[0.3em]
Khorov et al. (2018) \cite{khorov2018tutorial} 
& 11ax 
& Detailed technical overview and early tutorial on potential Wi-Fi~6 features 
& No — Based on early drafts; several features evolved or changed in final 802.11ax
& No 
\\
\addlinespace[0.3em]
Mozaffariahrar et al. (2022) \cite{mozaffariahrar2022survey} 
& 11ax 
& Comprehensive feature review and evaluation tools for Wi-Fi~6 
& Yes — Aligned with finalized 802.11ax amendment (Wi-Fi~6) 
& No
\\
\addlinespace[0.3em]
Khorov et al. (2020) \cite{khorov2020current}
& 11be 
& Tutorial on proposed Wi-Fi~7 features 
& No — Based on early \gls{ieee} 802.11be proposals before any draft existed; 
several proposed features were later dropped
& No 
\\
\addlinespace[0.3em]
Deng et al. (2020) \cite{deng2020ieee} 
& 11be 
& Feature summary and challenges in Wi-Fi~7 implementation
& No — Reviews draft-phase features; includes speculative and evolving proposals
& No \\
\addlinespace[0.3em]
Verma et al. (2024) \cite{verma2024multiap} 
& 11be 
& \gls{mapc} in dense deployments
& Partial — Focuses on \gls{mapc}, which was not finally addressed in 802.11be 
& No
\\
\addlinespace[0.3em]
Szott et al. (2022) \cite{szott2022wi} 
& 11ax–11be
& \Gls{ml} for Wi-Fi optimization (config., scheduling, etc.) 
& No — \gls{ml}-focused; indirectly covers some features of 802.11ax/11be but not the amendments themselves
& No 
\\
\bottomrule
\end{tabular}
\end{table*}
\endcommentfigtable

\subsubsection{Surveys on pre Wi-Fi~6}

The survey in \cite{karmakar2017impact} is one of the earliest survey papers on Wi-Fi, focusing on the high-throughput enhancements introduced in 802.11n and 802.11ac. This work uniquely examined the impact of \gls{phy} and \gls{mac} improvements---such as frame aggregation, channel bonding, and \gls{mimo}---on upper-layer protocols, including \gls{tcp} and \gls{udp}. It highlighted how \gls{phy}/\gls{mac} gains do not always translate to application-layer performance improvements, and emphasized the need for better cross-layer optimization. This upper-layer focus distinguishes it from more standard-centric surveys,
although it does not extend beyond Wi-Fi~5.

An early contribution to the coexistence discourse came from a survey in \cite{chen2017coex}, focused on the interaction between Wi-Fi and \gls{lte}-\gls{laa} in the 5\,GHz unlicensed band. Rather than providing a tutorial on Wi-Fi itself, the paper examined how different coexistence mechanisms---such as \gls{csat} and \gls{lbt}---operated under regional regulatory frameworks.
It also analyzed various deployment scenarios, and highlighted performance trade-offs between Wi-Fi and \gls{lte}-based systems. While not centered on standard evolution, it remained a valuable reference for understanding coexistence concerns that continued to affect Wi-Fi~6 and later generations.

\subsubsection{Surveys on Wi-Fi~6 (802.11ax)}

The survey in \cite{bellalta2016next} examined the trajectory of IEEE 802.11 amendments, including 802.11ac, 802.11ad, 802.11af, 802.11ah, and 802.11ax. 
Anchored around emerging use cases such as multimedia streaming, \gls{iot}, and spectrum sharing, the paper reviewed candidate technologies,
including packet aggregation, channel bonding, downlink and uplink \gls{mu}-\gls{mimo}, full-duplex wireless channels, and efficient transmission of small data packets. Notably, this survey was conducted during the early standardization phase of 802.11ax, prior to the finalization of its feature set. As such, while it provided valuable insights into anticipated trends and technical enablers,
it does not capture the complete set of capabilities that were eventually standardized in Wi-Fi 6.


The detailed tutorial in \cite{khorov2018tutorial},
also published during the development of 802.11ax, introduced the motivations behind Wi-Fi~6, and presented a comprehensive technical overview of its defining features.
The authors explained how \gls{ofdma}, downlink and uplink \gls{mu}-\gls{mimo}, spatial reuse, and improved power-saving mechanisms were designed to improve efficiency in dense deployments.
The tutorial also covered the amended \gls{phy} and \gls{mac} frame structures,
and included early performance evaluations. As one of the first structured analyses of Wi-Fi~6,
this work remained a foundational reference for understanding the standard’s architectural shift toward high-efficiency \glspl{wlan}.

The most recent and comprehensive survey on Wi-Fi~6 is based on the final version of the 802.11ax standard \cite{mozaffariahrar2022survey}.
It combines tutorial explanations of each major feature---\gls{ofdma}, downlink and uplink \gls{mu}-\gls{mimo}, 1024-\gls{qam}, spatial reuse, \gls{twt}, and ---with an extensive literature review organized by functionality. In addition to protocol features, the paper covers implementation aspects such as random access, scheduling, user selection, channel estimation, and rate control. It also surveys the simulation and evaluation tools commonly used in Wi-Fi~6 research, making it a valuable reference for both researchers and practitioners. While its breadth and structure make it the most complete and up-to-date reference on Wi-Fi~6 to date, it does not address the emerging developments in Wi-Fi~7 or Wi-Fi~8.

\subsubsection{Surveys on Wi-Fi~7 (802.11be)}

The early tutorial on 802.11be in \cite{khorov2020current} provided a structured overview of the proposed enhancements for Wi-Fi~7,
emphasizing its goal to support \gls{eht} and real-time applications. 
The authors described key features that were under consideration by the Task Group at the time,
including 320\,MHz channel bandwidths, support for 4096-\gls{qam}, \gls{mlo}, scalable channel sounding for \gls{mimo}, and \gls{mapc}. 
The paper also discussed potential directions such as implicit \gls{csi} acquisition, enhanced \gls{mac} aggregation, and support for \gls{tsn}.
While the work predated the first official draft of the standard,
it synthesized the proposals then under discussion and identified open technical challenges, 
serving as a reference point for early Wi-Fi~7 research.
However, many of the presented features were ultimately not developed in the final amendment.

The survey in \cite{deng2020ieee}, 
published in the same year, 
also focused on both \gls{phy} and \gls{mac} innovations discussed within the 802.11be Task Group.
The authors described how features such as wider channelization, multi-\gls{ru} allocation, \gls{harq}, \gls{mlo}, and enhanced \gls{mimo} were envisioned to support high-throughput and low-latency applications such as \gls{vr}/\gls{ar} and 8K video.
The paper reviewed challenges related to backward compatibility, \gls{csi} acquisition overhead, and efficient scheduling across multiple frequency bands.
Importantly, 
this survey combined technical feature descriptions with early academic and industry perspectives,
aiming to bridge the gap between standardization proposals and implementation realities.
However, as with the previous work, 
several of the discussed features did not make it into the final version of the amendment.

A more recent and domain-specific survey in \cite{verma2024multiap} examined \gls{mapc}, which emerged as a critical element in both Wi-Fi~7 and future \glspl{wlan}. The authors categorized coordination mechanisms, including joint transmission, coordinated scheduling, and dynamic association, and distinguished between centralized and distributed control approaches. The survey also reviewed channel sounding techniques, synchronization strategies, and practical deployment considerations. While \gls{mapc} was not finalized in Wi-Fi~7, the paper made clear that such techniques would be essential for scaling future \glspl{wlan} in dense and high-interference environments,
particularly as spatial reuse and coordinated scheduling became more prominent.

Finally, we highlight the survey in \cite{szott2022wi}, which explored how \gls{ml} techniques can be applied to improve 802.11 performance across multiple protocol layers and use cases. Although not limited to Wi-Fi~7, the paper covered research areas particularly relevant to next-generation \glspl{wlan}, including learning-based channel selection, interference mitigation, dynamic parameter tuning, and multi-band operation. The authors reviewed a large number of papers, and classified them by use case and \gls{ml} approach, with many works targeting performance issues in dense deployments and multi-link scenarios---both central to Wi-Fi~7. This survey filled a distinct gap by addressing the algorithmic complexity introduced by modern \glspl{wlan},
and argued for the integration of \gls{ai}-driven control into future standards.

\bigskip

Despite numerous valuable surveys/tutorials, the literature currently lacks a comprehensive, tutorial-style account that traces Wi-Fi’s full evolution from 802.11b to the emerging 802.11bn, Wi-Fi~8. Existing works tend to focus on specific generations or technical layers, often omitting broader system-level perspectives and tutorial-level explanations. Crucially, no survey to date covers the final set of Wi-Fi~7 features and Wi-Fi~8 overall. 
This paper addresses that gap by providing an integrated, accessible guide to the historical and technical development of Wi-Fi, offering both a foundation for new researchers and a forward-looking reference for ongoing work in the field.
\endcommentsection


\subsection{Contribution and Outline}
\label{sec:outline}

This article provides a comprehensive tutorial on the key innovations and challenges across Wi-Fi generations from the first release of the 802.11 to the anticipated 802.11bn (Wi-Fi~8). We aim to serve both researchers and practitioners by highlighting how Wi-Fi has adapted to meet growing demands for throughput, latency, reliability, and multi-device coordination. 
We deliberately use original illustrations to give the main concepts in Wi-Fi technology a clear visual identity. Through carefully designed figures and diagrams, our tutorial consolidates knowledge across standards, use cases, and research trends, and aims to position readers to understand and contribute to future developments in \gls{wlan}.

The primary contributions of this article are as follows:
\begin{itemize}
    \item 
    We detail the 802.11 evolution, focusing on \gls{phy} and \gls{mac} layer innovations and their corresponding use cases, from the early amendments to the upcoming 802.11bn.
    \item 
    We provide a tutorial overview of the core technologies introduced in 802.11ax/be (Wi-Fi~6/6E/7), including \gls{ofdma}, \gls{mu}-\gls{mimo}, and \gls{mlo}, supported by examples and simplified explanations for accessibility.
    \item 
    We examine emerging features in 802.11bn such as \gls{mapc} 
    and many others, and discuss their expected impact on future \gls{wlan} applications.
    \commentsection
    \item 
    We include system-level insights and performance comparisons, drawing on simulations and prior literature to illustrate the practical implications of new features.
    \endcommentsection
\end{itemize}

The remainder of this paper is organized as follows:
\begin{itemize}
    \item
    {Section~\ref{sec:StandardizationCertificationSpectrum}} provides an overview of the Wi-Fi spectrum allocation and standardization process, and summarizes the key features introduced in each amendment.
    \item 
    {Section~\ref{sec:PHY}} delves into \gls{phy} foundations and innovations, including symbol structures, modulation formats, and frame types.
    \item 
    {Section~\ref{sec:MAC}} discusses \gls{mac}-layer techniques including \gls{dcf} operation, channel bonding, \gls{ru}-based access, and power management, \blue{thereby setting the stage for advanced multi-user and coordination mechanisms.}
    \item 
    {Section~\ref{sec:Multiuser}} \blue{builds on these \gls{phy}/\gls{mac} foundations and} analyzes multi-user techniques such as \gls{ofdma} and \gls{mu}-\gls{mimo}, 
    detailing their design rationale and performance impacts.
    \item 
    {Section~\ref{sec:energy}} \blue{then describes Wi-Fi’s energy-saving mechanisms adopted over the generations, and discusses how they interact with the previously introduced multi-user features.}
    \item {Section~\ref{sec:MLO}} \blue{builds on the single-link operation of earlier sections and} explores \gls{mlo}, 
    with a focus on link aggregation strategies and their role in reducing latency and boosting throughput.
    \item 
    {Section~\ref{sec:MAPC}} \blue{finally surveys \gls{mapc} mechanisms, which extend the single-\gls{ap} and multi-link concepts to coordinated operation among multiple \glspl{ap}, covering schemes from distributed scheduling to centralized coordination and discussing trade-offs and open challenges.}
    \item 
    {Section~\ref{sec:conclusion}} concludes the paper.
\end{itemize}

\blue{
\subsubsection*{Appendix} Building on the advancements envisioned for 802.11bn, the Appendix presents parallel efforts shaping the future of \mbox{Wi-Fi}, including mmWave operations (Appendix~\ref{sec:11bq}), sensing (Appendix~\ref{sec:11bf}), security and privacy (Appendix~\ref{sec:bh_bi}), and the integration of \gls{ai} and \gls{ml} techniques (Appendix~\ref{sec:AIML}). 
}

A summary of the main acronyms used in this tutorial is provided in Table~\ref{tab:acronyms_main}.

\begin{table}[h!]
    \caption{List of the main acronyms used throughout this article.}
    \label{tab:acronyms_main}
    \centering
    \colorbox{BackgroundGray}{%
    \begin{tabular}{l|l}
    \toprule 
    \rowcolor{BackgroundLightBlue}
    \shortstack{\textbf{Acronym}} 
    & \shortstack{\textbf{Definition}} \\ \midrule 
\acrshort{ac} & \acrlong{ac}\\
\rowcolor{BackgroundLightBlue}
\acrshort{ack} & \acrlong{ack}\\
\acrshort{amsdu} & \acrlong{amsdu}\\
\rowcolor{BackgroundLightBlue}
\acrshort{ampdu} & \acrlong{ampdu}\\
\acrshort{ap} & \acrlong{ap}\\
\rowcolor{BackgroundLightBlue}
\acrshort{ba} & \acrlong{ba}\\
\acrshort{bss} & \acrlong{bss}\\
\rowcolor{BackgroundLightBlue}
\acrshort{ca} & \acrlong{ca}\\
\acrshort{cbf} & \acrlong{cbf}\\
\rowcolor{BackgroundLightBlue}
\acrshort{cca} & \acrlong{cca}\\
\acrshort{cfr} & \acrlong{cfr}\\
\rowcolor{BackgroundLightBlue}
\acrshort{csr} & \acrlong{csr}\\
\acrshort{csma} & \acrlong{csma}\\
\rowcolor{BackgroundLightBlue}
\acrshort{cw} & \acrlong{cw}\\
\acrshort{dcf} & \acrlong{dcf}\\
\rowcolor{BackgroundLightBlue}
\acrshort{difs} & \acrlong{difs}\\
\acrshort{dl} & \acrlong{dl}\\
\rowcolor{BackgroundLightBlue}
\acrshort{dru} & \acrlong{dru}\\
\acrshort{dso} & \acrlong{dso}\\
\rowcolor{BackgroundLightBlue}
\acrshort{dsss} & \acrlong{dsss}\\
\acrshort{edca} & \acrlong{edca}\\
\rowcolor{BackgroundLightBlue}
\acrshort{eht} & \acrlong{eht}\\
\acrshort{eifs} & \acrlong{eifs}\\
\rowcolor{BackgroundLightBlue}
\acrshort{elr} & \acrlong{elr}\\
\acrshort{icf} & \acrlong{icf}\\
\rowcolor{BackgroundLightBlue}
\acrshort{ieee} & \acrlong{ieee}\\
\acrshort{isi} & \acrlong{isi}\\
\rowcolor{BackgroundLightBlue}
\acrshort{ltf} & \acrlong{ltf}\\
\acrshort{mac} & \acrlong{mac}\\
\rowcolor{BackgroundLightBlue}
\acrshort{mapc} & \acrlong{mapc}\\
\acrshort{mcs} & \acrlong{mcs}\\
\rowcolor{BackgroundLightBlue}
\acrshort{mimo} & \acrlong{mimo}\\
\acrshort{mld} & \acrlong{mld}\\
\rowcolor{BackgroundLightBlue}
\acrshort{mlo} & \acrlong{mlo}\\
\acrshort{mpdu} & \acrlong{mpdu}\\
\rowcolor{BackgroundLightBlue}
\acrshort{mu} & \acrlong{mu}\\
\acrshort{nav} & \acrlong{nav}\\
\rowcolor{BackgroundLightBlue}
\acrshort{npca} & \acrlong{npca}\\
\acrshort{obss} & \acrlong{obss}\\
\rowcolor{BackgroundLightBlue}
\acrshort{ofdm} & \acrlong{ofdm}\\
\acrshort{ofdma} & \acrlong{ofdma}\\
\rowcolor{BackgroundLightBlue}
\acrshort{pch} & \acrlong{pch}\\
\acrshort{phy} & \acrlong{phy}\\
\rowcolor{BackgroundLightBlue}
\acrshort{ppdu} & \acrlong{ppdu}\\
\acrshort{rf} & \acrlong{rf}\\
\rowcolor{BackgroundLightBlue}
\acrshort{rts} & \acrlong{rts}\\
\acrshort{ru} & \acrlong{ru}\\
\rowcolor{BackgroundLightBlue}
\acrshort{sch} & \acrlong{sch}\\
\acrshort{sig} & \acrlong{sig}\\
\rowcolor{BackgroundLightBlue}
\acrshort{sifs} & \acrlong{sifs}\\
\acrshort{snr} & \acrlong{snr}\\
\rowcolor{BackgroundLightBlue}
\acrshort{sp} & \acrlong{sp}\\
\acrshort{sr} & \acrlong{sr}\\
\rowcolor{BackgroundLightBlue}
\acrshort{sta} & \acrlong{sta}\\
\acrshort{stf} & \acrlong{stf}\\
\rowcolor{BackgroundLightBlue}
\acrshort{tb} & \acrlong{tb}\\
\acrshort{twt} & \acrlong{twt}\\
\rowcolor{BackgroundLightBlue}
\acrshort{txop} & \acrlong{txop}\\
\acrshort{ueqm} & \acrlong{ueqm}\\
\rowcolor{BackgroundLightBlue}
\acrshort{uhr} & \acrlong{uhr}\\
\acrshort{ul} & \acrlong{ul}\\
\rowcolor{BackgroundLightBlue}
\acrshort{wlan} & \acrlong{wlan}\\
    \bottomrule 
    \end{tabular}
    }
\vspace{-2pt}
\end{table}



\section{Wi-Fi Spectrum Allocation, Standardization and Certification}
\label{sec:StandardizationCertificationSpectrum}

\begin{figure}
\centering
\includegraphics[width=\figwidth]{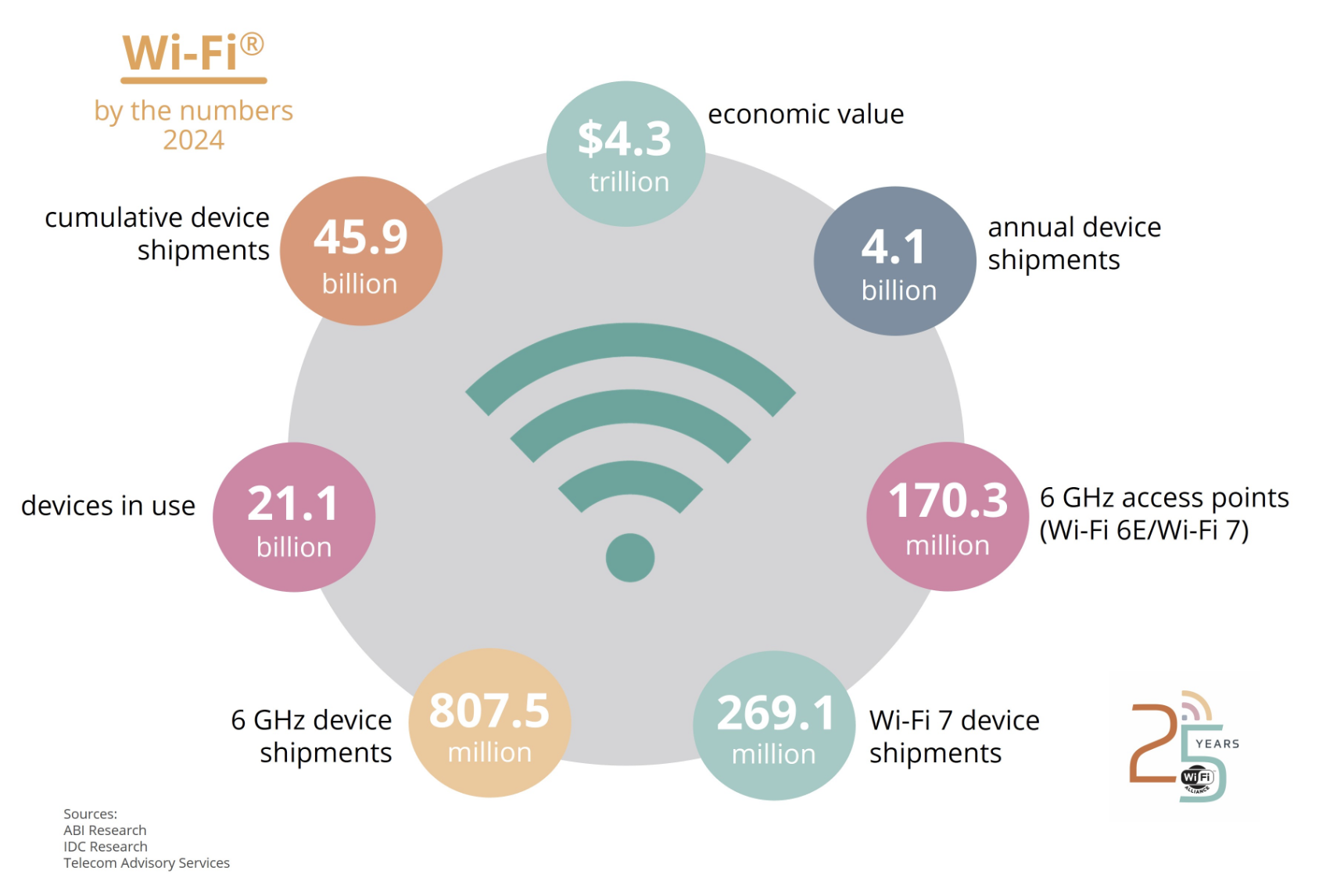}
\caption{Wi-Fi by the numbers in 2024 (source: \gls{wfa}).}
\label{fig:TheValueOfWi-Fi}
\end{figure}

\commentfigtable
as depicted in Fig.~\ref{fig:TheValueOfWi-Fi}.

\endcommentfigtable

\blue{As a cornerstone of today's digital ecosystem, Wi-Fi operates in unlicensed spectrum, enabling ease of deployment and free and widespread access across the globe.} Among other things, Wi-Fi's widespread success can be attributed to the collaborative efforts of key industry stakeholders, industry forums such as \gls{wfa}, and standards organizations, most notably the \gls{ieee}, which have continuously developed and refined the 802.11 family of standards and associated \gls{wfa}'s certification programs. These standards specify the \gls{phy} and \gls{mac} protocols that underpin Wi-Fi, ensuring interoperability and robust performance across a diverse range of devices and environments.
This section provides an overview of the Wi-Fi standardization cycle, from the latest advances in unlicensed spectrum allocation and regulation to the development of IEEE 802.11 standards and the subsequent Wi-Fi certification process led by \gls{wfa}.


\subsection{Spectrum Allocation and Coexistence}

Spectrum allocation has been a key driver of Wi-Fi's success, enabling it to operate in unlicensed bands, primarily the 2.4\,GHz and 5\,GHz bands, which are allocated worldwide for \gls{ism} applications. \blue{There are only three non-overlapping 20~MHz channels in 2.4~GHz and a single non-overlapping 40~MHz channel spanning the available bandwidth of approximately 80~MHz, severely limiting high-throughput applications. The 5 GHz bands offer wider bandwidths and more channels designated in the U.S. into four \gls{unii} bands as follows: \gls{unii}-1 (5.15 - 5.25 GHz), \gls{unii}-2A (5.25 - 5.35 GHz), \gls{unii}-2C (5.47 - 5.725 GHz), \gls{unii}-3 (5.725 - 5.85 GHz)  and \gls{unii}-4 (5.85 - 5.895 GHz). The \gls{unii}-2A and \gls{unii}-2C bands have to implement \gls{dfs} to protect airport radars that operate in those bands, and hence are sparsely used.}

As Wi-Fi has evolved to meet growing demands for high-speed applications such as cloud storage, video streaming, and real-time communications, these bands have become increasingly congested. Even though 802.11ax brought improvements in spectrum efficiency, high-density scenarios still face significant challenges due to spectrum congestion in 5\,GHz\blue{~\cite{barrachina2021wi}}. 

The introduction of Wi-Fi 6E (based on 802.11ax) in the unlicensed 6\,GHz band marked a major advancement, offering up to 1.2\,GHz of additional spectrum in some countries. This new band provides significant relief from congestion, allowing for up to seven 160\,MHz channels, as compared to only two in the 5\,GHz band. Wi-Fi 6E also introduced tri-band \glspl{ap} that simultaneously support the 2.4\,GHz, 5\,GHz, and 6\,GHz bands, enabling devices to operate in the frequency band with the best available quality at a given time. The latest 802.11be amendment, with \gls{mlo} and 320\,MHz channels, further enhances this by allowing devices to connect over two or more bands simultaneously, improving overall capacity, reducing latency, and boosting reliability. The 6\,GHz band is partitioned into four \gls{unii} bands as follows: \gls{unii}-5 from 5.925--6.425\,GHz, \gls{unii}-6 from 6.425--6.525\,GHz, \gls{unii}-7 from 6.525--6.875\,GHz and \gls{unii}-8 from 6.875--7.125\,GHz. Most regulatory regions around the globe have opened either all these bands for unlicensed operation or only \gls{unii}-5~\cite{WiFi_LPI}. 

\commentfigtable
\begin{figure}
\centering
\includegraphics[width=\figwidth]{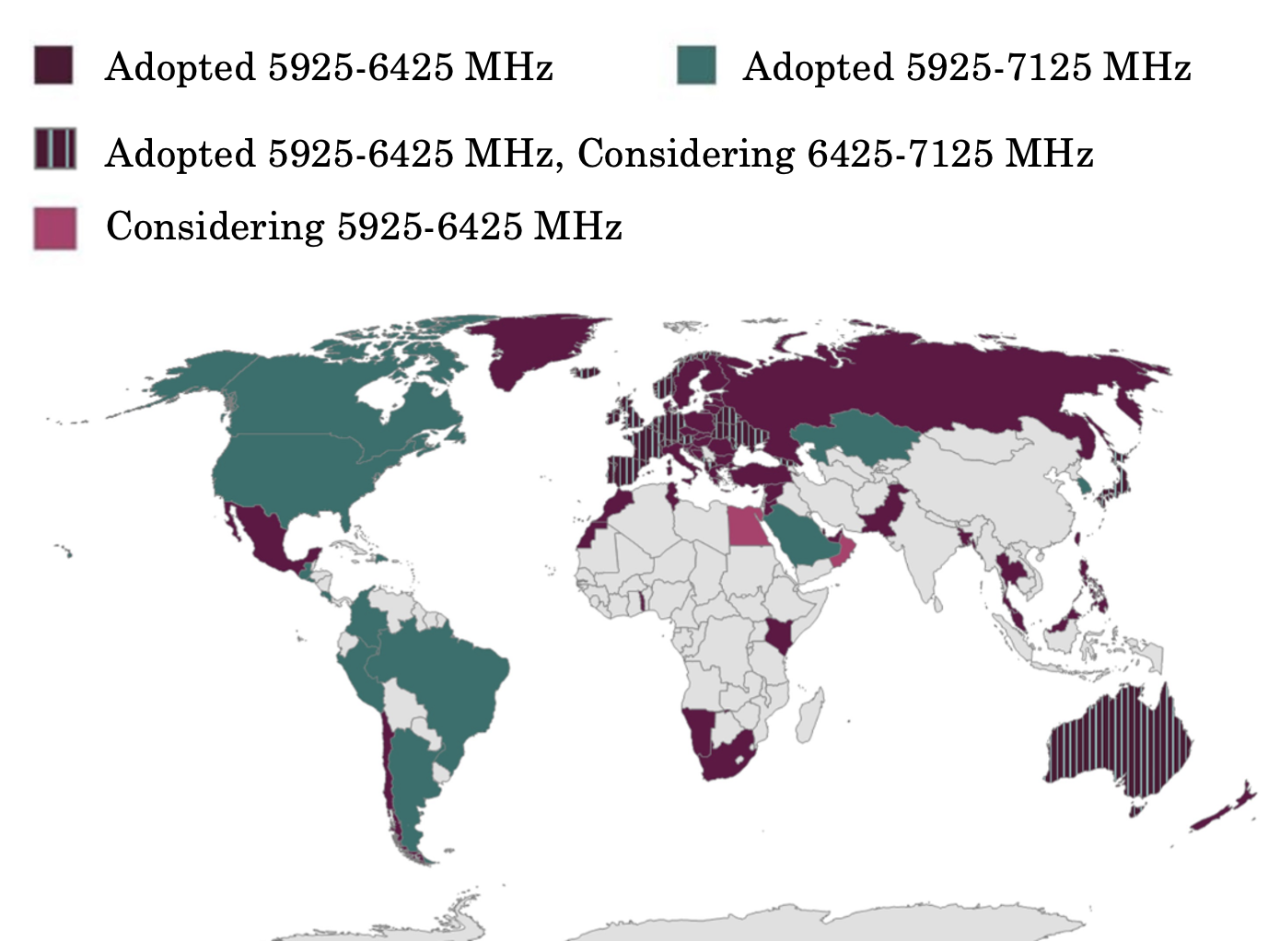}
\caption{Worldwide regulations enabling 6 GHz Wi-Fi (source: \gls{wfa}).}
\label{fig:6GHzAdoption}
\end{figure}
\endcommentfigtable


\subsubsection{The 6 GHz Unlicensed Band Allocation}

The 6\,GHz band is a shared but unlicensed band. The primary incumbents in \gls{unii}-5 and \gls{unii}-7 are fixed links while \gls{bas} are the main incumbent in \gls{unii}-6 and \gls{unii}-8. In addition, satellite uplink and downlink allocations exist across the band. Allocating the band on a low-power unlicensed basis instead of high-power, exclusively licensed basis permits these incumbents to continue operating without interference. For Wi-Fi, the 6\,GHz band is particularly valuable because it is free from contention with and interference by legacy Wi-Fi devices operating in the 2.4\,GHz and 5\,GHz bands. This spectrum expansion is critical for accommodating the exponential increase in demand for Wi-Fi capacity. Following the US \gls{fcc} decision in 2020 to open the 6\,GHz band for unlicensed operation, allocation of the 6\,GHz band has become a focal point of regulatory efforts worldwide, particularly during the World Radiocommunication Conference 2023 (WRC-23). One of the most debated issues is the future of the upper 6\,GHz band (6425--7125\,MHz). Regulatory bodies across the three \gls{itu} regions are evaluating how to balance its use between Wi-Fi and future 6G applications:
\begin{itemize}
    \item \emph{Region 1:} Covering Europe, the Middle East, Africa, and the CIS, with regulators such as CEPT, ASMG, ATU, and RCC.
    \item \emph{Region 2:} The Americas, regulated by CITEL.
    \item \emph{Region 3:} The Asia-Pacific region, regulated by APT.
\end{itemize}

Early studies suggest the upper 6\,GHz band shows promise for extending indoor cellular coverage, but coexistence with Wi-Fi remains a challenge, particularly regarding interference. Coexistence of indoor and outdoor Wi-Fi in a dense stadium deployment has been studied in~\cite{dogantusha2025evaluationindooroutdoorsharingunlicensed}, demonstrating that coexistence even within Wi-Fi systems may be a challenge. While reducing the power output of cellular systems might mitigate interference with Wi-Fi, this would compromise the indoor penetration and overall effectiveness of cellular networks. In addition to this \gls{wlan}-cellular spectrum sharing challenge, the 6\,GHz band already supports several incumbent services such as satellite communications, fixed links, and passive Earth observation systems. Effective coordination between these services and new Wi-Fi deployments are critical to successful spectrum sharing in this band~\cite{dogan2023evaluating, dougan2025spectrum}.


\subsubsection{6\,GHz Band Operating Rules}\label{sec:operating_rules}
Strict regulatory frameworks govern the use of the 6\,GHz spectrum to ensure coexistence with incumbent services operating in this band. These rules vary between different countries. Depending on location (indoor or outdoor), devices operating in the 6\,GHz band must adhere to specific power emission limits to avoid interference. 
To protect incumbent services and manage potential interference, the 6\,GHz band has been divided into categories of Wi-Fi devices, each with distinct operational rules and power limits:
\begin{itemize}
    \item \emph{\Gls{lpi} devices} are restricted to indoor use and are subject to lower power limits to minimize interference risks. These devices do not require \gls{afc} but must meet stringent \gls{psd} limits, use fixed antennas, and cannot operate on battery power. In the USA, \gls{lpi} devices are limited to a PSD of 5\,dBm/MHz and maximum \gls{eirp} of 30\,dBm, which is achieved over 320\,MHz channels. However, in the EU and UK the \gls{eirp} limits are 23\,dBm and 24\,dBm respectively, irrespective of channel bandwidth.
    \item \emph{\Gls{vlp} devices}, designed for portable and mobile applications, can be used indoors and outdoors with minimal power output, making them suitable for applications like wearables and \gls{iot} sensors. They can operate in all \gls{unii} bands, with a PSD limit of $-5$\,dBm/MHz. 
    \item \emph{Client devices} can operate indoors and outdoors across all \gls{unii} bands but must comply with power limits that are 6\,dB lower than their associated \glspl{ap}.  
    \item \emph{Standard power devices} are permitted to operate both indoors and outdoors in the \gls{unii} 5 and \gls{unii} 7 bands. However, they are required to use  \gls{afc}---discussed in the sequel---to dynamically adjust power levels and frequency usage, preventing interference with incumbent services. Additionally, their transmissions are restricted to angles below 30 degrees to avoid interfering with services deployed at higher altitudes, such as satellite links. Standard power has been fully authorized in North America and other countries are considering similar rules~\cite{WiFi_SP}.
\end{itemize}


\subsubsection{Automated Frequency Coordination (AFC)}
\Gls{afc} plays a crucial role in the regulatory framework for the 6\,GHz band, ensuring that standard power Wi-Fi devices can operate without disrupting incumbent services. This system works by dynamically allocating frequencies and power levels based on the location of the Wi-Fi device and nearby incumbent systems. Before transmitting, a standard power device queries the \gls{afc} system to determine which frequencies are safe to use at its location and what power levels are permissible. The \gls{afc} system takes into account the geographical coordinates, elevation, and antenna characteristics of the device, as well as the presence of other users in the band. In the USA, standard power devices operating under an \gls{afc} do not need to implement a contention-based mechanism unlike LPI devices: this allows the use of the band by standard power systems that do not use the Wi-Fi protocol. 
This centralized approach ensures that Wi-Fi devices stay within regulatory boundaries, while still making full use of the available 6\,GHz spectrum. By managing interference in real-time, \gls{afc} allows Wi-Fi devices to operate alongside existing services without causing disruption.

\commentfigtable
\begin{figure}
\centering
\includegraphics[width=\linewidth]{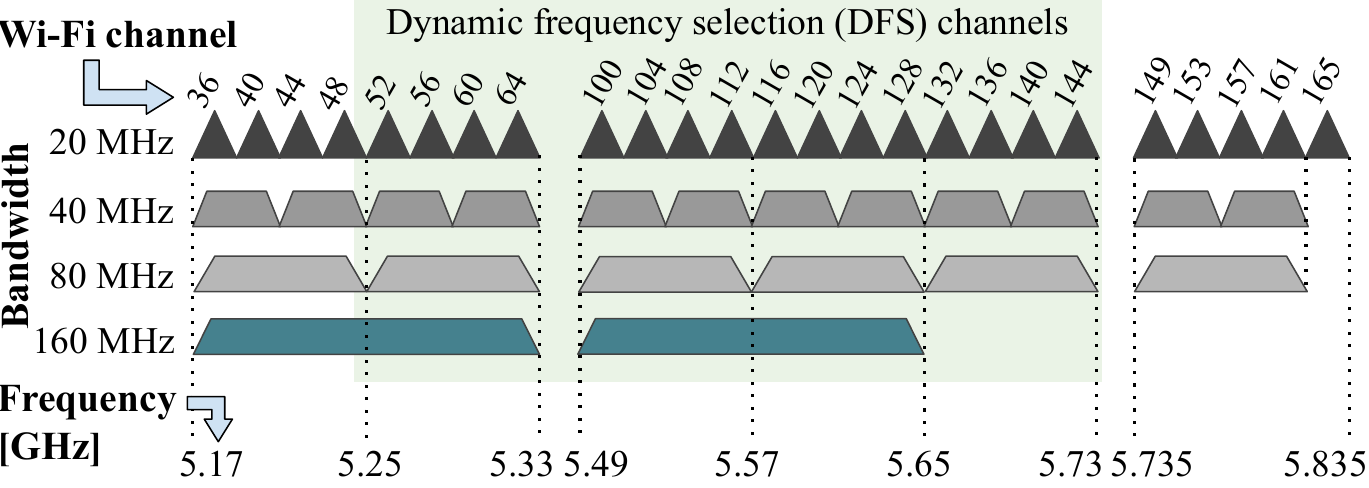}
\caption{Channelization in the 5\,GHz band and dynamic frequency selection (DFS) channels.}
\label{fig:5GHz_band_DFS}
\end{figure}
\endcommentfigtable



\subsection{Wi-Fi Amendments: From Drafting to Certification}

The IEEE 802.11 standard forms the backbone of Wi-Fi, but Wi-Fi technology goes far beyond and involves a complex set of standardization procedures in which different players have their own role. \blue{Fig.~\ref{fig:wifi_standardization_overview} summarizes the procedures and stakeholders involved in the elaboration of Wi-Fi specifications, from the creation of 802.11 standards to their certification by the \gls{wfa}. However, for Wi-Fi to be part of the broader picture that is the Internet, its interworking with the protocol suite (TCP/IP) is enabled thanks to the efforts of organizations like \gls{ietf} and protocols like \gls{dhcp}, \gls{dns}, or \gls{ip}, which are deeply integrated into Wi-Fi devices like \glspl{ap}.}

\commentfigtable
\begin{figure*}
\centering
\includegraphics[width=.7\textwidth]{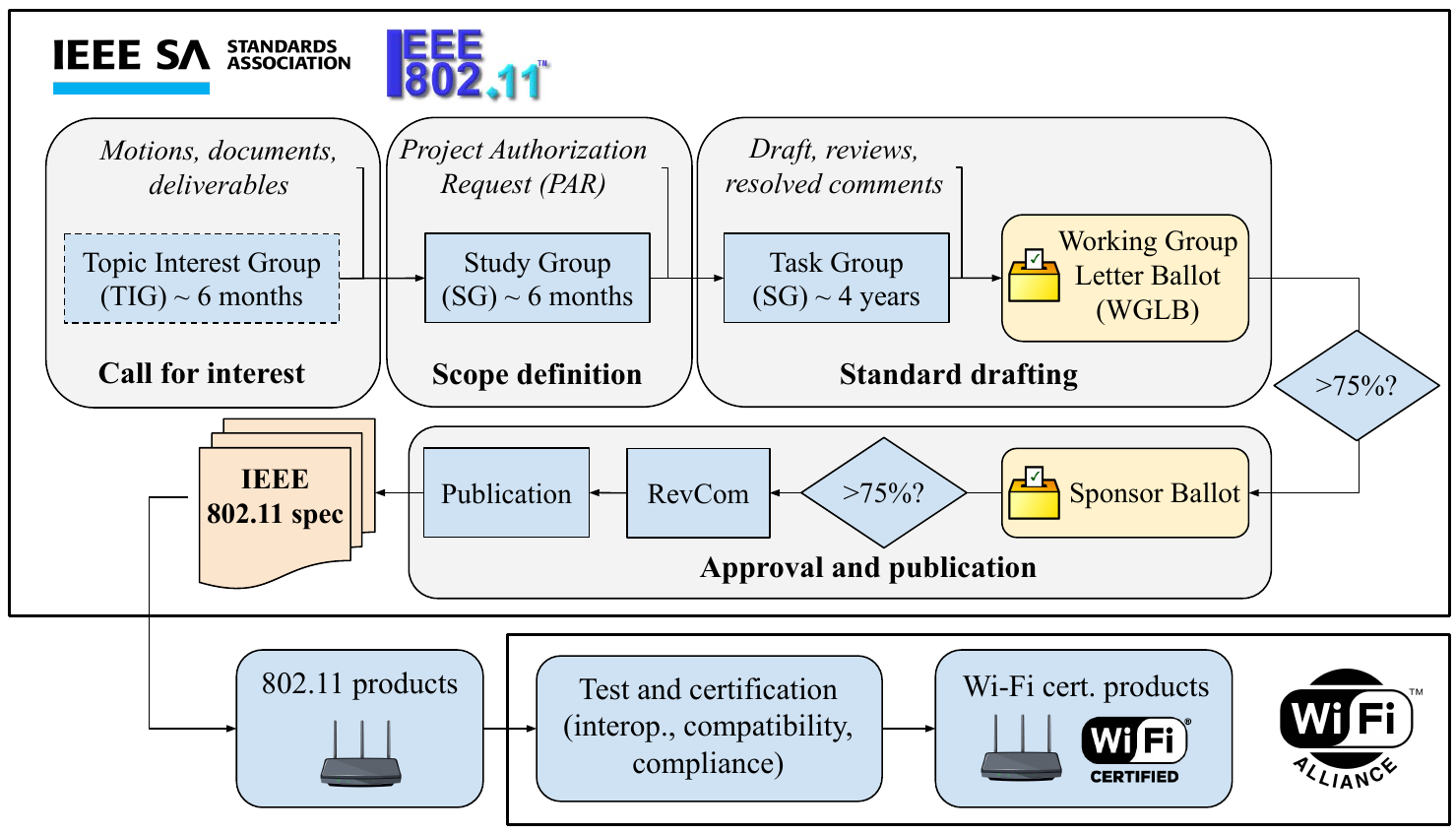}
\caption{Overview of the processes involved in Wi-Fi standardization and certification.}
\label{fig:wifi_standardization_overview}
\end{figure*}    
\endcommentfigtable

\begin{figure}
\centering
\includegraphics[width=\columnwidth]{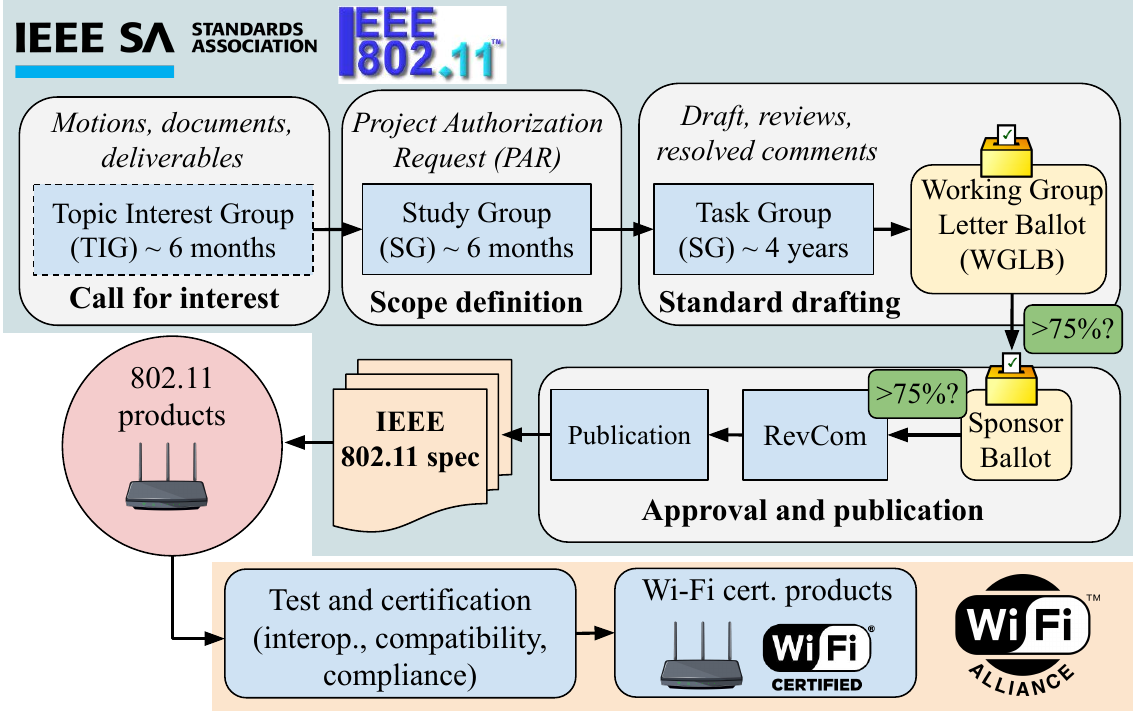}
\caption{\blue{Overview of the Wi-Fi standardization and certification process. The figure shows the main steps within IEEE 802.11, from the initial call for interest and Topic Interest Group, through Study and Task Groups, balloting, and final approval and publication of the IEEE 802.11 specification, and then how the published standard feeds into Wi-Fi Alliance interoperability testing and certification of commercial Wi-Fi products.}}
\label{fig:wifi_standardization_overview}
\end{figure}

\subsubsection{IEEE 802.11 Standardization}

The \gls{ieee} 802.11 standard defines the \gls{mac} and \gls{phy} layers of Wi-Fi, but does not go above in the TCP/IP stack. At the core of 802.11 is the concept of \gls{bss}, which represents a basic network unit managed by an \gls{ap} and to which \glspl{sta} (or non-\gls{ap} \glspl{sta}, according to 802.11 terminology) can connect and access the Internet thanks to built-in routing capabilities on the \gls{ap} devices, e.g., \gls{dhcp} and \gls{nat}. Based on the \gls{bss} concept, multiple solutions and architectures are derived. While mesh and device-to-device modes also exist within Wi-Fi, in this paper, we will mainly focus on \gls{ap}-\gls{sta} communication within \glspl{bss}.

The 802.11 \gls{wg} oversees all the standardization activities within the group, which occur at different levels through the \glspl{tg}, \glspl{sg}, \glspl{sc}, \glspl{ahc}, and \glspl{tig}. While the development cycle is specific to each 802.11 amendment, it can be roughly divided into three phases:
\begin{itemize}
    \item \emph{Scope definition:} After identifying new market demands and the need for a new amendment to the 802.11 standard---typically as a result of multiple discussions within the 802.11 \gls{wg} or from the activities of a \gls{tig}---a \gls{par} is prepared to narrow down the goals that the upcoming amendment will need to fulfil. The \gls{par}, typically prepared by an appointed \gls{sg} within six months, serves as the green light to initiate the development of the amendment.
    \item \emph{Standard amendment drafting:} On approving a \gls{par}, a \gls{tg} is appointed to develop the drafting of the specification, for which four years are granted. Drafting an amendment to the standard involves delving into the technical requirements and solutions to achieve the goals specified in the \gls{par}. For that, \gls{tg} participants discuss features and contributions until a stable draft of the amendment is achieved. At that stage, 
    the draft is put to vote among all members of the 802.11 \gls{wg}. This procedure is known as \gls{wglb}, and it requires a minimum of 75\% positive votes to move forward. During this process, comment resolution typically entails several back and forth interactions between the \gls{tg} and the \gls{wg}. 
    \item \emph{Standard approval and publication:} When the \gls{wglb} is approved, a second voting level (the Sponsor Ballot) \blue{needs to be passed} with a minimum support of the 75\%. In this case, the vote is extended to the \gls{ieee-sa} and the resulting text is forwarded to the \gls{ieee} \gls{revcom} for final approval. \blue{Once obtained,} the final specification is published, so that 802.11 vendors and manufacturers can develop their products accordingly.    
\end{itemize}

\subsubsection{Wi-Fi\textsuperscript{\textregistered} Certification}

The Wi-Fi certification process helps guarantee that certified devices (e.g., network equipment, computers, smartphones, appliances) implement the latest 802.11 features, security protocols, and more, and ensure a good user experience. The \gls{wfa} plays a vital role in ensuring device compatibility and interoperability by certifying products that adhere to 802.11 standard amendments. \gls{wfa}'s history began in 1999, when a group of companies founded the \gls{wfa}---at that time called the \gls{weca}. 
In 2000, the term ``Wi-Fi'' was coined, and since then the \gls{wfa} controls the use of the ``Wi-Fi Certified'' logo, a registered trademark. This logo is only stamped on devices that pass the certification tests defined by the \gls{wfa}. 

The \gls{wfa} defines multiple certification programs and tracks (\textit{FlexTrack}, \textit{QuickTrack}, and \textit{Derivative}), which focus not only on physical transmissions and data formats, but also on other functionalities such as those related to security or quality of service. The \gls{wfa} provides certification programs directly aligned with 802.11 amendments (e.g., Wi-Fi CERTIFIED 7\texttrademark
~\cite{wfa_wifi7_certification}) and specific to other functionalities such as WPA3~\cite{wfa_wpa3_certification} or \mbox{Wi-Fi} direct~\cite{wfa_wifidirect_certification}. Apart from the certification tests performed 
by authorized test laboratories, the \gls{wfa} organizes interoperability events (or plugfests), where Wi-Fi vendors, developers, and manufacturers meet to conduct tests that demonstrate robust performance of multiple Wi-Fi implementations under real-world conditions, interoperability with other certified devices, and backward compatibility.


\subsection{Overview of IEEE 802.11 Amendments}

The \gls{ieee} 802.11 standard amendments form the backbone of Wi-Fi technology, with each iteration introducing new features and improvements to address the evolving needs of wireless communications. The intricate interaction between an \gls{ap}, its \glspl{sta}, and neighboring \glspl{bss} is key to understanding Wi-Fi’s operational framework, the challenges it faces, and the features developed to address those challenges.


\subsubsection{From Legacy 802.11 to 802.11ax}

The original \gls{ieee} 802.11 standard (1997), following the \gls{fcc}'s 1985 opening of the \gls{ism} band for \gls{wlan} usage, laid the groundwork for data transmission over unlicensed spectrum bands. Back in the 1990s and early 2000s, Wi-Fi could offer up to 2\,\blue{Mb/s} to devices operating in the 2.4\,GHz band, but its meteoric pace of development has led to reaching up to 9.6\,\blue{Gb/s} just two decades in the 802.11ax amendment (Wi-Fi 6/6E). This dramatic increase has been driven by the adoption of new bands (5\,GHz and 6\,GHz) and technologies such as \gls{ofdm} and \gls{mimo}. The evolution of Wi-Fi from its inception to Wi-Fi 6/6E can be summarized as follows:
\begin{itemize}
    \item Wi-Fi’s journey began in 1997 with the original 802.11 standard, which was based on \gls{csma}/\gls{ca} and single-carrier modulations such as \gls{dsss} or \gls{fhss}. In addition, the foundations in terms of association and authentication were laid at this point.
    \item In 1999,  802.11b amended the initial specification from 1997 to enhance data rates to up to 11 \blue{Mb/s}, in part due to the adoption of \gls{cck} modulation. It was at this time that Wi-Fi products started to be commercially attractive, and thus its massive adoption began to provide residential wireless Internet access, wireless networking in offices, and public hotspots. The growth in popularity of Wi-Fi concides with the creation of the \gls{wfa}, the growth of broadband Internet usage, and the need for wireless mobility by the increased accessibility to laptops by the bulk of the population.    
    \item That same year, 802.11a was introduced to operate in the 5\,GHz band. This, together with the adoption of \gls{ofdm}, allowed reaching peak data rates of 54\,\blue{Mb/s}. 
    In 2001, the \gls{fcc} permitted the use of \gls{ofdm} in the 2.4\,GHz band, leading to the 802.11g amendment in 2003. The 802.11g amendment became very important thanks to its backward compatibility with 802.11b.
    \item The release of 802.11n (Wi-Fi 4~\cite{WiFiCertifiedN}) in 2009 marked another significant leap in the evolution of Wi-Fi~\cite{gast2012802}. The adoption of features such as \gls{mimo} spatial multiplexing and channel bonding enabled peak data rates of up to 600\,\blue{Mb/s}, and spatial diversity led to improved reliability.  802.11n expanded Wi-Fi beyond casual web browsing and email to support \gls{hd} video streaming and richer web applications.
    \item In 2013, 802.11ac (Wi-Fi 5~\cite{WiFiCertifiedAC}) built upon the advancements of 802.11n and introduced even larger channel bandwidths (up to 160\,MHz), higher modulations (256-\gls{qam}), and \gls{mu}-\gls{mimo} in the downlink, enabling multiple devices to receive data simultaneously~\cite{gast2013802-2}. The 802.11ac reached multigigabit speeds, cementing Wi-Fi as the de facto standard for home entertainment, cloud services, and mobile offload.
    \item In 2021, 802.11ax (Wi-Fi~6~\cite{WiFiCertified6}) \cite{sankaran2021wi} marked a shift from raw speed to efficiency. Recognizing the challenge of high-density environments such as stadiums, airports, and apartment complexes, it introduced features such as \gls{ofdma}, uplink \gls{mu}-\gls{mimo}, and spatial reuse (\gls{bss} coloring), significantly improving spectral efficiency, latency, and fairness. Wi-Fi~6E \cite{WiFiCertified6E}, 
    an extension of Wi-Fi~6, 
    unlocked 1.2\,GHz of bandwidth in the 6\,GHz band, providing additional spectrum to reduce congestion and enable cleaner, high-throughput links.
\end{itemize}

\begin{figure}
\centering
\includegraphics[width=0.99\linewidth]
{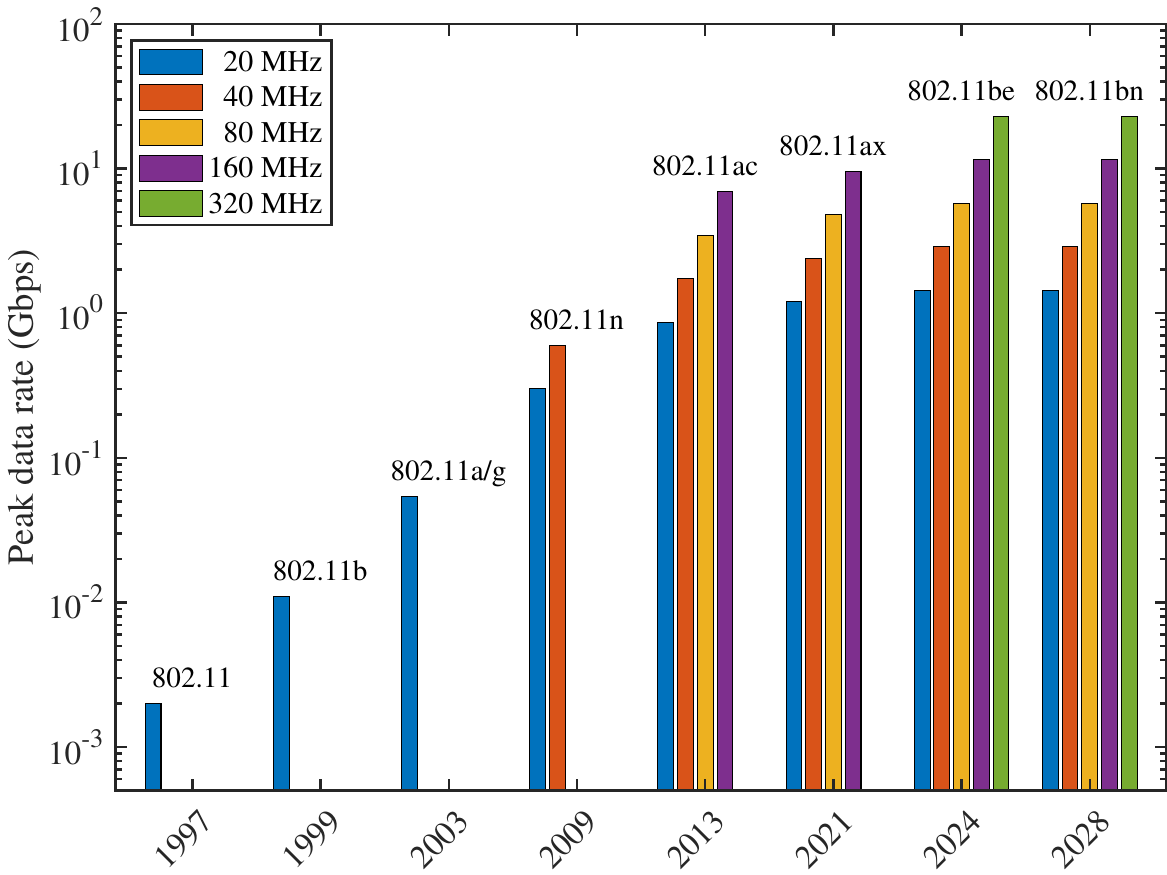}
\caption{\blue{Peak data rate and bandwidth utilization across IEEE 802.11 standard amendments. The bars show the evolution of the theoretical peak PHY data rate (\blue{Gb/s}, logarithmic scale) for different channel bandwidths (20–320 MHz, color‐coded) as new 802.11 amendments are introduced over time (from 802.11 to 802.11bn).}}
\label{fig:80211standards}
\end{figure}

Fig.~\ref{fig:80211standards} provides an overview of the evolution of Wi-Fi amendments in terms of their achievable peak data rates (see Section~\ref{subsec:peakrate} for their computation). 

\subsubsection{802.11be Extremely High Throughput}

Commercialized in 2024, the 802.11be amendment (Wi-Fi 7) introduces even more advanced features to do justice to its full name, \emph{\gls{eht}}. The performance goals of the 802.11be are achieved through new \gls{mac} and \gls{phy} modes of operation, as well as the use of the 6\,GHz band.

The 802.11be's \gls{phy} defines support for wider bandwidths (up to 320\,MHz channels) and higher modulation schemes (up to 4096-\gls{qam}). 
To exemplify the potential of 802.11be's \gls{phy}, a link using a 320\,MHz channel, 8$\times$8 MIMO, a short guard interval of 0.8\,$\mu$s, and 4096-\gls{qam} with 5/6 coding can achieve a data rate of 23\,\blue{Gb/s}. A more conservative setup, e.g., a 160\,MHz channel with 2$\times$2 \gls{mimo} and 1024-\gls{qam} with 3/4 coding, would still lead to a 2.1\,\blue{Gb/s} data rate.

As for the 802.11be's \gls{mac}, the main breakthrough has been achieved by \gls{mlo}, which we will discuss in depth in Section~\ref{sec:MLO}. In a nutshell, \gls{mlo} further increases throughput by using multiple physical links in parallel through a single association. Apart from that, the 802.11be includes appealing features such as support for multiple \glspl{ru} in \gls{ofdma}. 

Although 802.11be makes important strides toward lower latency and higher reliability, these aspects were not fully quantified and rather deferred to 802.11bn, which will prioritize reliability in addition to throughput.

\commentfigtable
\begin{table*}[h]
\vspace{1mm}
\caption{Representative emerging use cases for Wi-Fi beyond 2030 and their corresponding requirements~\cite{P80211bnPAR,HexaD13}.}
\label{table:use_cases}
\centering
\colorbox{BackgroundGray}{%
\begin{tabular}{ |m{5.0cm}|m{1.8cm}|m{1.8cm}|m{3.2cm}|m{4.0cm}| } 
\toprule
\rowcolor{BackgroundLightBlue}
 \textbf{Use case} & \textbf{Reliability} & \textbf{Latency} & \textbf{Data rate} & \textbf{Notes} \\ \midrule
\textbf{Immersive communications:} \newline
{Physically present and holographically telepresent consumers with AR glasses and body sensors/actuators.} 
& 
99.9\%
&
<\,20\,ms
&
1-10\,\blue{Gb/s} downlink (DL) \newline (AR stream + spatial map)
\newline
0.1\,\blue{Gb/s} uplink (UL) \newline (spatial map + user data)
&
End-to-end roundtrip UL+DL <\,100\,ms
for 99.99\% of packets to avoid nausea and user distress and
discomfort.
\\ \midrule
\textbf{Digital twins for manufacturing:} \newline
{Machine-to-machine traffic between sensors, alarms, fixed machinery, and moving autonomous vehicles. Human traffic from operators monitoring the factory.}
& 
99.9\% \newline to \newline 99.999999\%
&
0.1\,--\,100\,ms
&
1\,--\,10\,\blue{Gb/s} (average)
\newline to \newline
10\,--\,100\,\blue{Gb/s} (peak)
&
Lower reliability for process and
asset monitoring and higher reliability for
motion control and alarms.
\\ \midrule
\textbf{e-Health for all:} \newline
{Connectivity for users (healthcare professionals, patients, administrators) and devices (smart medical instruments, wearables, on- and in-body sensors/actuators).}
& 
99.999\% \newline to \newline 99.9999999\% &
0.1\,--\,100\,ms
&
100\,\blue{kb/s} (sensor data) \newline to  \newline 25\,\blue{Mb/s} (4K video)
&
Lower latency required in robotic-assisted surgery operations. Peak bit rates can be much higher for specific applications, e.g., XR remote diagnostics.
\\ \midrule
\textbf{Cooperative mobile robots:} \newline
{Communication between robots and static machinery. Direct (through XR devices) or intent-based (trajectory crossing) human-machine interaction on a shop floor.}
& 
up to \newline 99.9999999\%
&
0.5\,--\,25\,ms
&
< 0.1\,\blue{Mb/s} (for control)
&
Deterministic communication required for control applications.
\\ \bottomrule 
\end{tabular}
}
\end{table*}
\endcommentfigtable

\subsubsection{802.11bn Ultra High Reliability}

As Wi-Fi continues to evolve, emerging use cases and applications require not only increased throughput and reduced latency, as addressed by Wi-Fi 7, but also a higher level of reliability. These evolving demands are shaping the development of the 802.11bn, with a strong focus on \gls{uhr}. Some of the most critical emerging use cases for 2030 are driven by the advances in immersive communications, digital twins, and real-time control systems, among others (see Table~\ref{tab:wifi_evolution}). 
Each of these applications requires low-latency, highly reliable communication. To meet these requirements, Wi-Fi is moving toward more deterministic performance.

At the time of writing, the UHR Task Group (TGbn) has released a draft of the 802.11bn amendment (D1.1)~\cite{80211bnD1.1}. As included in the \gls{par}, 802.11bn will target a 25\% improvement in data rates, even at lower \gls{sinr} levels, along with a 25\% reduction in tail latency and packet loss rates, particularly in environments with mobility and \gls{obss}. Additional advancements in power-saving features and peer-to-peer operation are expected, with 802.11bn extending its operation across all sub-7\,GHz bands.

The TGbn, which was formally established in November 2023, will define the technical objectives and protocol functionalities of Wi-Fi 8 by 2028, when the final amendment is expected to be published. In the meantime, the first draft (D1.0) is planned to be released in July 2025, whereas the final draft is expected to be submitted for Sponsor Ballot in May 2027.\footnote{\url{https://www.ieee802.org/11/Reports/802.11\_Timelines.htm\#tgbn}}

To guarantee low-latency, high-reliable communication in unlicensed bands, one of the key areas of development for the 802.11bn will be \gls{mapc}, extensively discussed in Section~\ref{sec:MAPC}, which will allow multiple \glspl{ap} to cooperate in managing interference, coordinating channel access, and improving overall spectrum efficiency. This approach is expected to significantly improve performance, especially in dense network environments where many devices compete for limited spectrum. \gls{mapc} will entail a true paradigm shift in Wi-Fi, which has historically relied on the inherently non-deterministic \gls{csma}/\gls{ca}, to comply with the regulation on the unlicensed spectrum that mandates the usage of \gls{lbt}~\cite{etsi_en_301_893}. 

\begin{figure}
\centering
\includegraphics[width=\linewidth]{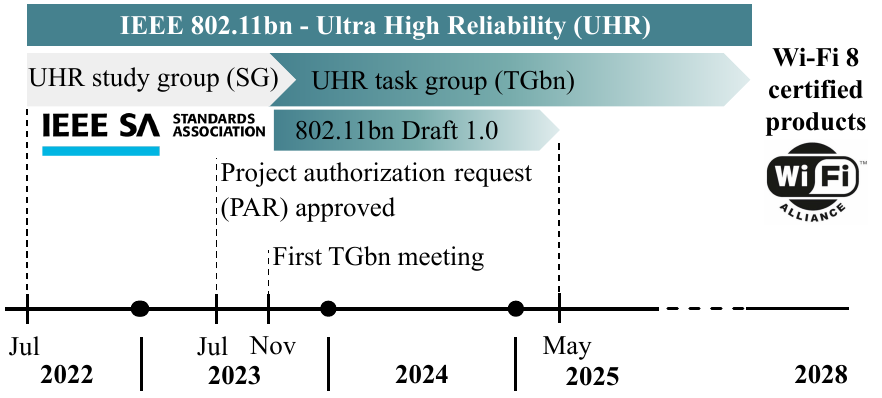}
\caption{Wi-Fi 8 standardization (\gls{ieee}) and certification (\gls{wfa}) efforts timeline.}
\label{fig:80211be_80211bn}
\end{figure}

\commentfigtable
Figure~\ref{fig:80211be_80211bn} summarizes the ongoing standardization effort for Wi-Fi 8, including the development of 802.11bn by the 802.11 \gls{wg} and its certification program by the \gls{wfa}. 
\endcommentfigtable

\commentfigtable
\endcommentfigtable


\subsubsection{Other Ongoing Standardization Activities}

While 802.11bn will focus primarily on improving Wi-Fi performance in the sub-7\,GHz bands, the future of Wi-Fi also includes expansion into higher frequency ranges. The 60\,GHz band, in particular, presents new opportunities for ultra-high-throughput and low-latency applications. The 802.11bq Task Group (TGbq) is tasked with exploring operational expansion into the 60\,GHz band, building upon the work done in previous amendments like 802.11be and 802.11bn. The 60\,GHz band offers significant advantages for short-range, high-bandwidth applications, but also introduces challenges such as propagation loss and sensitivity to obstacles, as discussed in Appendix~\ref{sec:11bq}. Solving these challenges will be key to enabling the next generation of Wi-Fi applications, such as uncompressed video streaming and high-speed data transfer for industrial use.

As discussed in Appendix~\ref{sec:11bf}, Wi-Fi sensing, standardized \blue{in 2025} under the 802.11bf amendment \blue{(``Wireless LAN Sensing'')}, leverages Wi-Fi signals to enable applications such as motion detection, human activity recognition, and proximity sensing. By tracking changes in channel, devices can infer environmental factors without requiring the target to carry a Wi-Fi device. Wi-Fi sensing has promising applications in residential, enterprise, and industrial settings, offering capabilities like security monitoring, gesture control, and smart device management. The 802.11bf amendment \blue{unifies} sensing procedures across devices, paving the way for interoperable and scalable sensing solutions using existing Wi-Fi infrastructure. \blue{As a complementary capability, the IEEE 802.11-2016 standard revision incorporated the \gls{ftm} protocol to enable precise indoor localization through \mbox{Wi-Fi}. \Gls{ftm} estimates the distance between devices by computing the \gls{rtt} of sounding frame transmissions. Building on this, in 2022, the 802.11az amendment (``Enhancements for Positioning'') standardized a procedure to enable an \gls{sta} to identify its position relative to a set of \glspl{ap} by combining \gls{ftm} measurements collected from each of them. The update also improved the accuracy from 1-2 meters of 802.11-2016 to sub-meter precision. In 2025, this further evolved into the IEEE 802.11bk amendment (``320~MHz Positioning'') which extends the \gls{ftm} positioning mechanisms to make use of 320~MHz wide channels available with 802.11be.}

In terms of privacy and security, the TGbh and TGbi are currently working on randomized \gls{mac} addresses (so that tracking devices can be more difficult for potential attackers) and \gls{edp} (including mechanisms to protect transmissions and management frames), respectively. The 802.11bh amendment has already passed the Sponsor Ballot and is about to be published officially, while 802.11bi is currently at D1.0. Both 802.11bh and 802.11bi are further discussed in Appendix~\ref{sec:bh_bi}.

Another critical development in Wi-Fi standardization is the integration of \gls{ai} and \gls{ml} features, aimed at enhancing network management and operational efficiency and discussed in Appendix~\ref{sec:AIML}. As \mbox{Wi-Fi} networks grow increasingly complex, \gls{ai}/\gls{ml} will play a pivotal role in real-time decision-making and optimization. The efforts to incorporate \gls{ai}/\gls{ml} were initiated by the \gls{ieee} 802.11 \gls{ai}\gls{ml}  \gls{tig} and are now being followed up by the \gls{ieee} 802.11 \gls{ai}\gls{ml} \gls{sc}, which is exploring how these technologies can be leveraged in novel use cases that allow automating network management tasks, optimizing resource allocation, or predicting traffic patterns to improve overall network performance, among others.

\begin{figure*}
\centering
\includegraphics[width=\linewidth]{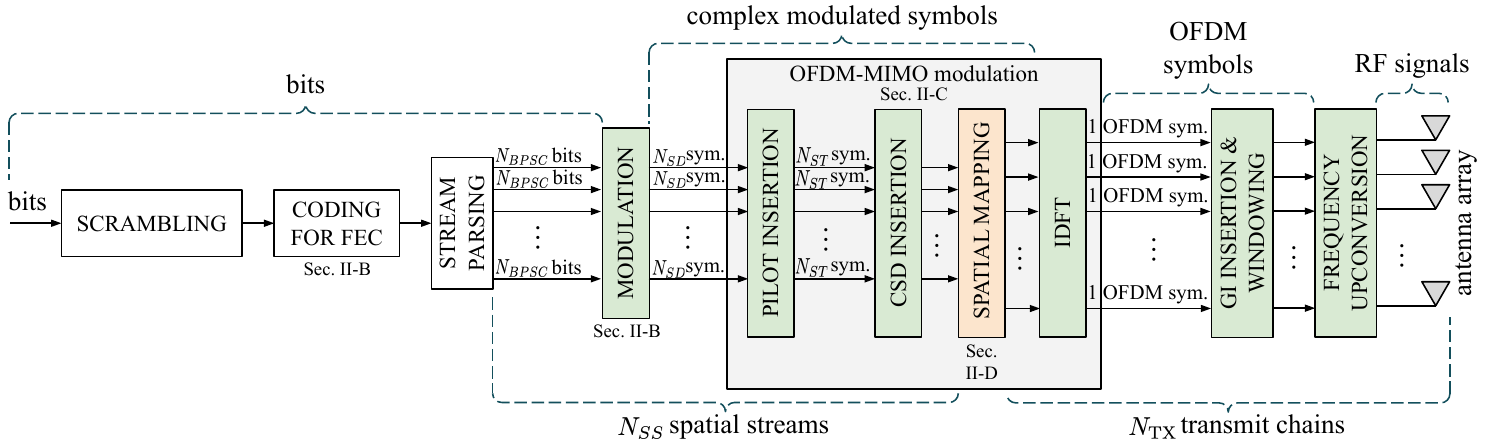}
\caption{Main signal processing blocks used by 802.11-compliant devices to convert bits into RF signals. The green blocks refer to processing steps applied to each spatial stream or transmit chain independently, e.g., $N_{\rm SS}$ modulation blocks are used, one for each of the $N_{\rm SS}$ spatial streams. The spatial mapping (yellow block) simultaneously processes all the data streams. Depending on the coding strategy adopted (BCC or LDPC), a few additional processing blocks are included. The notation `sym.' stands for `symbol(s)'.\vspace{-0.5cm}}
\label{fig:PHY_signal_processing}
\end{figure*}

\section{Wi-Fi Physical Layer}
\label{sec:PHY}

Wi-Fi's \gls{phy} underpins the data transmission process over wireless channels. \blue{At the transmitter device, the \gls{phy} receives data from the \gls{mac} in a \gls{psdu}, decides the parameters for the \gls{phy} transmission (e.g., \gls{mcs}), and then builds a \gls{ppdu} by adding to the \gls{psdu} a \gls{phy} preamble, which specifies the selected parameters and other control fields for frequency offset correction and time synchronization at the receiver. The sequence of data and control bits in the \gls{ppdu} is then processed to obtain a \gls{rf} signal to be irradiated through the antennas. Note that while the parameters for the data field (\gls{psdu}) are decided at the \gls{phy}, the transmission parameters for the control fields in the preamble are fixed and specified in the standard. At the receiver, the \gls{phy} uses the control information in the preamble to detect the start of the \gls{ppdu} and decode it into a \gls{psdu}, which is forwarded to the \gls{mac}. This section provides an overview of the physical bit transmission pipeline,} deepening the explanation of the modulation and coding strategies together with the spatial and frequency multiplexing techniques. The \glspl{ppdu} formats are also detailed at the end of this section.

\subsection{\blue{Physical Layer Block Diagram}}\label{subsec:transmitter_block}

Fig.~\ref{fig:PHY_signal_processing} provides an overview of the processing steps to generate the Wi-Fi \gls{rf} signals to be radiated through the different antennas available at the transmitter side, starting from the \gls{ppdu} bits. In particular, the figure summarizes the processing steps applied to the \gls{ppdu} data field (see Section~\ref{subsec:ppdus}). The other \gls{ppdu} fields, e.g., \gls{stf} and \gls{ltf}, are processed following a similar pipeline where some of the blocks are removed. 

The procedure is as follows. \blue{At first, the data bits are scrambled using an additive scrambler, which applies an exclusive OR (XOR) of the bits with a scrambling sequence. This operation reduces the probability of long sequences of bits equal to zeros or ones, which helps time synchronization at the receiver. The receiver should use a matched descrambler to decode data.} Redundancy is then added to the scrambled bits for \gls{fec}. Two different \gls{fec} coding techniques can be used for this purpose, namely \gls{bcc} and \gls{ldpc}~\cite{benvenuto2021algorithms}. The amount of redundancy is a function of the coding rate, which is specified through the \gls{mcs} parameter, as detailed in Section~\ref{sec:mcs}. Note that padding should be added before the scrambler (pre-\gls{fec} \gls{phy}) and after the \gls{fec} encoder (post-\gls{fec} \gls{phy}). 

Depending on the coding strategy adopted (\gls{bcc} or \gls{ldpc}), an additional processing block is included in the diagram in Fig.~\ref{fig:PHY_signal_processing}. Specifically, \gls{bcc} requires an interleaver block, which consists of a two-step permutation of the coded bits to avoid long sequences of noisy bits on the \gls{bcc} decoder and thus improve transmission robustness to burst errors. The first permutation maps adjacent coded bits into non-adjacent \gls{ofdm} subcarriers (see Section~\ref{sec:ofdm}), while the second maps adjacent coded bits alternately onto less and more significant bits of the constellation. Instead, a tone mapper is used for \gls{ldpc}-coded bits after being modulated. The complex modulated symbols are permuted such that each two consecutive symbols will be transmitted on two \gls{ofdm} subcarriers that are separated by at least $D_{\rm TM}$ \gls{ofdm} data subcarriers, where $D_{\rm TM}$ is an integer parameter specified in the standard for each bandwidth: $4$, $6$, $8$, $9$ for $20$\,MHz, $40$\,MHz, $80$\,MHz, and $160$\,MHz respectively.

\blue{While the first version of 802.11 considered the transmission of a single modulated symbol per time resource, multiplexing techniques are key components of current \mbox{Wi-Fi} networks to improve spectrum efficiency.} Hence, the coded bits are associated with $N_{\rm SS}$ different spatial streams for \gls{mimo} transmission in a round-robin fashion, creating groups of $N_{\rm BPSC}$ bits, where $2^{N_{\rm BPSC}}$ is the modulation order. The bits on each of the $N_{\rm SS}$ spatial streams are modulated independently, using the modulation scheme defined by the \gls{mcs} parameter (see Section~\ref{sec:mcs}). Hence, the modulated symbols are forwarded to the \gls{ofdm}-\gls{mimo} modulator for frequency and spatial multiplexing. The complex modulated symbols in each spatial stream are processed in groups of $N_{\rm SD}$ symbols via \gls{ofdm}, where $N_{\rm SD}$ depends on the bandwidth and the \gls{ofdm} subcarrier spacing, as it will be detailed in Section~\ref{sec:ofdm}. Additional known \gls{ofdm} symbols called pilots are added within the set of $N_{\rm SD}$ modulated data symbols in specific positions for receiver-transmitter synchronization purposes to make the data detection robust against frequency offsets and phase noise. The indices of the pilot \gls{ofdm} subcarriers are defined by the 802.11 standard for the different operational bandwidths. \blue{\Gls{csd} is used to prevent unintentional beamforming---which occurs when correlated signal are transmitted over multiple antennas---by applying slightly different cyclic shift to each of the space-time streams. This technique was introduced in the 802.11n amendment and is also essential for backward compatibility with single-antenna receivers.}

Spatial mapping is applied next, as it will be explained in Section~\ref{sec:MIMO}. Following this, an \gls{ifft} operation allows converting the set of modulated symbols plus pilots into an \gls{ofdm} symbol, as it will be described in Section~\ref{sec:ofdm}. \blue{A \Gls{gi} is added at the beginning of each \gls{ofdm} symbol} to reduce \gls{isi} during transmission, and a time-windowing function is applied to set the boundary of the transmitted signal. \blue{The specific time-windowing function to be applied is not specified in the standard and is therefore a choice of the manufactures.} Finally, the \gls{rf} signals to be transmitted over the different antennas are obtained through a frequency up-conversion at the specific operating frequency defined by the selected Wi-Fi channel.

\paragraph*{Latest 802.11bn Enhancements}

Building upon the transmitter block logic described above, recent developments under the forthcoming 802.11bn amendment introduce significant enhancements aimed at improving performance and reliability. One of the key updates focuses on \gls{fec} through the inclusion of a new, longer block-length \gls{ldpc} code of 3888 bits (2$\times$1944). This new code complements the existing options of 648, 1296, and 1944 bits, and is designed to enhance error correction, particularly in high-reliability scenarios like \gls{xr} applications and dense wireless environments.
The longer code provides greater coding gain, improving performance across different \gls{mcs} levels (see Section~\ref{sec:mcs}). It follows a \gls{qc} structure derived from the base 1944-bit code and supports code rates of 1/2, 2/3, 3/4, and 5/6. The 3888-bit code integrates smoothly into the updated \gls{ppdu} encoding table while staying compatible with legacy codeword lengths for smaller payloads.
Thanks to its modular structure, this design allows for parallel encoding and decoding, enabling higher throughput with minimal added complexity. Performance evaluations show gains of 0.5 to 1.0\,dB, especially at higher \gls{mcs} levels and when multiple spatial streams are used.
That said, using 3888-bit codes can be challenging for short payloads, where puncturing (see Section~\ref{sec:wideband_op}) may hurt performance. Successful implementation depends on the ability to parallelize decoding. In the 5\,GHz and 6\,GHz bands, this is often feasible due to the presence of multiple decoders. However, in the 2.4\,GHz band, where channels are narrower, this capability may be lacking. As a result, it may be best to restrict support for 3888-bit \gls{ldpc} codes to the 5\,GHz and 6\,GHz bands to avoid added complexity.


\subsection{Modulation and Coding Schemes}\label{sec:mcs}

The \gls{mcs} index is a positive integer number that defines the parameters for modulation and coding. With modulation, we refer to the techniques adopted in wireless systems to convert digital data (bits) into analog signals by changing the shape of a reference signal, called carrier, within a set of possible modifications. The resulting modulated signal carries the bit stream. Devices adopting the 802.11 standard support \gls{psk} and \gls{qam} schemes, where the former entails changing the phase of the carrier to represent data while the latter does it by also changing the carrier's amplitude. The decoder detects the specific changes to retrieve the transmitted bits. Each modification can be identified as a point in the complex plane (I/Q) and, in turn, is referred to as constellation point. The modulation order identifies the number of constellation points in the modulation scheme ($2^{N_{\rm BPSC}}$), which in turn defines the number of bits that can be carried by each signal modulated differently ($N_{\rm BPSC}$). For this, the stream of coded bits is divided into groups of $N_{\rm BPSC}$ bits, each of which is mapped into a constellation point generating a modulated symbol. For example, a 256-QAM modulated symbol is obtained by combining 16 bits. 

The second information provided by the \gls{mcs} is the coding rate, which defines the redundancy level introduced by \gls{bcc} or \gls{ldpc} coding to enable \gls{fec} at the receiver. For example, a 1/2 coding rate means that, for each information bit, there is another bit for \gls{fec} purposes. This rate should be used under less favorable channel conditions where more redundancy is required, while coding rates of 3/4 or 5/6 can be adopted when the \gls{snr} is sufficiently high.

\commentfigtable
\begin{table}
\vspace{1mm}
\caption{Modulation and coding schemes in 802.11be and minimum required sensitivity on a $20$\,MHz channel.}
\label{tab:MCS}
\centering
\colorbox{BackgroundGray}{%
\begin{tabular}{ |m{0.7cm}|m{1.7cm}|m{1.7cm}|m{2.0cm}|} 
\toprule
\rowcolor{BackgroundLightBlue}
 \textbf{MCS} & \textbf{Modulation} & \textbf{Coding rate} & \textbf{Min. sensitivity}   \\ \midrule
 {0} & {BPSK} & {1/2} & {-82\,dBm} \\ \midrule
 {1} & {QPSK} & {1/2} & {-79\,dBm} \\ \midrule
 {2} & {QPSK} & {3/4} & {-77\,dBm} \\ \midrule
 {3} & {16QAM} & {1/2} & {-74\,dBm} \\ \midrule
 {4} & {16QAM} & {3/4} & {-70\,dBm} \\ \midrule
 {5} & {64QAM} & {2/3} & {-66\,dBm} \\ \midrule
 {6} & {64QAM} & {3/4} & {-65\,dBm} \\ \midrule
 {7} & {64QAM} & {5/6} & {-64\,dBm} \\ \midrule
 {8} & {256QAM} & {3/4} & {-59\,dBm} \\ \midrule
 {9} & {256QAM} & {5/6} & {-57\,dBm} \\ \midrule
 {10} & {1024QAM} & {3/4} & {-54\,dBm} \\ \midrule
 {11} & {1024QAM} & {5/6} & {-52\,dBm} \\ \midrule
 {12} & {4096QAM} & {3/4} & {-49\,dBm} \\ \midrule
 {13} & {4096QAM} & {5/6} & {-46\,dBm} \\ \bottomrule 
\end{tabular}
}
\end{table}
\endcommentfigtable

The \gls{mcs} to be used for data transmission is defined by the transmitter by evaluating the channel conditions. 
Low \gls{mcs} values, i.e., low modulation order and high coding redundancy, lead to more robust transmissions at the expense of reduced throughput and should be preferred in low \gls{snr} conditions. 
High \glspl{mcs} require increasingly higher \gls{snr} to maintain robustness against noise and interference but increase data rates by modulating more bits together per symbol and reducing coding redundancy, benefiting bandwidth-intensive applications such as 4K video streaming, virtual reality, and large file transfers.
\commentfigtable
The \gls{mcs} values defined in 802.11be and their specifications are included in Table~\ref{tab:MCS}. In addition to the modulation scheme and the coding rate, the table also includes the minimum sensitivity level, i.e., the minimum signal power, for which the data is received with a \gls{per} of less than 10\%, when transmitting over a 20\,MHz bandwidth.
\endcommentfigtable
For this, in addition to the \glspl{mcs} [0, $\dots$, 9] available in 802.11ac, 802.11ax introduced 1024-QAM with two possible coding rates (3/4 for MCS 10 and 5/6 for MCS 11), improving link efficiency in high-density deployments~\cite{liu2023first}.
The 802.11be amendment took this further by introducing 4096-QAM, tied to \gls{mcs} 12 and 13~\cite{henry2024wi}. However, 4096-QAM has stricter \gls{evm} requirements, about 3\,dB higher than 1024-QAM, necessitating exceptionally clean signals for successful demodulation: the required \gls{snr} is around 40\,dB, which may only be achievable through techniques like beamforming. 

802.11be also introduced \gls{dcm} to improve robustness to fading in signal reception by transmitting each modulated symbol into two different \gls{ofdm} subcarriers~\cite{henry2024wi}. \Gls{dcm} is used in conjunction with BPSK and coding rate 1/2 and defines two new \glspl{mcs}, i.e., \gls{mcs} 14 and \gls{mcs} 15, both of which can only be used for single spatial stream transmissions. The difference between these two new \glspl{mcs} is that the first requires the use of \gls{dup} mode, where the data is also duplicated in the lower and upper parts of the channel (e.g., lower and upper 484 subcarriers for 80~MHz), thus doubling the duplication performed through \gls{dcm} and, in turn, further increasing robustness. \Gls{dup} mode is only applicable in the 6\,GHz band with \gls{ldpc} coding.
Note, however, that these two additional \glspl{mcs} have not been certified by the Wi-Fi Alliance as part of Wi-Fi~7, thus limiting their adoption into commercially available products.  

\paragraph*{Latest 802.11bn Enhancements}

In earlier 802.11 devices employing multi-stream \gls{mimo} transmissions (see Section~\ref{sec:MIMO}), all spatial streams were configured with the same modulation order and a unified coding rate. However, when beamforming is used, the wireless channel conditions can vary significantly across different spatial streams, leading to unequal \gls{snr}.
To address this, 802.11bn introduces \gls{ueqm}, a feature that allows assigning different modulation orders to each stream within a single transmission. Despite this variability, a common \gls{ldpc} code is applied to all streams to maintain a consistent error correction.
By adapting modulation to the real-time conditions of each stream, \gls{ueqm} can improve both spectral efficiency and link robustness, making it especially useful in high-interference or spatially heterogeneous environments.
Complementing this, 802.11bn also adds several new \gls{mcs} levels to allow for finer-grained link adaptation. Examples include QPSK with a 2/3 coding rate and 5/6 16-QAM, offering greater flexibility to optimize performance in diverse deployment scenarios.
However, these enhancements also increase the complexity of rate adaptation. With more combinations of \gls{mcs} and number of spatial streams ($N_{\rm SS}$) available, the transmitter must navigate a larger decision space. To manage this effectively, improved feedback mechanisms may be needed, guiding the transmitter in selecting the most suitable \gls{mcs} and transmission configuration under current channel conditions.


\subsection{Orthogonal Frequency Division Multiplexing (OFDM)}\label{sec:ofdm}

\Gls{ofdm} was introduced in 802.11a and allows multiplexing modulated symbols in the frequency domain.
To do this, $N_{\rm ST}$ partially overlapping orthogonal subcarriers are obtained from the available bandwidth $B$,
and used to simultaneously transmit $N_{\rm SD}$ modulated symbols. The remaining $N_{\rm ST}\!-\!N_{\rm SD}$ subcarriers are occupied by $N_{\rm SP}$ pilot symbols and additional unused subcarriers~\cite{benvenuto2021algorithms}. \blue{The unused ones include the DC subcarrier (center subcarrier) and the guard band subcarriers at the band edges. Moreover other null (unused) subcarriers located near the DC provide protection from transmit center frequency leakage, \gls{dac} and \gls{adc} offsets at the receiver, and interference from neighboring \glspl{ru} in \gls{ofdma} transmissions (see Section~\ref{sec:OFDMA})~\cite{goldsmith2005wireless}. Their number depends on the operating bandwidth and the subcarrier spacing. For example, in 802.11ax, there are 3 null subcarriers around the DC for 20~MHz transmissions, and 5 for 40~MHz and 80~MHz transmissions.} Note that \gls{ofdm} subcarriers are also referred to as \textit{\gls{ofdm} sub-channels} or \textit{tones} in the literature. 

\begin{figure}[!t]
\centering
\subfloat[\gls{phy} features in IEEE 802.11ac and earlier amendments.]{\includegraphics[width=\figwidth]{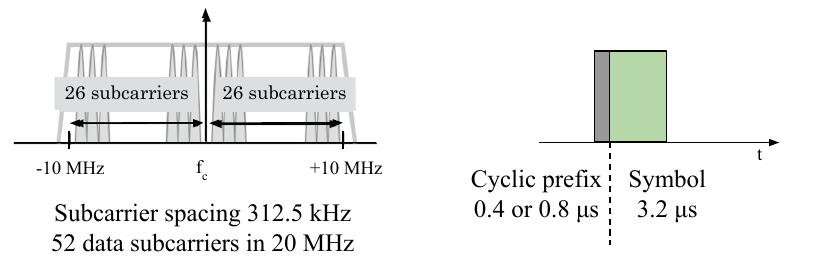}}
\hspace{0mm}\\
\vspace*{-3mm}
\subfloat[\gls{phy} features in IEEE 802.11ax and beyond.]{\includegraphics[width=\figwidth]{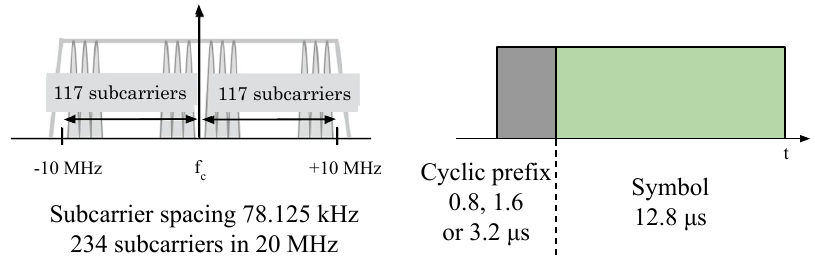}}
\caption{\blue{OFDM subcarrier spacing with an example of subcarrier distribution on a 20\,MHz channel (left), and corresponding \gls{ofdm} symbol duration (right). The top (a) illustrates IEEE 802.11ac and earlier amendments with 52 data subcarriers at 312.5\,kHz spacing and a 3.2\,µs symbol (with 0.4 or 0.8\,µs cyclic prefix). The bottom (b) shows IEEE 802.11ax and beyond with 234 subcarriers at 78.125\,kHz spacing and a 12.8\,µs symbol and longer cyclic prefix options.}}
\label{fig:OFDM_symbol_subcarrier}
\end{figure}

The spectrum width of each subcarrier is defined by the subcarrier spacing $\Delta_{\rm F}\!=\!B/N_{\rm ST}$, which is a parameter specified in 802.11. The spacing has been historically set to $\Delta_{\rm F}\!=$ 312.5\,KHz, and has been reduced to $\Delta_{\rm F}\!=$ 78.125\,KHz in 802.11ax. For a fixed channel bandwidth, this entails that the number of modulated symbols that can be simultaneously transmitted over different subcarriers increases by four times, as depicted in Fig.~\ref{fig:OFDM_symbol_subcarrier}.

\Gls{ofdm} transmissions can be implemented by upconverting each modulated symbol to the frequency of the $k$-th subcarrier ($k\! \in \! \{0, \dots, N_{\rm ST}\!-\!1\}$).
However, to improve the efficiency of the procedure, 
the signal is usually obtained by forwarding the $N_{\rm ST}$ modulated symbols---referred to as \gls{ofdm} samples---though an \gls{ifft} block.
The set of $N_{\rm ST}$ samples at the output of the \gls{ifft} block is referred to as an \gls{ofdm} symbol,
which is the fundamental unit of a Wi-Fi transmission. 
Importantly, a \gls{cp} is added at the beginning of each \gls{ofdm} symbol by repeating the last portion of the same symbol. This acts as a \gls{gi} between symbols, helping to reduce \gls{isi} even in environments with significant delay spread. 
The operations are performed in the base-band domain,
and the signal is then upconverted to the center carrier frequency $f_c$.
Being $a_{k}$ the $k$-th modulated symbol (\gls{ofdm} sample), 
the $m$-th \gls{ofdm} symbol is obtained as\vspace{-0.25cm}
\begin{equation}
    x_m(t) = \sum_{k=-N_{\rm ST}/2}^{N_{\rm ST}/2-1} a_{m, k} \, e^{j2\pi (f_c+k\Delta_{\rm F})t}.\vspace{-0.2cm}
\end{equation}

\Gls{ofdm} effectively mitigates the frequency-selective fading responsible for \gls{isi}, i.e., the phenomenon for which delayed copies of a transmitted signal carrying previous symbols overlap at the receiver with the signal carrying the current symbol.
By design, the \gls{ofdm} symbol duration is $T \! = \! 1/\Delta_{\rm F}$, resulting in an \gls{ofdm} symbol time that is $N_{\rm ST}$ times longer than that of a single-carrier system.
By appropriately choosing $N_{\rm ST}$, the symbol duration can be made significantly larger than the typical delay spread of the channel, thereby reducing \gls{isi} and improving the overall system's robustness. 
The change in the subcarrier spacing introduced by 802.11ax goes in this direction. Indeed, it results in a symbol time of 12.8\,µs compared to the 3.2\,µs in previous amendments, thus enhancing performance in environments with high delay spreads, such as outdoor scenarios. Moreover, longer symbol duration reduces cyclic prefix overhead relative to the symbol length, increasing efficiency, especially indoors. However, longer symbols require more precise \gls{cfo} correction to avoid signal degradation. 

For what concerns the cyclic prefix duration,
the legacy value is 0.8\,µs, 
ideal for indoor environments where delay spread is minimal,
as it minimizes overhead. 
In addition to this, 802.11ax introduced two more options: 1.6\,µs and 3.2\,µs. The 1.6\,µs cyclic prefix strikes a balance between efficiency and robustness, making it suitable for outdoor use and uplink \gls{mu}-\gls{mimo}/\gls{ofdma} transmissions. 
The 3.2\,µs cyclic prefix is specific for extreme delay spread environments, such as outdoor uplink \gls{mu}-\gls{mimo}/\gls{ofdma}, where longer \glspl{gi} prevent \gls{isi} by ensuring that reflections of one \gls{ofdm} symbol do not overlap with the subsequent one.

\paragraph*{Latest 802.11bn Enhancements}

Rate adaptation selects the best transmission parameters based on current channel and noise conditions. However, it may react too slowly to effectively handle rapidly time-varying interference.
To address this limitation, 802.11bn is introducing receiver-based \gls{im} techniques. 
\gls{im} enhances interference estimation and suppression by embedding additional pilot tones within the data portion of the \gls{ppdu}. These pilots provide insight into the characteristics of the interference observed during reception.
When the number of receiver antennas exceeds the number of spatial streams, \gls{im} pilots can support noise covariance estimation and receiver-side beamforming, allowing the receiver to suppress portions (or even the entirety) of the interference.
A trade-off must be made in pilot tone allocation: denser pilot insertion improves interference tracking but reduces data throughput. 
Implementing \gls{im} also raises configuration challenges.
Fixed pilot patterns are simple but inflexible.
Flexible configurations offer a limited set of predefined layouts, balancing adaptability and complexity.
Fully configurable modes provide maximum flexibility but at the cost of high signaling overhead and implementation difficulty.
In practice, fixed or flexible configurations are preferred due to their reduced complexity. It is also important to note that \gls{im} may be less effective in downlink \gls{mu}-\gls{mimo} scenarios, where receiver constraints can limit its interference suppression capabilities.


\subsection{Multi-Antenna Systems}\label{sec:MIMO}

The availability of multiple antennas at the \gls{ap}, and possibly, at the \glspl{sta} allows improving the robustness to fading (\textit{diversity gain}), the received \gls{snr} (\textit{beamforming gain}), and the capacity (\textit{multiplexing gain}) of a Wi-Fi system by manipulating the data streams before and/or after their transmission over the \gls{mimo} channel (\textit{precoding})~\cite{heath2018foundations}.
The key idea behind these processing techniques is that, if the antennas in a \gls{mimo} system are spaced by at least half a wavelength ($\lambda/2$), they provide space diversity, i.e., they generate independent fading channels, which can be leveraged to enhance the robustness of the system to noise and increase the capacity~\cite{goldsmith2005wireless}.

The diversity gain is linked to the fact that signals transmitted over such independent fading channels are unlikely to experience deep fades simultaneously.
In turn, robustness to fading can be achieved by transmitting the same signal over the different antennas at the transmitter and combining the collected signals at the receiver, obtaining a signal where the fading is reduced. 
This technique is useful in scenarios with low \gls{snr}~\cite{bjornson2023twenty}. 

Beamforming gain is achieved by modifying the complex modulated symbols before transmission to compensate for the channel impairments and increase the \gls{snr} at the receiver(s) antennas. Note that when a technique leads to an increase in the \gls{snr} plus an increase in robustness and/or in capacity, it is referred to as beamforming in 802.11.

Finally, the independent fading channels generated by the multiple antenna pairs can be used for multiplexing, i.e., to transmit multiple data streams simultaneously using the same time-frequency resource. The number of data streams that can be simultaneously transmitted is constrained by the channel rank and the condition number. The rank indicates the number of linearly independent rows (or columns) of the $N_{\rm RX}\!\times\!N_{\rm TX}$-dimensional channel matrix, where $N_{\rm RX}$ and $N_{\rm TX}$ are respectively the total numbers of antennas at the receiver and transmitter sides. The condition number relates to their quality. From a physical standpoint, the rank depends on the number of uncorrelated transmission paths available between the transmitter and the receiver. In \gls{su}-\gls{mimo}, the multi-path propagation in \gls{nlos} scenarios offers the degrees of freedom needed for multiplexing, while \gls{los}-dominated settings reflect in high channel matrices' condition numbers, making it challenging to transmit multiple streams. In addition to this, the 802.11 standard defines the maximum number of spatial streams that can be simultaneously transmitted: 802.11be and 802.11bn-compliant devices shall support up to 8 streams~\cite{henry2024wi}. 

The mapping and spatial processing between the stream(s) to be transmitted ($N_{\rm SS}$) and the transmitted antennas ($N_{\rm TX}$) is represented by the yellow box in Fig.~\ref{fig:PHY_signal_processing} and is performed by applying specific spatial matrices to the modulated and coded symbols obtained from the input bits, as detailed in Section~\ref{sec:mcs}. 
The 802.11 standard provides some examples of the spatial matrices that can be used, but the implementation is not restricted to them. The simplest approach consists of applying a direct mapping, which can be an identity matrix mapping or a diagonal matrix where the diagonal elements are the cyclic shifts in the time domain. When the number of data streams to be transmitted is less than the transmitting antennas, a spatial expansion matrix can be used to replicate the streams over different antennas. The columns of the spatial expansion matrix should be orthogonal among them, and the matrix can also be different for the different \gls{ofdm} subcarriers. If \gls{mimo} channel information is available at the beamformer, a \textit{beamforming steering matrix} designed based on such knowledge can be applied to improve the decoding at the receiver and transmit multiple streams simultaneously. This approach is referred to as \gls{txbf} in 802.11, and the transmitter and receiver are referred to as the beamformer and beamformee, respectively. When multiple streams are to be transmitted, they are simultaneously output through all the beamformer antennas using the combination (precoding) factors defined by the beamforming steering matrix. This technique is identified as \gls{su}-\gls{mimo}. The beamforming steering matrix may be devised to minimize the interference among data streams and improve the decoding quality at the different beamformee antennas. However, note that other types of precoders are not precluded, as the beamforming steering matrix is implementation-dependent.

\begin{figure}
\centering
\includegraphics[width=\figwidth]{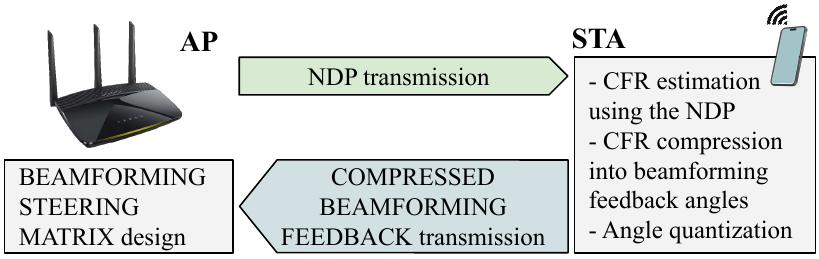}
\caption{\blue{Explicit sounding procedure for beamforming steering matrix calculation at the \gls{ap} (beamformer). The \gls{ap} first transmits a null data packet (NDP) to the \gls{sta}, which estimates the channel frequency response (CFR), compresses it into beamforming feedback angles with quantization, and returns this information as a compressed beamforming feedback frame, enabling the \gls{ap} to design the steering matrix for subsequent beamformed transmissions.}}
\label{fig:beamforming_sounding}
\end{figure}

The beamforming procedure is detailed in the following. We will focus on \gls{txbf}/\gls{su}-\gls{mimo}, while the extension to the \gls{mu}-\gls{mimo} case is detailed in Section~\ref{sec:MU-MIMO}.
At first, a preliminary \textit{channel sounding} phase allows the \gls{ap} (beamformer) to obtain the \gls{cfr} over the different antennas and \gls{ofdm} subcarriers. As introduced before, this channel knowledge at the beamformer is needed to shape the beamforming steering matrix. Channel sounding is continuously performed---at a suggested rate of 10\,ms~\cite{gast2013802}---to provide the \gls{ap} with the most updated channel information to generate proper precoding. Theoretically, two approaches can be followed to acquire the channel information, namely, implicit sounding and explicit sounding. Implicit sounding leverages the channel reciprocity between uplink and downlink channels, and is more convenient from a spectrum efficiency perspective as it requires less control data transmissions with respect to explicit sounding. However, it requires precise hardware calibration to avoid mismatches that could compromise channel reciprocity and degrade performance. For this reason, explicit sounding, which entails obtaining the estimate of the downlink channel from the \gls{sta}, is the only sounding mechanism standardized in 802.11 so far. This approach directly provides the \gls{ap} with the downlink channel information, and does not require relying on channel reciprocity, thus avoiding calibration. The procedure is summarized in Fig.~\ref{fig:beamforming_sounding}, and proceeds as follow~\cite{meneghello2022deepcsi,zhang2023implementation}.
\begin{itemize}
    \item 
        The \gls{ap} transmits a sounding frame, 
        named \gls{ndp}, to allow the \gls{sta} measuring the response of the downlink channel (green arrow in Fig.~\ref{fig:beamforming_sounding}). This is preceded by a \gls{ndpa} frame, which informs the \gls{sta} about the start of the sounding. \blue{To enable channel estimation at the \gls{sta}, the \gls{ndp} contains a number of \glspl{ltf} equal to the number of streams to be transmitted ($N_{\rm SS}$). These \glspl{ltf} are included as different subsequent fields in the \gls{ppdu} and, in turn, are transmitted sequentially in time. Specifically, each \gls{ltf} is transmitted through the $N_{\rm TX}$ \gls{ap} antennas through \gls{csd} (see Section~\ref{subsec:transmitter_block}) and applying an orthogonal mapping over the antennas and \glspl{ltf}. This orthogonal mapping allows the \gls{sta} to obtain a number of independent equations sufficient for the estimation of the $N_{\rm ss} \times N_{\rm TX}$-dimensional \gls{cfr} matrix for each \gls{ofdm} subcarrier.} 
    \item 
        The \gls{sta} uses the known \glspl{ltf} in the \gls{ndp} to estimate the \gls{cfr}, which describes how the signal propagates from the transmitter to the receiver for each pair of transmitter and receiver antennas over the different \gls{ofdm} subcarriers. In turn, the \gls{cfr}, denoted by $\mathbf{H}$, is an $N_{\rm RX} \! \times N_{\rm TX} \! \times \! (N_{\rm SD}\!+\!N_{\rm SP})$-dimensional complex-valued matrix, where $N_{\rm RX}$ and $N_{\rm TX}$ are the numbers of receiver and transmitter antennas, respectively, while $N_{\rm SD}\!+\!N_{\rm SP}$ is the number of \gls{ofdm} subcarriers including data and pilots, as described in Section~\ref{sec:ofdm}. 
    \item    
        This information should then be fed back to the \gls{ap}. To do it quickly and efficiently, the $N_{\rm RX} \! \times N_{\rm TX}$ \gls{cfr} matrices for each \gls{ofdm} subcarrier $k$ are first compressed through \gls{svd}, and only the right unitary matrices are considered for the feedback, as it has been shown that the performance of using them for beamforming is comparable to using the complete \gls{cfr}. In particular, only the sub-matrices consisting of the first $N_{\rm SS}$ columns of the right unitary matrices, referred to as $\mathbf{V}_k$, $\forall k$, are retained. As a further compression, each $\mathbf{V}_k$ matrix is decomposed into rotation matrices, which are fully defined by a set of angular values denoted by $\phi$ and $\psi$. In turn, feeding back such angles is sufficient for the bemaformer to retrieve the $\mathbf{V}_k$ matrix. The number of angles depends on $N_{\rm TX}$ and $N_{\rm SS}$~\cite{haque2023wi}. As a final step, the angles are quantized with a number of bits specified in the standard, and fed back inside a compressed channel feedback frame (blue arrow in Fig.~\ref{fig:beamforming_sounding})~\cite{meneghello2024evaluating}. To further reduce overhead, the 802.11 standard defines grouping strategies to transmit the feedback for a subset \blue{$\hat{N}_{\rm SD}\!=\!N_{\rm SD}/N_g$} of the \gls{ofdm} subcarriers, \blue{where $N_g$ is the grouping factor,} instead of feeding back the angles computed for all of them. Moreover, the compressed channel feedback frame is transmitted unencrypted to reduce latency, which would let the channel estimate be outdated when reaching the \gls{ap}~\cite{meneghello2022deepcsi}. 
    \item 
        Upon receiving the compressed beamforming feedback frame, the \gls{ap} retrieves the angles and uses them to reconstruct the $\mathbf{V}_k$ matrices for all the subcarriers, which are used to define the beamforming steering matrix.
\end{itemize}
The procedure just described is known as non-\gls{tb} sounding and can only be used for \gls{su}-\gls{mimo}. Another channel reporting procedure, namely \gls{tb} sounding, will be detailed in Section~\ref{sec:Multiuser}. Such a procedure is mostly used for \gls{mu}-\gls{mimo} feedback but can also be used to simultaneously collect multiple \gls{su}-\gls{mimo} feedback frames.

\begin{figure}
\centering
\includegraphics[width=.88\figwidth]{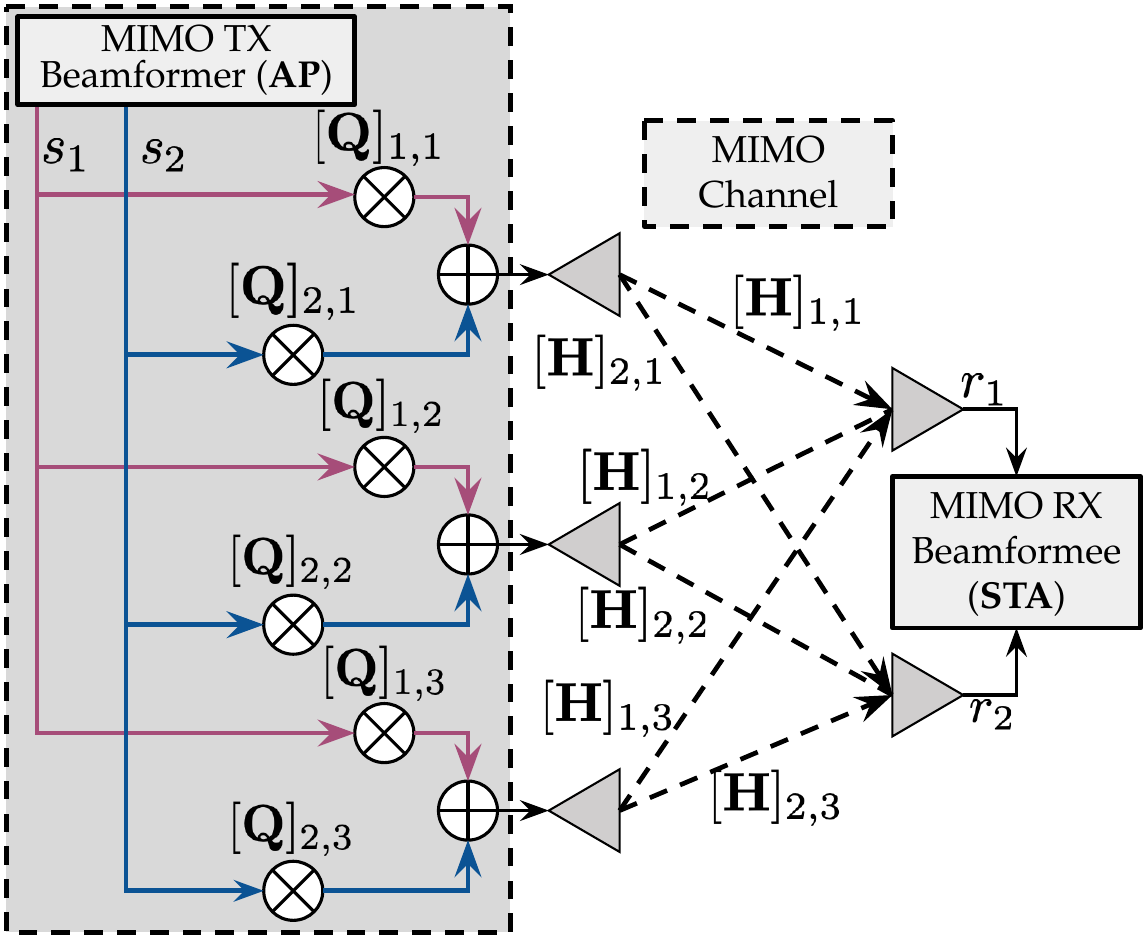}
\caption{\blue{Example of SU-MIMO beamforming with one STA (MIMO RX Beamformee, on the right) requesting two data streams $s_1$ and $s_2$. The AP (MIMO TX Beamformer, on the left) obtains the $\mathbf{Q}$ precoding weight matrix using the CFR of the MIMO channel $\mathbf{H}$. $[\mathbf{A}]_{x,y}$ indicates the entry of the matrix in row $x$ and column $y$. The two data streams $s_1$ and $s_2$ are multiplied by the corresponding antenna weights before transmission. $r_1$ and $r_2$ represent the signals collected at the two receiving antennas.}}
\label{fig:SU-MIMO}
\end{figure}

In addition to the beamforming feedback matrix, 802.11ax has introduced the so-called \gls{cqi} feedback, entailing the transmission to the beamformer of the singular values obtained from the \gls{svd} decomposition of the \gls{cfr} matrix. This data is useful for \gls{mcs} selection. Once the channel feedback is obtained, the beamformer derives the beamforming steering matrix through an implementation-dependent procedure. Possible strategies are \gls{zf} and \gls{mmse} precoding. Through \gls{zf}, the steering matrix $\mathbf{Q}_k$ is obtained by applying channel inversion: $\mathbf{Q}_k \!=\!\! \mathbf{\Tilde{V}_k}\!\left(\mathbf{\Tilde{V}_k}^\dag\mathbf{\Tilde{V}_k}\right)^{\!-1}$\!. The \gls{mmse} method includes an additional regularization term that improves reception performance in low \gls{snr} scenarios~\cite{peel2005vector}. An example of applying the beamforming steering matrix is depicted in Fig.~\ref{fig:SU-MIMO}, where the \gls{sta} is requesting two data streams.
After data transmission, the beamformee may use \gls{mmse} equalization to retrieve and decode the received streams. 
The decoding matrix is usually obtained using the \gls{cfr} estimated on the \gls{ltf} included in the transmitted frame, and allows separating the different streams simultaneously transmitted~\cite{perahia_stacey_2008,meneghello2024whack}. 

Overall, while explicit sounding is the best strategy to provide the beamformer with precise information to shape the steering matrix, it comes with significant overhead due to the feedback transmission, which increases with the number of transmit antennas, spatial streams, and bandwidth~\cite{reshef2022future}. For example, to support 16 data streams (a number considered for 802.11be but not included in the amendment), with 16 antennas at the beamformer and 320~MHz bandwidth, the feedback size---applying the most aggressive compression and quantization supported by the 802.11 standard---would be about \blue{22.4\,kB (120 $\phi$ angles quantized with 4 bits each and 120 $\psi$ angles quantized with 2 bits each, for each of the $\hat{N}_{\rm SD}\!=\!249$ reported subcarriers, i.e., $N_{\rm SD}\!=\!3984$ data subcarriers with a grouping of $N_g\!=\!16$~\cite{rumman2025shrink})} corresponding to an airtime overhead of about \blue{7.5\,ms (considering a typical feedback transmission rate of $24$~Mb/s for high reliability)}, which is a large amount considering the typical sounding rate of 10\,ms. As such, an open research avenue focuses on reducing the overhead of the explicit sounding process. This includes innovations in compressed feedback matrices and optimized channel sounding protocols. As for the former, researchers are proposing to use deep learning strategies in place of the standard feedback compression algorithm to further reduce the airtime overhead linked with feedback transmission~\cite{sangdeh2020lb,hu2023learnable,Bahadori2023splitbeam,wu2025mimo} (see Appendix~\ref{sec:AIML}). Regarding the optimization of the channel sounding procedure, a possible approach is to adapt the sounding interval to the variability of the channel as initially proposed in~\cite{bejarano2014mute, sangdeh2021deepmux}. Another strategy consists of using some side information to reduce the sounding rate~\cite{mei2024learning}. None of these solutions is currently implemented in the 802.11 standard, but their integration will be key to increase the number of antennas of a \gls{mimo} system and in turn its multiplexing gain. 


\subsection{Maximum Achievable PHY Data Rate}\label{subsec:peakrate}

Taking as an example the most recent 802.11be amendment, the maximum data rate achievable at the physical layer for the specific combination of \gls{mcs}, \gls{ofdm} and \gls{mimo} parameters introduced before is obtained as
\begin{equation}
\text{Max. PHY data rate} = \frac{N_{\rm SD} \cdot N_{\rm CBPS} \cdot R \cdot N_{\rm SS}}{T_{\rm SYM}}
\label{eqn:formula_rate}
\end{equation}
where:
\begin{itemize}
\item $N_{\rm SD} \!\in\!$ \{234, 468, 936, 1960, 3920\} is the number of data subcarriers for channel bandwidths  \{20\,MHz, 40\,MHz, 80\,MHz, 160\,MHz, 320\,MHz\}, respectively.
\item $N_{\rm CBPS} \in$ \{1, 2, 4, 6, 8, 10, 12\} is the number of coded bits per \gls{ofdm} symbol for modulations \{BPSK, QPSK, 16QAM, 64QAM, 256QAM, 1024QAM, 4096QAM\}, respectively.
\item $R \in \{\frac{1}{2}, \frac{2}{3}, \frac{3}{4}, \frac{5}{6}\}$ is the coding rate.
\item $N_{\rm SS} \in \{1, 2, 4, 8\}$ is the number of spatial streams.
\item $T_{\rm SYM} \!=\!$ 12.8\,$\mu$s$+T_{\text{GI}}$ is the \gls{ofdm} symbol duration, including a \gls{gi} $T_{\text{GI}} \in $\{0.8\,\text{us}, 1.6\,\text{us}, 3.2\,\text{us}\}.
\end{itemize}
While Eq.~(\ref{eqn:formula_rate}) refers to the peak data rate on a single link, \gls{mlo} described in Section~\ref{sec:MLO} can further increase the data rate in 802.11be by using several physical links in parallel.

\subsection{Physical Layer Protocol Data Units (PPDUs)}\label{subsec:ppdus}

The \gls{ppdu} represents the single unit of information to be transmitted over the wireless channel and consists of two main parts: the \gls{phy} preamble and the data payload. The preamble is critical for facilitating successful communication, as it contains information needed at the receiver to perform automatic gain control, timing synchronization, channel estimation, decoding, and demodulation. The data payload includes one or more \glspl{psdu} (as described later, in Section~\ref{sec:packet_aggregation}), where a \gls{psdu} entails the Application data and the \gls{pci} (header) for the Transport, Network and \gls{mac} layers functionalities. Various \gls{ppdu} formats have been defined in the different versions of the 802.11 standard, each optimized for specific transmission conditions and user scenarios. In particular, the design of the preamble and field structures has evolved across Wi-Fi generations to accommodate new features, while maintaining backward compatibility with earlier amendments.


\subsubsection{PPDU Formats}\label{subsec:ppduformat}

\begin{figure}[t]
\centering
\includegraphics[width=\linewidth]{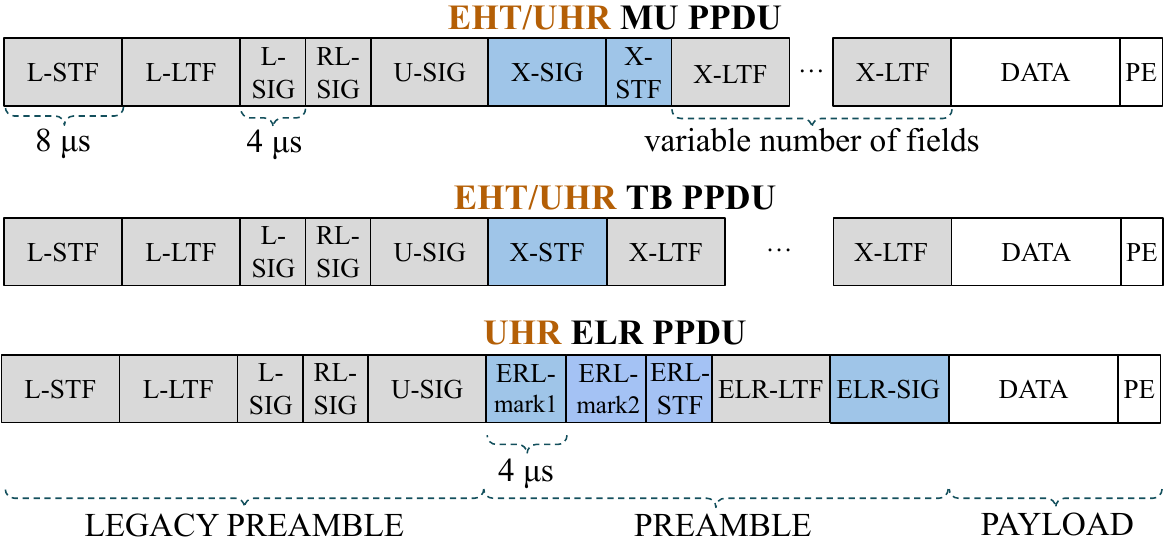}
\caption{PPDU formats in IEEE 802.11be/bn. The fields that differ in the EHT MU and EHT TB PPDU formats are highlighted in blue. The width of the \blue{gray and blue} boxes is associated with their time duration. `X' stands for either `EHT' or `UHR'. \blue{The length of the `DATA' field is variable.}}
\label{fig:PPDU}
\end{figure}

Two \gls{ppdu} formats are defined in 802.11be, building upon two formats originally defined in 802.11ax. Additionally, as part of the 802.11bn standardization process, an additional \gls{ppdu} format is being considered for long-range communications. The three formats are depicted in Fig.~\ref{fig:PPDU}, and described next. Note that, as part of the payload, after the Data field, a Packet Extension (PE) field is added to take the channel occupied and provide the receiver with more time to process the received data. 
\begin{itemize}
    \item \textit{Multi-user \gls{ppdu} (\gls{eht}/\gls{uhr} MU)}:
        Used for both single-user and multi-user downlink transmissions and non-TB single-user uplink transmissions. 
        Multi-user transmissions can be implemented through either \gls{ofdma} or \gls{mu}-\gls{mimo},
        depending on the \gls{ru} allocation (see Section~\ref{sec:Multiuser}).
        The \gls{eht} MU \gls{ppdu} includes an \gls{eht}-\gls{sig} field, 
        which provides details for decoding this specific \gls{ppdu} format.
    \item \textit{Trigger-based \gls{ppdu} (\gls{eht}/\gls{uhr} TB)}: 
        Used for uplink transmissions in response to a trigger frame sent by the \gls{ap}.
        This format is crucial for coordinating uplink \gls{ofdma} and \gls{mu}-\gls{mimo} (see Section~\ref{sec:Multiuser}), 
        reducing contention and increasing efficiency when multiple \glspl{sta} transmit simultaneously. 
        Additionally, the \gls{eht}-\gls{stf} is twice as long as in the \gls{eht} MU \gls{ppdu} version.
        This allows improving performance, 
        particularly in environments with multiple simultaneous uplink transmissions.
        The \gls{eht}-\gls{sig} field is not present as the information has already been communicated to the \gls{sta} inside the trigger~\cite{mentorBEu-sig}.
    \item \textit{\Gls{elr} \gls{ppdu} (\gls{uhr} ELR)}:
        Introduced in 802.11bn to improve the range of uplink transmissions and overcome the link budget imbalance between downlink and uplink channels, which occurs because \glspl{ap} often transmits at higher power levels than \glspl{sta}~\cite{mentorELRppdu,mentorELRppdu2}. The \gls{elr} \gls{ppdu} is configured for single-user transmission over a 20\,MHz channel, supporting \gls{mcs} 0 and 1 with four repetitions of frequency domain duplications across a 52-tone \gls{ru}. It operates in the 2.4\,GHz band for both uplink and downlink, and in the 5\,GHz and 6\,GHz bands for uplink only. This configuration mitigates uplink signal limitations while supporting data rates higher than legacy \gls{dsss}. Hence, \gls{elr} improves uplink range and reliability, and is thus particularly useful in scenarios where Wi-Fi coverage extends beyond typical indoor environments, such as outdoor campuses, industrial sites, and low-data-rate applications like \gls{iot} and telemetry. To allow for better detection at the receiver, the transmitter of an \gls{elr} \gls{ppdu} is expected to synchronize its transmit clock to that of the \gls{ap}. This can be done by compensating the frequency offset relative to packets received from the \gls{ap}.
\end{itemize}

\subsubsection{Preamble Design}\label{subsubsec:pramble}
A critical objective of 802.11be and now 802.11bn is maintaining backward compatibility with earlier Wi-Fi amendments, while operating across the 2.4\,GHz, 5\,GHz, and 6\,GHz bands. The preamble plays a key role in facilitating this compatibility, providing key transmission parameters such as the \gls{mcs} and the number of spatial streams. For this reason, their preambles include legacy (L) fields, namely L-\gls{stf}, L-\gls{ltf}, and L-\gls{sig}, which enable some common initial processing steps on the signal(s) collected at the receiver. To start with, the L-\gls{stf} allows detecting the starting of the \gls{ppdu}, and performing \gls{agc} and coarse carrier recovery. After that, the L-\gls{ltf} enables fine carrier and timing recovery and channel estimation for the equalization of the subsequent SIG fields in the preamble. Once equalized, the L-\gls{sig} together with its replication, referred to as repeated legacy \gls{sig} (RL-\gls{sig}), provides the information to recognize if the \gls{ppdu} is in one of the 802.11be formats (MU or TB). If so, it also indicates the specific format. \blue{Note that before the IEEE 802.11n amendment, only the L-\gls{stf} and L-\gls{ltf} filed were indicated as `PHY preamble', while the `L-\gls{sig} was referred to as `PHY header'. Instead, the current standard terminology makes less use of `header' for the PHY layer fields and refer to all the fields added to the \gls{psdu} as `preamble'.}

Following the legacy fields in the preamble, 802.11be introduced a universal \gls{sig} (U-\gls{sig}) field to simplify the detection of different \gls{phy} versions, and ensure compatibility with both older and future devices. The U-\gls{sig} spans two OFDM symbols, and contains both version-independent and version-dependent bits~\cite{mentorBEu-sig}. Among the version-dependent bits, the U-\gls{sig} indicates whether \gls{csr} or \gls{cbf} are used (see Section~\ref{sec:mapc_features}). 

The subsequent EHT/UHR-\gls{sig} field is only present in EHT/UHR MU \glspl{ppdu}, and provides information regarding the use of \gls{ofdma} or \gls{mu}-\gls{mimo} as the multi-user technology, including the frequency and spatial resources allocated to each user, and their assigned \gls{mcs} \blue{(see Section~\ref{sec:Multiuser})}.

Finally, both the MU and TB \gls{ppdu} preambles include 802.11 version-specific short and long training fields: EHT/UHR-\gls{stf} and EHT/UHR-\gls{ltf}. The former is used to refine the \gls{agc}, carrier, and timing. The latter is used to obtain a channel estimate to equalize the Data field. The number of EHT/UHR-\glspl{ltf} is at least equal to the number of spatial streams per \gls{ru}. 

The UHR \gls{elr} \gls{ppdu} maintains the same structure of the two EHT-\gls{ppdu} formats for the legacy part of the preamble, including the U-\gls{sig} field. In addition, two \gls{elr}-Mark symbols are added for \gls{elr} mode classification~\cite{mentorBNspecificationframework}. \Gls{elr}-\gls{stf} and \gls{elr}-\gls{ltf} are included as in 802.11be. After the \gls{elr}-\gls{ltf}, a \gls{elr}-\gls{sig} is included to indicate the transmission direction, the \gls{mcs} and coding strategy, and the length of the payload.


\commentsection
\subsubsection{Trigger Frames}\hfill

\blue{remove this section and introduce the different trigger frames only when needed}

The trigger-based PPDU format, introduced already in 802.11ax, significantly enhances uplink performance by coordinating multi-user transmissions. Trigger frames are used by the access point (AP) to allocate resources to client devices for uplink transmissions, reducing contention and improving efficiency. A trigger frame contains essential information that instructs client devices on how and when to transmit, including:
\begin{itemize}
\item
\emph{Uplink transmission window}: Defines the duration of the transmission window for client devices. 
\item 
\emph{Assigned client devices}: Lists which devices are permitted to transmit during the allocated window. 
\item
\emph{Resource Unit (RU) allocation}: Assigns specific OFDMA RUs to each client, defining the frequency resources each can use. 
\item
\emph{Spatial stream allocation}: Specifies the number of spatial streams assigned to each client, particularly in MU-MIMO operations. 
\item
\emph{Modulation and Coding Scheme (MCS)}: Informs each client of the MCS level they should use, based on channel conditions.
\item
\emph{Power control information}: Allows clients to adjust their transmission power based on the AP's transmit power and the received signal strength indicator (RSSI). 
\end{itemize}

As shown in Fig.~\ref{fig:TriggerFrame}, 802.11ax introduced support for different types of trigger frames for various functions:
\begin{itemize}
\item
\emph{Basic Trigger Frame}: A simple frame that instructs clients when to transmit, without any additional features.
\item
\emph{Beamforming Report Poll (BRP)}: Requests beamforming feedback from clients, with specific fields to collect the necessary beamforming data.
\item
\emph{Multi-user Block Acknowledgment Request (MU-BAR)}: Enables the AP to request acknowledgment from multiple clients simultaneously, reducing overhead in multi-user environments.
\item
\emph{Multi-user Request to Send (MU-RTS)}: Similar to traditional RTS-CTS (Request to Send-Clear to Send), but for multi-user scenarios, ensuring the air is clear before transmission.
\item
\emph{Buffer Status Report Poll (BSRP)}: Allows the AP to query the clients about their current data buffer status, which helps optimize scheduling for uplink traffic.
\item
\emph{Group Cast with Retries MU BlockAck Request (GCR MU-BAR)}: Used in multicast group transmissions to request BlockAcks from all group members.
\item
\emph{Bandwidth Query Report Poll (BQRP)}: Asks clients to report on the occupancy of 20 MHz RF channels, enabling the AP to manage channel use more efficiently.
\end{itemize}

Upon receiving a trigger frame, client devices must transmit within their assigned \glspl{ru} or spatial streams during the designated window. The AP typically acknowledges these transmissions using individual or multi-user \gls{ba} frames. This coordination reduces collisions and maximizes throughput, making it highly efficient in high-density environments.
The flexibility of trigger frames allows Wi-Fi to handle diverse multi-user scenarios, optimizing uplink transmissions in environments with numerous client devices, as discussed in the following section.

\begin{table}
\vspace{1mm}
\caption{Trigger frame variants and primary use.}
\label{tab:triggerFrame}
\centering
\colorbox{BackgroundGray}{%
\begin{tabular}{ |m{2.5cm}|m{5.0cm}|} 
\toprule
\rowcolor{BackgroundLightBlue}
 \textbf{TF variant} & \textbf{Primary use}   \\ \midrule
 {Basic TF} &  
{Solicit HE TB PPDU in the UL for data transmission
}   \\ \midrule
 {Beamforming Report Poll (BFRP)} &  
{Solicit beamforming sounding feedback}   \\ \midrule
 {MU-BAR} &  
{Solicit Block Acks from multiple STAs}   \\ \midrule
 {MU-RTS} &  
{Allow an AP to initiate a TXOP and protect the TXOP frame exchanges}   \\ \midrule
 {Buffer Status Report Poll (BSRP)} &  
{Solicit buffer status report from STAs}   \\ \midrule
 {Group Cast with Retries (GCR) MU-BAR} &  
{Solicit Block Acks from multiple STAs in a GCR group}   \\ \midrule
 {Bandwidth Query Report Poll (BQRP)} &  
{Solicit bandwidth query reports from STAs to help AP allocate DL/UL MU resources}   \\ \midrule
 {NDP Feedback Report Poll (NFRP)} &  
{Allow an HE AP to collect feedback that is not channel sounding from multiple STAs}
\\ \bottomrule 
\end{tabular}
}
\end{table}
\endcommentsection


\section{Wi-Fi Medium Access Control}
\label{sec:MAC}


Wi-Fi generations have consistently introduced substantial enhancements at the \gls{mac} layer to improve network efficiency, reduce latency, and ensure fair access to the wireless medium, especially in dense environments. However, to understand how Wi-Fi works and why its various enhancements are designed the way they are, it is essential to master the core components of its channel access protocol and understand their broader implications.

This section begins with an overview of \gls{dcf}, the foundational \gls{mac} protocol on which all subsequent enhancements are built. One such enhancement is \gls{edca}, which serves as the default access mechanism in current 802.11 devices.\footnote{Note that we do not cover the \gls{pcf}, now considered obsolete, and the \gls{hcca}, as these mechanisms have seen limited adoption in commercial deployments.} We also introduce \gls{pedca}, a new feature in 802.11bn that addresses a key limitation of \gls{edca}: its inability to effectively manage multiple high-priority accesses within a single \gls{bss} and multiple \gls{obss}. Next, we examine packet aggregation and more deterministic access mechanisms, such as \gls{ap} scheduled transmissions, which extend \gls{edca} by granting the \gls{ap} greater control over \gls{ul} and \gls{p2p} scheduling, and \gls{rtwt}, a feature for latency control. Finally, we present wideband access techniques---including channel bonding, preamble puncturing, and two additional features introduced in~802.11bn, namely \gls{npca} and \gls{dso}---and \gls{sr} for coexistence among \glspl{obss}.


\subsection{Enhanced Distributed Channel Access}
\label{sec:dcf}

The \gls{dcf} protocol has been a fundamental part of 802.11 since its inception, enabling devices to share the unlicensed spectrum fairly and efficiently. It implements \gls{csma} with \gls{ca} through a random backoff countdown and an \gls{arq} to handle packet errors and retransmissions. 

\subsubsection{Distributed Coordination Function}

The basic principle of \gls{csma}/\gls{ca} is simple: 
before transmitting, a device listens to the channel to determine if it is free.
If the channel is busy, the device waits until the channel becomes free to transmit.
\blue{Devices utilize two methods to check whether the medium is free:
(i) physical carrier sensing, which detects energy above -62\,dBm to determine channel occupancy; 
and (ii) virtual carrier sensing, in which the device first decodes the \gls{phy} header when the detected energy is above -82\,dBm, and then decodes the \gls{mac} header to read the Duration/ID field that carries the medium occupancy information. 
For pre-\gls{he} devices, 
the Duration field in the \gls{mac} header is utilized to update the \gls{nav} counter,
which reduces the need for continuous physical carrier sensing and helps conserve power.
Importantly, this update can only occur after the complete \gls{mpdu} has been received and the \gls{fcs} has been validated.
In \gls{he} devices and subsequent amendments,
the HE-SIG-A preamble introduces the \gls{txop} Duration field,
which specifies the remaining time in the current \gls{txop}. 
This mechanism enables expedited \gls{nav} updates, 
as only the \gls{ppdu} preamble needs to be decoded. 
A similar approach is employed in \gls{eht} and \gls{uhr} devices, 
where the U-SIG preamble field provides equivalent functionality.
}

However, relying solely on checking whether the channel is free before transmitting does not prevent multiple devices, concurrently sensing the channel, from transmitting simultaneously, which can result in a collision. To minimize such collisions, 802.11 employs a \textit{collision avoidance} strategy based on the \gls{beb} mechanism. With \gls{beb}, after detecting that the medium is idle, a device waits for a random backoff time, uniformly selected between $0$ and a \gls{cw}. The \gls{cw} is initially set to a minimum value (CW$_\min$) to enable fast channel access, and it doubles after each failed transmission attempt---up to a maximum limit (CW$_\max$)---to improve network stability and throughput under high contention. Importantly, during the backoff process, a device pauses its countdown whenever it detects that the channel is busy, resuming only after the channel has been idle again for at least a \gls{difs}. As a result, the actual backoff duration is uncertain, not only due to the initial random value, but also because of potential interruptions caused by transmissions from other devices.

When a transmission fails due to channel errors or collisions, current 802.11 systems rely on \gls{arq} to enable retransmissions and ensure reliable communication. Retransmissions are triggered when the expected \gls{ack} is not received by the transmitter of the \gls{mpdu} after an \gls{ack} timeout. The transmitter of the \gls{mpdu} then repeats the process of sending the packet and waiting for the \gls{ack} until it is successfully received. This mechanism allows 802.11 devices to recover from errors through multiple retransmission attempts, providing robustness against adverse channel conditions and collisions caused by simultaneous transmissions from multiple devices. However, \gls{arq} retransmissions often necessitate conservative link adaptation, typically involving a reduction in the \gls{mcs} to maintain reliable communication (see Section~\ref{sec:mcs}).

\begin{figure}[t]
\centering
\includegraphics[width=\linewidth]{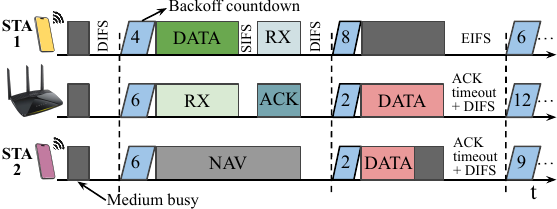}
\caption{\blue{Distributed Coordination Function (DCF) operation for two contending STAs and one AP. The timeline shows DIFS/SIFS/EIFS intervals, DATA and ACK exchanges, NAV periods, ACK timeouts, and intervals when the medium is sensed busy (grey segments); the numbers in the blue boxes represent the backoff countdown values at each station.}}
\label{fig:DCF}
\end{figure}

An example of the \gls{dcf} operation is illustrated in Fig.~\ref{fig:DCF}, where one \gls{ap} and two \glspl{sta} share the same channel. We observe that the first transmission is successful (only \gls{sta}~1 accesses the medium), while the second results in a collision (both the \gls{ap} and \gls{sta}~2 access the medium). 

The interval between receiving an \gls{mpdu} and transmitting the corresponding \gls{ack} is denoted as \gls{sifs} and represents the time required to switch the transceiver circuitry between transmit and receive modes. The \gls{sifs} is shorter than the \gls{difs} to ensure that immediate transmissions such as \glspl{ack} have priority over other stations contending for the medium. Devices that detect the channel as busy but are unable to decode the frame header, and therefore cannot update their \gls{nav}, must wait for an \gls{eifs} instead of a \gls{difs} before resuming their backoff countdown. The \gls{eifs} duration roughly corresponds to the time required to account for the potential \gls{ack} timeout plus a \gls{difs}, protecting ongoing transmissions that the device could not properly decode.

Analyzing the performance of the \gls{dcf} under various scenarios has been a key research focus since the inception of Wi-Fi, primarily due to the complex interactions between traffic load, the number of contending devices, and the \gls{csma}/\gls{ca} mechanism. A seminal contribution in this domain is the work of Bianchi~\cite{bianchi2000performance}, which has served as the foundation for most 802.11 performance analyses over the past 25 years. 


\subsubsection{RTS/CTS}

\begin{figure}[!t]
    \centering
    \subfloat[]
    {\includegraphics[width=0.325\columnwidth]{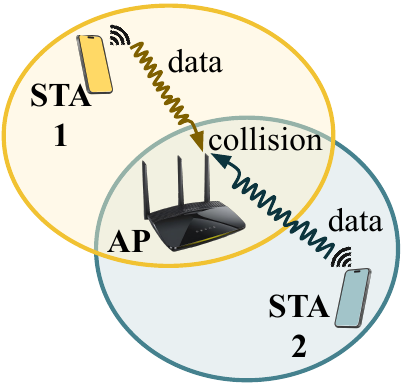}\label{fig:LBT_RTS_CTS_a}} 
    \subfloat[]
    {\includegraphics[width=0.325\columnwidth]{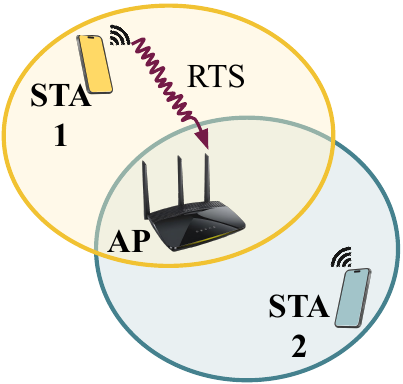}\label{fig:LBT_RTS_CTS_b}} 
    \subfloat[]
    {\includegraphics[width=0.325\columnwidth]{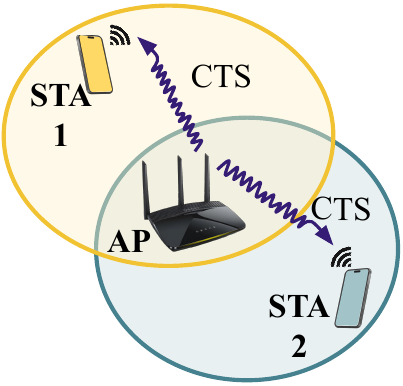}\label{fig:LBT_RTS_CTS_c}}     
    \caption{\gls{rts}/\gls{cts} mechanism: (a) \gls{sta} 1 and \gls{sta} 2 cannot hear each other, \gls{ap} hears both; (b) \gls{sta} 1 sends \gls{rts} to \gls{ap}; (c) \gls{cts} message prevents a collision between \gls{sta} 1 and \gls{sta} 2.}
    \label{fig:LBT_RTS_CTS}
\end{figure}

While \gls{dcf} effectively manages medium access in most scenarios, it suffers in the presence of hidden nodes. For example, as shown in Fig.~\ref{fig:LBT_RTS_CTS_a}, if \gls{sta}~1 and \gls{sta}~2 cannot detect each other's transmissions but both communicate with the same \gls{ap}, collisions may occur at the \gls{ap}. These collisions cannot be resolved using \gls{dcf}, as the devices are unaware of each other’s activity. To address this issue, 802.11 employs the \gls{rts} and \gls{cts} mechanism. When initiating a transmission, an \gls{sta} or \gls{ap} can send an \gls{rts} message, which is shorter and more robust as it is transmitted using a low \gls{mcs} than a typical data frame. The recipient responds with a \gls{cts} message, signaling that the channel is reserved by the \gls{rts} sender. The \gls{rts} and \gls{cts} messages include the duration of the subsequent data transmission, allowing all devices receiving the \gls{rts}, the \gls{cts}, or both, to set up their \glspl{nav}. This virtual carrier sensing mechanism ensures that other devices in the vicinity are informed that the channel is in use, thereby preventing them to further try to access the medium until current transmission ends, as illustrated in Fig.~\ref{fig:LBT_RTS_CTS_b} and Fig.~\ref{fig:LBT_RTS_CTS_c}.

\subsubsection{Traffic Differentiation and Prioritization}

\commentfigtable
\begin{figure}[ht!!]
\centering
\includegraphics[width=0.8\linewidth]{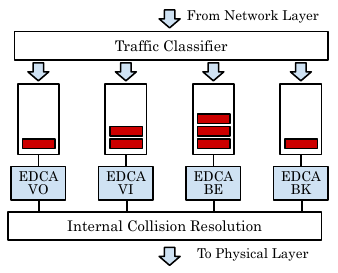}
\caption{EDCA Architecture}
\label{fig:EDCA_architecture}
\end{figure}
\endcommentfigtable

In the \gls{ieee} 802.11e amendment, published in 2005 with the goal of providing traffic differentiation and prioritization capabilities, \gls{edca} was introduced to replace \gls{dcf}, becoming the default channel access mechanism in 802.11 networks~\cite{sanabria2018traffic}. The \gls{edca} architecture introduced several new features\commentfigtable (Fig.~\ref{fig:EDCA_architecture}) \endcommentfigtable
:

\begin{itemize}
    \item \emph{Four \glspl{ac}} are defined: \gls{vo}, \gls{vi}, \gls{be}, and \gls{bk}, ordered from highest to lowest priority. Each \gls{ac} is represented by an independent buffer and \gls{edca} function.
    \item \emph{Traffic classification:} Packets arriving from the network layer are classified into different \glspl{ac} based on information in the IP headers, e.g., the \gls{dscp} field.
    \item \emph{Parallel \gls{edca} functions:} Channel access contention is handled in parallel by all \glspl{ac} with queued packets. Distinct \gls{aifs}, CW$_\min$, and CW$_\max$ values are defined for each \gls{ac}. The \gls{aifs} parameter in \gls{edca} generalizes the \gls{difs} used in \gls{dcf}; in fact, \gls{difs} is equivalent to the \gls{aifs} used for the video access category in default configurations. A maximum \gls{txop} duration is also specified per \gls{ac}, limiting the duration of channel occupancy. Higher-priority traffic (e.g., voice and video) benefits from shorter inter-frame spaces and smaller \glspl{cw}, enabling reduced access delays and improved performance for time-sensitive applications. 
    \item \emph{Internal collision resolution:} If two or more \glspl{ac} complete their backoff at the same time, an internal collision occurs. In such cases, an internal resolution mechanism selects the highest priority \gls{ac} involved in the collision to proceed with transmission.
\end{itemize}

\commentfigtable
Table~\ref{tab:edca_params_ax} presents the default \gls{edca} parameter values recommended by \gls{ieee} 802.11ax, which are optimized for high-density deployments. These differentiated settings ensure that voice and video traffic can meet their delay and jitter requirements even under heavy network load. Additionally, they allow coexistence with legacy devices while maintaining fairness and throughput optimization. 
\endcommentfigtable

While \gls{edca} provides effective traffic differentiation when high-priority traffic is sparse, it struggles under high traffic load. For instance, when multiple uplink flows use the voice or video \glspl{ac}, the resulting competition can lead to frequent collisions and performance degradation due to the use of low CW$_\min$ and CW$_\max$ values. In such scenarios, these flows may perform better if assigned to lower-priority access categories. 

\commentfigtable
\begin{table}[h!!!]
\centering
\begin{tabular}{|c|c|c|c|c|}
\hline
\textbf{Access Category} & \textbf{AIFSN} & \textbf{CWmin} & \textbf{CWmax} & \textbf{TXOP Limit (µs)} \\
\hline
AC\_VO (Voice)        & 2    & 3    & 7    & 2080 \\
AC\_VI (Video)        & 2    & 7    & 15   & 3120 \\
AC\_BE (Best Effort)  & 3    & 15   & 1023 & 0 \\
AC\_BK (Background)   & 7    & 15   & 1023 & 0 \\
\hline
\end{tabular}
\caption{Default \gls{edca} Parameters for \gls{ieee} 802.11ax STAs. \red{Why tables are pushed to the end of the document?}}
\label{tab:edca_params_ax}
\end{table}
\endcommentfigtable

\paragraph*{Latest 802.11bn Enhancements}

\blue{\gls{pedca} is an enhancement of the \gls{edca} mechanism that reduces the access delay distribution tail for VO access category traffic \cite{80211bnD1.1,mentorIntel_1144r1}}. The use of \gls{pedca} by an 802.11bn \gls{sta} shall be balanced to reduce the impact on \glspl{sta} that do not support this mechanism, focusing on the fairness guarantees. Consequently, \gls{pedca} is selectively activated for \glspl{sta} that have repeatedly failed to transmit VO AC traffic using the legacy \gls{edca} procedure. The basic operating principle of \gls{pedca} involves transmitting a Defer Signal (\gls{cts} frame) after the \gls{pedca} \glspl{sta} verify that the channel has remained idle, bypassing the legacy backoff procedure. This mechanism grants \gls{pedca} \glspl{sta} a priority advantage over other \glspl{sta} in accessing the channel. As a result, any ongoing backoff processes from non-\gls{pedca} \glspl{sta} are interrupted, allowing the \gls{pedca} contention process to initiate without competition. Following the Defer Signal, the \gls{pedca} \gls{sta} initiates a new random backoff using the default AIFS and CW parameters for the VO access category. Once the \gls{sta} wins the contention, it must protect the channel using an RTS/CTS exchange before transmitting any data frame, following the legacy procedure.


\subsection{Packet Aggregation}
\label{sec:packet_aggregation}

\begin{figure}
\centering
\includegraphics[width=\linewidth]{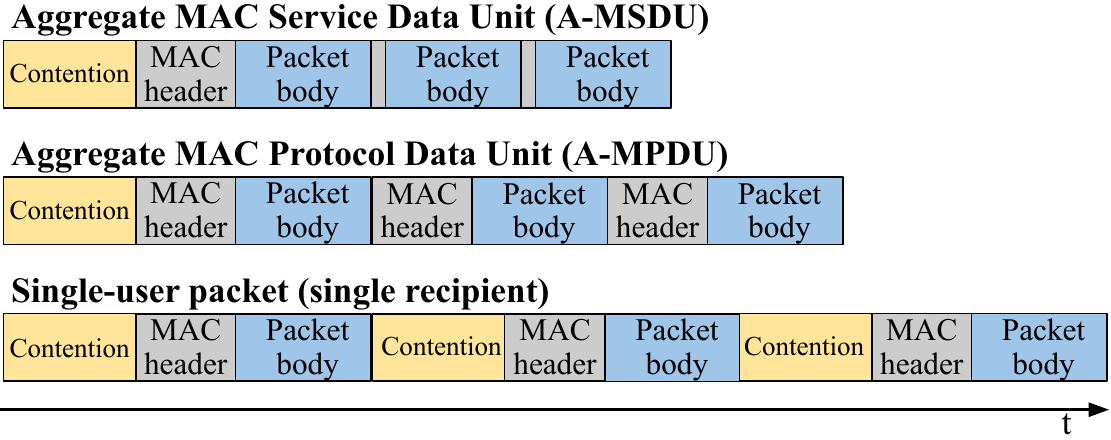}
\caption{\blue{\gls{amsdu} and \gls{ampdu} packet aggregation strategies versus single-packet transmission, highlighting how aggregation reduces contention overhead and improves MAC efficiency. Top: A-MSDU, where multiple packet bodies are encapsulated under a single MAC header after one contention period. Middle: A-MPDU, where several complete MAC Protocol Data Units (each with its own header) are aggregated into one transmission opportunity. Bottom: separate single-user packets requiring repeated contention.}}
\label{fig:PacketAggregation}
\end{figure}

Packet aggregation is a fundamental technique in 802.11 networks to improve airtime efficiency by bundling multiple data packets into a single transmission~\cite{skordoulis2008ieee}. Originally introduced in 802.11n, it has become essential for supporting high-throughput applications such as large file transfers and high-quality video streaming. Indeed, without packet aggregation, each data packet requires its own \gls{phy} preamble and \gls{mac} header, resulting in considerable protocol overhead. In addition, control frames such as \gls{rts}/\gls{cts} and \glspl{ack}, interframe spaces, and the initial backoff duration must all be considered in the total airtime. By assembling multiple packets into a single transmission, packet aggregation minimizes these overheads, as they are included only once per transmission. This not only reduces contention and the likelihood of collisions (as fewer transmission attempts are needed) but also improves overall system efficiency, leading to higher network throughput and lower latency. 802.11 employs two primary types of packet aggregation, namely \gls{ampdu} and \gls{amsdu}, illustrated in Fig.~\ref{fig:PacketAggregation} and detailed next.

\textit{\Gls{ampdu}} aggregation combines multiple \glspl{mpdu} into a single \gls{ppdu}, separating them through \gls{mpdu} delimiters. Each \gls{mpdu} retains its own \gls{mac} header and trailer, but the aggregated transmission uses a single \gls{phy} preamble and interframe space, minimizing \gls{phy} overhead. The receiver processes the aggregated \glspl{mpdu} and sends a \gls{ba} to confirm reception, which reduces the number of required \glspl{ack} and the associated airtime overhead. \Gls{ampdu} aggregation is particularly effective in scenarios involving large data transmissions, such as video streaming or bulk data transfers.

\textit{\Gls{amsdu}} aggregation operates at a higher level than \gls{ampdu} by combining multiple \glspl{msdu} into a single \gls{mpdu}. Unlike \gls{ampdu}, \gls{amsdu} aggregation shares a single \gls{mac} header and trailer for all the aggregated \glspl{msdu}, which further reduces overhead. All aggregated \glspl{msdu} share one \gls{mac} header and trailer, resulting in lower overhead compared to \gls{ampdu}. \gls{amsdu} aggregation is efficient for smaller packet sizes, \blue{where reducing MAC overhead significantly impacts performance.} However, \gls{amsdu} is less resilient to errors: if a single \gls{msdu} in the aggregated frame is corrupted, the entire \gls{mpdu} must be retransmitted. 

In recent 802.11 amendments, packet aggregation continues to evolve to support the growing demands of modern wireless networks, particularly when combined with the use of wideband channels, \gls{mu}-\gls{mimo}, and \gls{ofdma} technologies. Each 802.11 amendment has progressively increased the number of \glspl{mpdu} that can be aggregated. In 802.11be, this number has been expanded to 1024 \glspl{mpdu}, representing a substantial enhancement in throughput and transmission efficiency.


\subsection{AP Scheduled Transmissions}\label{sec:AP_scheduled}

In dense and highly congested environments, the contention among devices to access the medium results in significant inefficiencies, including frequent collisions and transmission delays, particularly when numerous devices attempt to communicate simultaneously. 
\commentsection
Enhanced \gls{qos} management is a key focus of 802.11 to address the increasing demand for reliability, latency guarantees, and spectrum efficiency. 802.11 is introducing advancements to optimize \gls{ul} access, facilitate peer-to-peer communication, support aperiodic low-latency traffic (see Section~\ref{sec:dcf}), and leverage secondary channel access (see Section~\ref{sec:wideband_op}).
\endcommentsection
\gls{ap}-triggered transmissions, introduced in 802.11ax, change this dynamic by allowing the \gls{ap} to take control of when and how \glspl{sta} transmit. Instead of devices initiating communication on their own based on the regular backoff mechanism, the \gls{ap} manages and schedules \gls{ul} and \gls{p2p} transmissions, reducing contention and improving overall network performance.

\blue{In this \gls{ap}-scheduled mode, trigger-based transmissions allow the \gls{ap} to enforce \gls{tspec} and \gls{scs} requirements directly---bypassing \gls{edca} contention---and perform per-flow aware scheduling within the \gls{bss}. While this provides IntServ-like control at the \gls{mac} level, it remains a local (intra-\gls{bss}) mechanism rather than a full end-to-end IntServ architecture.}

\subsubsection{Trigger-based UL Transmission}

This is a key feature introduced in 802.11ax and enhanced in 802.11be with \gls{scs} and \gls{qos} management to improve \gls{ul} transmission efficiency~\cite{bellalta2019ap}. Trigger-based \gls{ul} allows the \gls{ap} to schedule \gls{ul} transmissions from multiple \glspl{sta} simultaneously using \gls{ofdma} (see Section~\ref{sec:OFDMA}). To do so, the \gls{ap} sends a trigger frame that allocates \glspl{ru} to specific \glspl{sta}, which transmit their data simultaneously using the assigned \glspl{ru} immediately after receiving the trigger frame. The \gls{ap} answers by sending a Multi-\gls{sta} \gls{ba} frame.



\subsubsection{SCS and QoS Management}

Triggered access optimization has been introduced in 802.11be to improve \gls{qoe} for real-time applications such as gaming, industrial control systems, and other delay-sensitive use cases that follow a bidirectional and periodic traffic pattern. These applications require predictable latency and guaranteed minimum data rates, which can be challenging in contention-based networks like Wi-Fi. The core idea behind triggered \gls{ul} access optimization is the establishment of a desired \gls{ul} access interval. This interval provides deterministic scheduling of \gls{ul} transmissions, effectively bounding the delay while ensuring a minimum data rate. The mechanism minimizes jitter and delay by aligning transmission opportunities with traffic flow requirements, which is critical for time-sensitive applications.  

In 802.11be, \glspl{sta} can communicate their \gls{qos} expectations to the \gls{ap}. This is achieved through the \gls{qos} Characteristics element, a standardized structure that conveys the key \gls{qos} parameters for a traffic flow. The \gls{qos} Characteristics element includes the following: parameters that define the \gls{qos} requirements of specific traffic flows (\gls{ul} and \gls{dl}), support for \gls{scs} which enables classification of traffic streams for optimized handling, and integration with \gls{rtwt} for additional scheduling flexibility and latency control (see Section~\ref{sec:rtwt}).


\subsubsection{Triggered P2P Transmission}

The concept of \gls{txop} sharing was first introduced in 802.11ax, where multiple \glspl{sta} were allowed to transmit simultaneously within a shared \gls{txop} during multi-user transmissions such as in \gls{ofdma}. 802.11be extends this concept further with triggered \gls{p2p} transmission, where the \gls{ap} can share its \gls{txop} with \glspl{sta}, allowing them to use the allocated time for direct \gls{sta}-to-\gls{sta} communication, using the \gls{txop} Sharing (Mode 2) mechanism. This mechanism supports direct \gls{sta}-to-\gls{sta} data exchanges, which are essential for applications such as wireless file transfers between devices (e.g., phone-to-printer communication), video streaming (e.g., laptop-to-monitor wireless display), and \gls{vr} applications requiring low-latency peer communication.

\begin{figure}
\centering
\includegraphics[width=\linewidth]{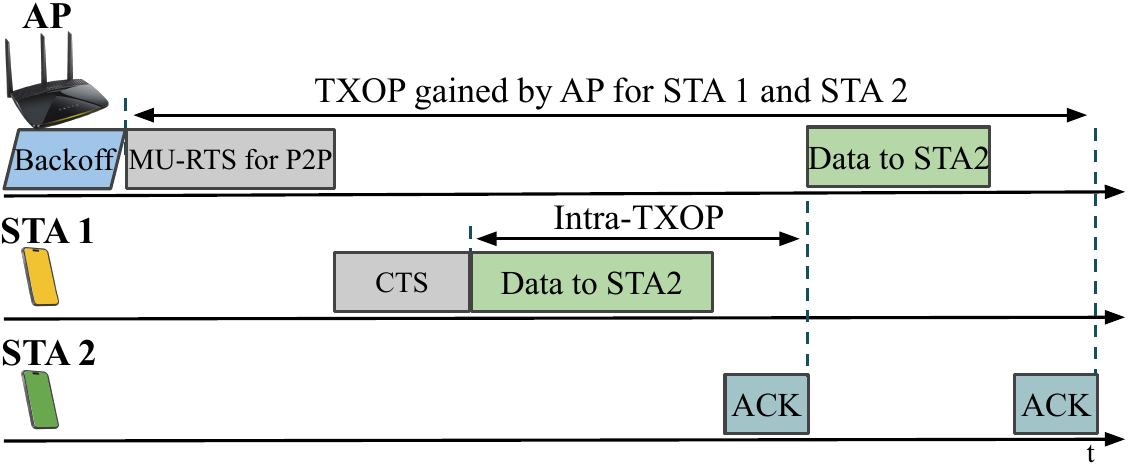}
\caption{\blue{Illustration of triggered \gls{p2p}: \gls{txop} Sharing Mode 2. The \gls{ap} first wins the channel after backoff and gains a \gls{txop} for \gls{sta}~1 and \gls{sta}~2 using an MU-RTS/CTS exchange, then, within the same intra-\gls{txop}, triggers a direct data transmission from STA~1 to STA~2, which is completed with individual ACKs from STA~2 and the \gls{ap}.}}
\label{fig:triggered_P2P}
\end{figure}

In triggered \gls{p2p} transmission, the \gls{ap} can use favorable contention parameters (e.g., small \gls{cw} values) to increase its chances of winning a \gls{txop}. The \gls{ap} transmits a \gls{mu-rts} \gls{txop} Sharing Trigger frame Mode 2, allocating a portion of the \gls{txop} for \gls{p2p} transmission. Other devices defer their transmissions, which improves spectrum utilization and reduces contention within the \gls{bss}. Moreover, direct communication between two \glspl{sta} enhances airtime efficiency, as the transmitted data does not need to be duplicated when it passes through the \gls{ap}. For example, in the scenario illustrated in Fig.~\ref{fig:triggered_P2P}, the \gls{ap} gains a \gls{txop} and allocates a portion of it to \gls{sta}~1. \gls{sta}~1 responds with a \gls{cts} frame and then uses the allocated time to send data directly to \gls{sta}~2. \gls{sta}~2, in turn, sends an \gls{ack} to \gls{sta}~1. If the shared TXOP period ends within the remaining time, the \gls{ap} can reclaim the rest of the \gls{txop} duration to send its data.

\paragraph*{Latest 802.11bn Enhancements}

802.11bn aims to extend the concept of \gls{p2p} to a \gls{p2p} group, composed of two or more \glspl{sta}. Similarly, the \gls{ap} shares the obtained \gls{txop} with a previously created \gls{p2p} group, and these \glspl{sta} will communicate with each other within the shared \gls{txop} time length. The idea behind this mechanism is to offload the \gls{ap} scheduling task for each \gls{p2p} communication link, easing the \gls{ap} workload, and maximizing the communication efficiency. Within the shared \gls{txop}, various self-organizing mechanisms can be employed to coordinate channel access among the designated \gls{p2p} group members, as detailed next.
\begin{itemize}
\item 
\emph{Contention-based approach:} Each device that has a non-empty buffer for \gls{p2p} traffic selects a random backoff counter, similar to the standard \gls{csma}/\gls{ca} process, and once a device wins the contention, it starts transmitting its data to its intended peer within the group.
\item
\emph{Rule-based approach:} It predefines an order or pattern for devices within the \gls{p2p} group to access the \gls{txop} allocated by the \gls{ap}. This approach eliminates the need for contention and focuses on deterministic scheduling.
\item
\emph{Hybrid approach:} In some scenarios, a hybrid approach combining contention and rule-based methods may be employed. For example, devices with higher-priority traffic (e.g., latency-sensitive packets) could be given a shorter backoff window in a contention-based approach, ensuring they win more often. Alternatively, the \gls{p2p} group could follow a rule-based order initially, switching to a contention-based approach only when there are changes in group membership or unexpected traffic patterns.
\end{itemize}

While triggered \gls{p2p} transmission improves spectrum utilization and reduces collisions, it requires the \gls{ap} to have visibility into the buffer state of the participating \glspl{sta} or rely on additional information to make efficient decisions about \gls{txop} sharing. By enabling direct communication between \glspl{sta}, this mechanism enhances network efficiency, reduces transmission delays, and supports latency-sensitive peer-to-peer applications.


\subsection{Restricted Target Wake Time (R-TWT)}
\label{sec:rtwt}

802.11be introduced \gls{rtwt} to support latency-sensitive applications by providing greater control over transmission timing. 
\gls{rtwt} builds upon \gls{twt} (described in Section~\ref{sec:twt}),
which was originally designed to reduce energy consumption.

To introduce determinism in transmissions, 
\gls{rtwt} addresses a key limitation of traditional \gls{twt},
namely,
the lack of coordination among \glspl{sta} regarding scheduled wake times.
Indeed, in conventional \gls{twt}, 
other \glspl{sta} might be unaware of an ongoing agreement, 
leading to potential collisions or inefficiencies. \Gls{rtwt} overcomes this by ensuring that only \glspl{sta} with an established \gls{rtwt} membership may access the channel during a designated restricted service period. 

While \gls{rtwt} retains the foundational mechanisms of \gls{twt}
(explained in detail in Section~\ref{sec:twt}), 
it introduces several key enhancements for deterministic latency:
\begin{itemize}
    \item
    During schedule negotiation,
    the \gls{ap} and associated \glspl{sta} establish an \gls{rtwt} agreement. 
    This agreement defines the start times, durations, and periodicity of the \glspl{sp}. 
    It also includes a list of \glspl{tid}, 
    allowing prioritization of traffic types within the \glspl{sp}.
    
    \item
    To enforce the restricted access during an \gls{sp}, 
    the \gls{ap} initiates explicit channel protection via control signaling,
    such as \gls{cts}-to-Self or Quiet Element frames,
    at the beginning of each \gls{sp}.
    This signaling silences both compliant and non-compliant \gls{twt}/\gls{rtwt} \glspl{sta},
    ensuring predictable and interference-free transmissions.
\end{itemize}

\begin{figure}
\centering
\includegraphics[width=\linewidth]{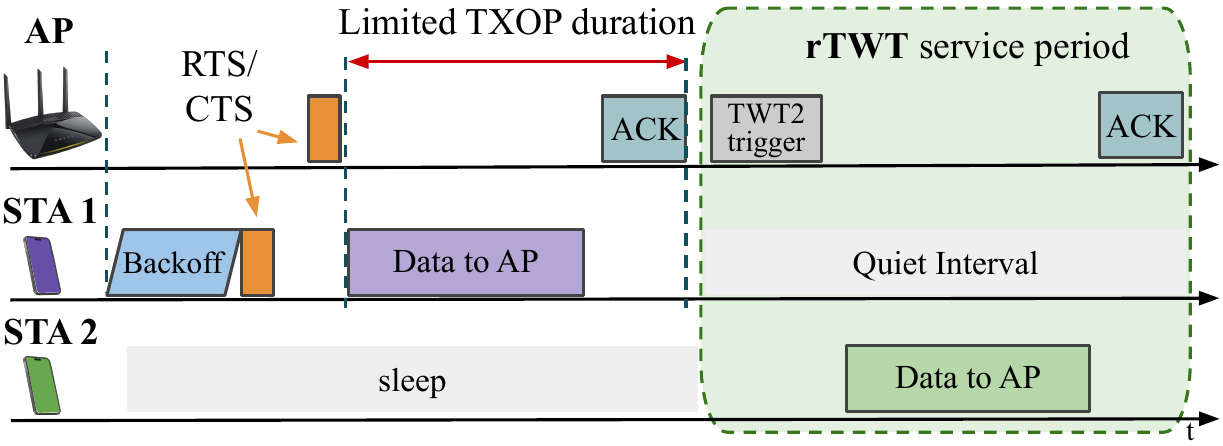}
\caption{\blue{Illustration of R-TWT operation. The timeline shows an \gls{ap} first granting a limited \gls{txop} to \gls{sta}~1 via an RTS/CTS exchange, followed by an R-TWT service period during which STA~2 wakes up from sleep and transmits scheduled uplink data to the AP, while STA~1 remains silent during the quiet interval.}}
\label{fig:R-TWT}
\end{figure}

Fig.~\ref{fig:R-TWT} illustrates the operation of \gls{rtwt}, 
where STA\,2 adheres to an \gls{rtwt} schedule (rTWT2), 
while STA\,1 follows a conventional \gls{twt} schedule (TWT1) for energy savings. 
This coexistence demonstrates the flexibility of \gls{rtwt}, 
which can operate in parallel with other \gls{twt}-based mechanisms to optimize both latency and power consumption.

Although \gls{rtwt} enhances latency control, 
it introduces certain trade-offs that may impact spectrum efficiency. 
Specifically, to meet the strict timing of service periods, 
ongoing transmissions---such as the one from \gls{sta}~1 in Fig.~\ref{fig:R-TWT}---may need to terminate early, 
reducing the number of packets that can be aggregated.
In other cases, 
the requirement that all \glspl{txop} finish before the start of an \gls{rtwt} service period may prevent new transmissions from being initiated, 
resulting in underutilized channel time.
Moreover, a key limitation of \gls{rtwt} is that its operation is confined to the \gls{bss}, 
meaning it cannot regulate transmissions from STAs associated with neighboring \glspl{obss}. 
To address this issue, 
multi-\gls{ap} coordinated \gls{rtwt} is being specified (see Section~\ref{sec:twt}), 
enabling the propagation of \gls{rtwt} schedules to neighboring \glspl{obss}.

We note that \gls{rtwt} prioritizes deterministic behavior over maximum throughput. Despite its limitations, \gls{rtwt} enables predictable and bounded latency,
substantially improving the suitability of 802.11be for delay-sensitive applications, such as real-time communication, AR/VR, and industrial automation.


\subsection{Wideband Operations}\label{sec:wideband_op}

Accessing the spectrum efficiently is key to achieving higher transmission rates, resulting in increased throughput and lower latency~\cite{barrachina2019dynamic}. This property is fundamental to modern 802.11 amendments, enabling them to meet the demands of bandwidth-intensive applications such as 4K video streaming, virtual reality, and other high-data-rate services~\cite{michaelides2025lessons}.


\subsubsection{Channel Bonding}

\begin{figure}
\centering
\includegraphics[width=\linewidth]{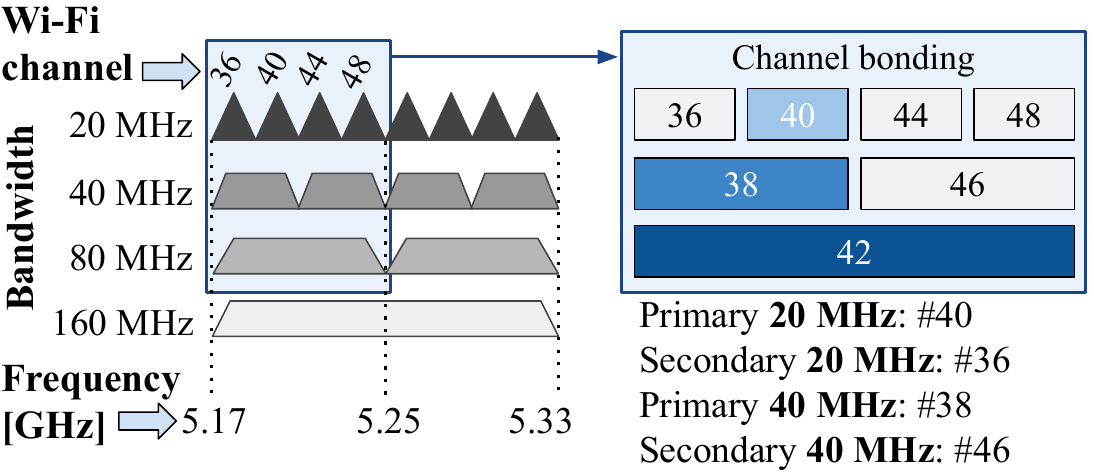}
\caption{\blue{Illustration of channel bonding in the 5\,GHz band. Left: four adjacent 20\,MHz channels (36, 40, 44, 48) can be combined into wider 40, 80, and 160\,MHz bandwidths. Right: corresponding channel-bonding configuration, with the selected primary and secondary 20\,MHz channels (\#40 and \#36) and the resulting primary and secondary 40\,MHz channels (\#38 and \#46).}}
\label{fig:ChannelBonding}
\end{figure}

In 802.11ax, \gls{ap} and \glspl{sta} can dynamically select the channel width for each frame transmission, bonding 20\,MHz channels into wider configurations such as 40, 80, or 160\,MHz~\cite{barrachina2019dynamic}. This adaptability allows the system to maximize throughput while accounting for real-time channel conditions. For example, an \gls{sta} that gains access to the medium on a primary 20\,MHz channel can expand the bandwidth by adding secondary channels, provided they are idle. The extension from 20\,MHz to wider channels follows a hierarchical process as shown in Fig.~\ref{fig:ChannelBonding}:
\begin{itemize}
\item 40\,MHz: Adds one 20\,MHz secondary channel.
\item 80\,MHz: Adds two more secondary channels if available.
\item 160\,MHz: Combines four 40\,MHz channels or eight 20\,MHz channels.
\end{itemize}
This process relies on mechanisms such as virtual carrier sense and backoff on the primary channel, followed by \blue{quick \gls{cca} checks (e.g., during a \gls{pifs} period)} on secondary channels just before transmission. \blue{Note that, for the sake of compatibility with legacy devices, channel access is performed on the primary channel only, and then extended to other channels if channel bonding is implemented.}

802.11ax supports channel bonding of up to 160\,MHz, achieved by aggregating eight 20\,MHz channels, while 802.11be further expands this capability to 320\,MHz channels in the 6\,GHz band. The 6\,GHz spectrum, initially accessible only to Wi-Fi 6E and Wi-Fi 7 devices, offers a pristine, low-interference environment, free from legacy devices and radar restrictions. This allows Wi-Fi 7 to unlock the full potential of channel bonding for next-generation applications. Specifically, the 802.11be-2024 amendment defines 320\,MHz channels by doubling the 160\,MHz tone plan used in 802.11ax~\cite{IEEE80211beD70_2024}. To maximize efficiency \blue{in various spectrum regulatory regimes, some of which may not have allocated the entire 1.2 GHz for unlicensed use}, overlapping 320\,MHz channels are defined, divided into two configurations: 320\,MHz-1 and 320\,MHz-2. 

\subsubsection{Preamble Puncturing}

While channel bonding provides substantial performance benefits, it also introduces challenges, particularly in environments with heavy spectrum utilization. In the 5\,GHz band, several factors constrain the availability of contiguous channels for bonding:
\begin{itemize}
\item \emph{Radar interference and DFS requirements:}
Some portions of the 5\,GHz spectrum overlap with radar systems, requiring 802.11 devices to vacate these channels when radar signals are detected. \Gls{dfs} mandates periodic scanning for radar activity, which can disrupt channel bonding and limit the availability of wider channels.
\item \emph{Non-contiguous spectrum:}
The 5\,GHz band is fragmented, leaving limited room for wider channels. For example, in the USA, only five non-overlapping 80\,MHz channels and one 160\,MHz channel are available when DFS is in use. This scarcity makes it challenging to fully exploit channel bonding in dense deployments or large outdoor environments.
\item \emph{Transmit power regulations:}
Power limitations in portions of the 5\,GHz band reduce the usability of certain channels for high-power deployments intended to cover large areas with many users, further restricting effective channel bonding.
\item \emph{Legacy 802.11 devices:}
Older devices operating on 20\,MHz or 40\,MHz channels can coexist with modern devices but often limit the ability of newer systems to fully utilize wide channels without interference or contention.
\end{itemize}

In scenarios where spectrum fragmentation is significant, such as in environments with legacy 802.11 networks operating on 20\,MHz channels, it becomes increasingly challenging to find contiguous 80 or 160\,MHz channels. For example, in a network capable of using an 80\,MHz channel, if one of the secondary 20\,MHz channels is frequently occupied, the network is restricted from utilizing the entire 80\,MHz channel, even if other secondary channels remain idle most of the time. This leads to underutilization of available spectrum, reducing overall network efficiency.

\begin{figure}
\centering
\includegraphics[width=.9\linewidth]{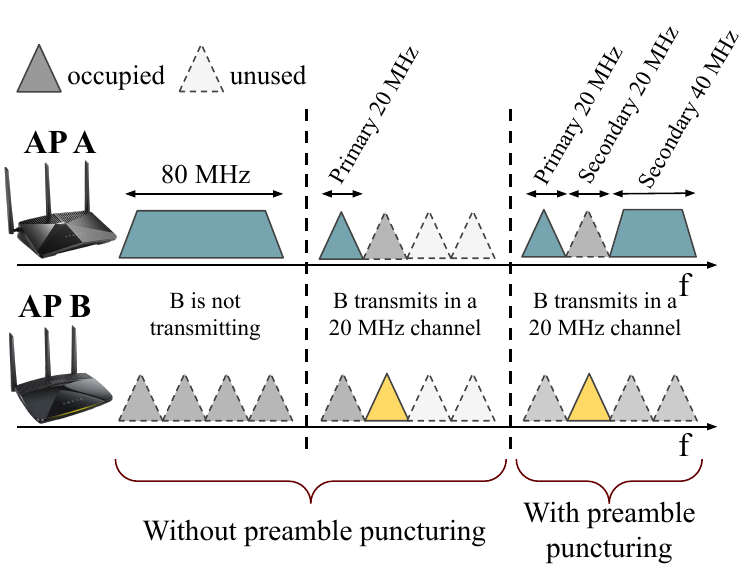}
\caption{\blue{Example of the benefits of preamble puncturing. Two neighboring APs share an 80\,MHz channel where AP~A uses the full bandwidth and AP~B may transmit on a single 20\,MHz subchannel. Without preamble puncturing (left), when AP~B becomes active in one 20\,MHz portion, AP~A can only utilize its primary 20\,MHz channel. With preamble puncturing (right), AP~A can mute (``puncture'') the overlapped 20\,MHz subchannel and continue transmitting over the remaining secondary 40\,MHz, improving channel utilization and spatial reuse.}}
\label{fig:PreamblePuncturing}
\end{figure}

To address this limitation, 802.11ax introduced preamble puncturing, which allows an \gls{ap} to transmit over non-contiguous channels by skipping busy portions of secondary channels. By relaxing the requirement for contiguous secondary channels, preamble puncturing improves spectrum utilization, particularly in dense deployments where wide channels (e.g., 80 or 160\,MHz) may not be fully available~\cite{barrachina2021wi}. Note that preamble puncturing operates in conjunction with \gls{ofdma}, as non-contiguous subchannels must be allocated to different \glspl{sta}. An example of the performance advantages of preamble puncturing is illustrated in Fig.~\ref{fig:PreamblePuncturing}. Without this feature, when secondary channels are frequently occupied, the \gls{ap} is restricted to the primary 20\,MHz channel, resulting in inefficient use of the spectrum and reduced transmission rates. With preamble puncturing, even if some secondary channels are occupied, the \gls{ap} can transmit over the remaining idle channels, such as utilizing a 40\,MHz secondary channel. 

While effective when multiple active \glspl{sta} are present, the benefits of preamble puncturing diminish when the number of active \glspl{sta} is low. To overcome this limitation, 802.11be introduces the ability to allocate multiple \glspl{ru} to a single \gls{sta}, as further discussed in Section~\ref{sec:Multiuser}.


\subsubsection{Non-Primary Channel Access (NPCA)}

\begin{figure}
\centering
\includegraphics[width=\linewidth]{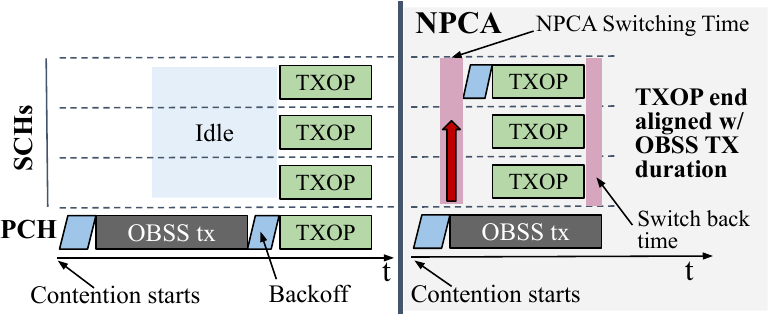}
\caption{\blue{\gls{npca} enables the use of the secondary channels, reducing channel access delay and improving the overall system performance. Left: baseline operation where an overlapping BSS (OBSS) transmission on the primary channel (PCH) forces the device to remain idle on its secondary channels (SCHs) until it can win a \gls{txop}. Right: \gls{npca}, where the device switches to idle SCHs during the NPCA switching time, initiates a \gls{txop} aligned with the OBSS transmission duration, and then switches back, thus exploiting otherwise unused spectrum.}}
\label{fig:NPCA}
\end{figure}

Traditional 802.11 channel access mechanisms rely on a \gls{pch} for contention-based backoff and frame transmission. Even if secondary channels are available, transmissions can only occur when the primary channel is idle. This limitation reduces spectrum efficiency, particularly in dense deployments with \gls{obss}, where the primary channel may often be occupied.

To address this challenge, 802.11bn is exploring \gls{npca}, a mechanism that allows \glspl{sta}  to leverage idle \glspl{sch}, also referred to as \gls{npca} primary channel, for frame transmission when the \gls{pch} is busy. 
When a Basic \gls{nav} is set on the \gls{pch} due to an ongoing \gls{obss} \gls{ppdu} transmission, the \gls{sta} can switch to an announced \gls{npca} primary channel to perform backoff and initiate frame transmission. This approach allows \glspl{sta} to utilize available resources on \glspl{sch}, avoiding idle time and improving overall throughput, as exemplified in Fig.~\ref{fig:NPCA}. 

    %
    %

For \gls{npca} to function effectively, the \gls{ap} must announce a set of parameters to the participating \glspl{sta}. These include the following:
\begin{itemize}
    \item \emph{\gls{npca} Primary Channel Announcement:} The \gls{ap} announces one \gls{npca} primary channel for backoff, including a puncturing bitmap to disable specific subchannels when necessary.
    \item \emph{Medium Synchronization:} Synchronization parameters such as \gls{ed} thresholds and maximum \gls{txop} durations are defined to ensure coordinated channel access among \glspl{sta}.
    \item \emph{Channel Switch Timing:} \blue{Switching between the \gls{pch} and \gls{sch} requires reconfiguring the device’s \gls{rf} front-end (e.g., \gls{pll} retuning, filter adjustment, \gls{agc} stabilization), which introduces a hardware-dependent delay. This effect is particularly clear in mixed-capability deployments. For instance, an \gls{ap} may support 160\,MHz channels, while some associated \glspl{sta} only 80\,MHz. When \gls{npca} triggers a switch to a channel segment outside the \gls{sta}’s native 80\,MHz range, the \gls{sta} must perform a full \gls{rf} retuning to a new center frequency, incurring a longer analog reconfiguration delay than the \gls{ap}. To guarantee that all \glspl{sta} complete this transition before any of them transmits on the new channel, the \gls{ap} announces a channel switching time large enough to accommodate the slowest device. Consequently, \gls{npca} adopts the maximum channel switching time among the associated \glspl{sta} to maintain proper synchronization and avoid collisions.}
\end{itemize}

Note that \gls{npca} may work at \gls{ppdu} or \gls{txop} level, while the second option relies on the initial received \gls{txop} length, creating unexpected situations if the effective \gls{txop} length changes. The first option of using one \gls{obss} \gls{ppdu} ensures predictable behavior at the expense of reducing transmission efficiency. 

\gls{npca} offers several benefits, especially in dense or congested environments. As shown in~\cite{bellalta2025performance}, it can substantially enhance spectral efficiency and throughput, while also reducing access delays under favorable conditions for supporting \glspl{bss}. Additionally, \gls{npca} helps alleviate the \gls{obss} performance anomaly, where low-rate transmissions from overlapping \glspl{bss} can degrade the performance of all nearby devices.


\subsubsection{Dynamic Spectrum Operation (DSO)}

\blue{While \glspl{ap} commonly support wide bandwidths such as 160~MHz or 320~MHz, many \glspl{sta} can only operate on narrower channels of 20, 40, or 80 MHz. This mismatch limits the \gls{ap}’s ability to fully utilize its wideband capabilities, since each transmission must accommodate the bandwidth constraints of the associated \gls{sta}. Furthermore, all \glspl{sta} are required to use the same 20~MHz primary channel defined by the \gls{ap}.}


\blue{For example, consider three \glspl{sta}, labeled A, B, and C, associated with a 160~MHz-capable \gls{ap}. Suppose \gls{sta}~A supports only 20~MHz, \gls{sta}~B supports 40~MHz, and \gls{sta}~C supports 80~MHz. Despite the \gls{ap}’s 160~MHz capability, in \gls{su} mode, it can only use its primary 80~MHz channel—and only when transmitting to \gls{sta}~C. When transmitting to \gls{sta}~A or \gls{sta}~B, the \gls{ap} must restrict the transmission to 20~MHz and 40~MHz, respectively. The use of \gls{ofdma} can increase bandwidth utilization efficiency by allowing simultaneous transmissions to all three \glspl{sta} within the primary 80~MHz channel (see Section~\ref{sec:OFDMA}). In such a case, the \gls{ap} could allocate 20~MHz to \gls{sta}~A (its primary channel), 20~MHz to \gls{sta}~B (its secondary 20~MHz), and 40~MHz to \gls{sta}~C (its secondary 40~MHz). However, even with \gls{ofdma}, 50\% of the \gls{ap}’s total bandwidth remains unused, and the wideband capabilities of \gls{sta}~B and \gls{sta}~C remain underutilized.}


\blue{\Gls{dso} offers a compelling solution to address this limitation. The concept is simple: remove the constraint of using the 20~MHz primary channel for data transmissions. By lifting this requirement, an \gls{sta} can be allocated to any portion of the \gls{ap}'s supported bandwidth, enabling more flexible and efficient use of the channel. To implement \gls{dso}, two mechanisms are necessary. First, the \gls{ap} and each \gls{sta} must exchange their capabilities information and agree (typically during the association phase) on one or several \gls{dso} subchannels, all with maximum width equal to the \gls{sta} bandwidth, to be used for future transmissions. Second, a signaling mechanism must be established to inform an \gls{sta} that it will be allowed to receive or transmit data on a specific \gls{dso} subchannel (out of the pre-agreed ones) during a given \gls{txop}. This can be achieved by extending existing \gls{bsrp} trigger frames to carry subchannel allocation information. Moreover, the network has to account for the required switching times to move to the assigned subchannel and back to the primary 20~MHz channel at the end of the transmission, so that default \gls{bss} operation can resume seamlessly.}

\begin{figure}
\centering
\includegraphics[width=\linewidth]{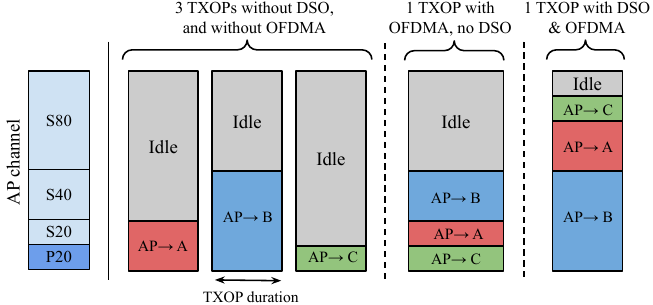}
\caption{Benefits of \gls{dso}: Combined with \gls{ofdma}, it improves spectrum utilization and reduces the \gls{txop} duration. \blue{Note that the figure illustrates only the data exchange. In the case of \gls{ofdma} without DSO, all stations’ channel allocations overlap with the P20 primary channel. With DSO, however, stations can be allocated across the AP’s full 160 MHz channel, removing the primary-channel overlap constraint and enabling them to use their full channel width for data exchange.}}
\label{fig:DSO}
\end{figure}

The example involving \glspl{sta}~A, B, and C is illustrated in Fig.~\ref{fig:DSO}. If the \gls{ap} transmits using \gls{su} transmissions, it must perform three separate transmissions, one for each \gls{sta}, each limited to the maximum bandwidth supported by the respective \gls{sta}. With \gls{ofdma}, a single transmission is possible, but each \gls{sta} receives only a portion of the available bandwidth, resulting in longer transmission durations. In contrast, by combining \gls{dso} with \gls{ofdma}, the \gls{ap} can fully exploit its wideband capability while simultaneously serving all \glspl{sta} using their maximum supported bandwidths, minimizing the portion of the spectrum that remains unused.

\subsection{Coexistence and Spatial Reuse}
\label{Sec:CoexistenceAndSpatialReuse}

802.11 devices belonging to independent \glspl{bss} need to coexist when operating on the same frequency channels. Ensuring smooth coexistence among \glspl{bss} is crucial to provide an acceptable performance to users, especially in dense deployments. 802.11 includes some coexistence mechanisms that aim to increase the efficiency and improve the performance of \gls{obss} deployments as detailed next.

\subsubsection{BSS Coloring}\label{sec:BSScolor}
\begin{figure}
\centering
\includegraphics[width=\linewidth]{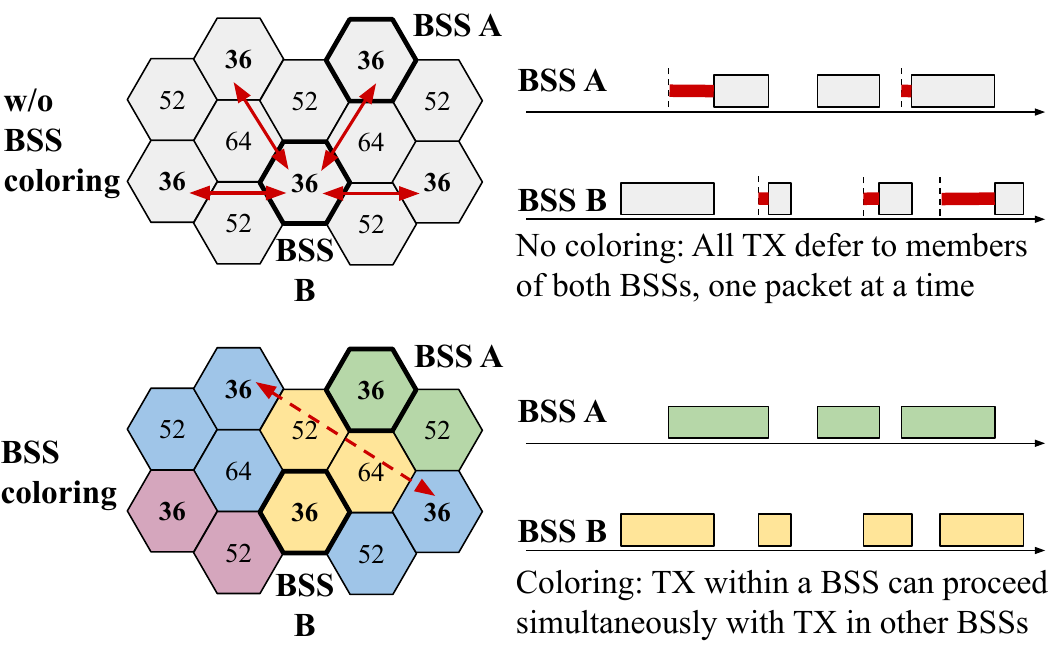}
\caption{\blue{\gls{bss} coloring improves spectrum utilization by enabling simultaneous transmissions in \glspl{bss} with different colors. Top: case without coloring, where all transmissions in overlapping \glspl{bss}~A and~B defer to each other and occur one at a time. Bottom: color-marked cells and timelines, where transmissions within each \gls{bss} can proceed concurrently with those in neighboring \glspl{bss} that use a different color.}}
\label{fig:BSS_coloring}
\end{figure}

This is a feature that was first introduced in 802.11ah for energy-saving purposes and then evolved in 802.11ax to improve efficiency in dense deployments. \gls{bss} coloring allows \glspl{ap} and \glspl{sta} to quickly differentiate between \emph{intra-\gls{bss}} (from their own \gls{bss}) and \emph{inter-\gls{bss}} (from neighboring \glspl{bss}) transmissions. To do so, each \gls{bss} is assigned a unique 6-bit identifier (\textit{color}) which is informed in the \gls{phy} header's signal preamble, part of the HE-SIG-A in the \gls{phy} header. The \gls{ap} is in charge of handling the \gls{bss} coloring (e.g., a color change is announced using BSS Color Change Announcement messages), but \glspl{sta} can also participate by sending \emph{\gls{bss} color collision reports} to the \gls{ap} when they notice that another \gls{bss} is using the same color. Fig.~\ref{fig:BSS_coloring} illustrates the potential of \gls{bss} coloring to enable simultaneous transmissions from multiple \glspl{bss}. As described next, \gls{bss} coloring is the foundation for specific mechanisms \blue{like \gls{obss} \gls{pd}-based \gls{sr} (described below), which} aim to make a more efficient use of the medium. \blue{In addition, by allowing the quick identification of the source of a transmission (read in the \gls{phy} headers rather than in the \gls{mac} ones), \gls{bss} coloring is an appealing feature for energy saving.}

\subsubsection{Basic vs. Intra-BSS NAV}

When a device detects the start of a \gls{ppdu}, it might activate a basic \gls{nav} upon certain conditions being met (see Section~\ref{sec:dcf}). Before 802.11ax, even if their signals were weak enough not to cause harmful interference, inter-\gls{bss} transmissions could provoke the activation of the \gls{nav} in other devices. Such a conservative approach is detrimental to efficiency, and that is why 802.11ax introduced a differentiation between intra-\gls{bss} and inter-\gls{bss} \glspl{nav} (or basic \gls{nav}). Thanks to \gls{bss} coloring, a device can differentiate between intra-\gls{bss} and inter-\gls{bss} transmissions and, as a result, maintain two different \glspl{nav}. For a device to mark the channel as idle, thus being able to initiate a transmission, both the basic and intra-\gls{bss} \glspl{nav} must be inactive.

\subsubsection{Spatial Reuse}

\begin{figure}[t]
\centering
\subfloat[OBSS PD-based SR.]
{\includegraphics[width=\columnwidth]{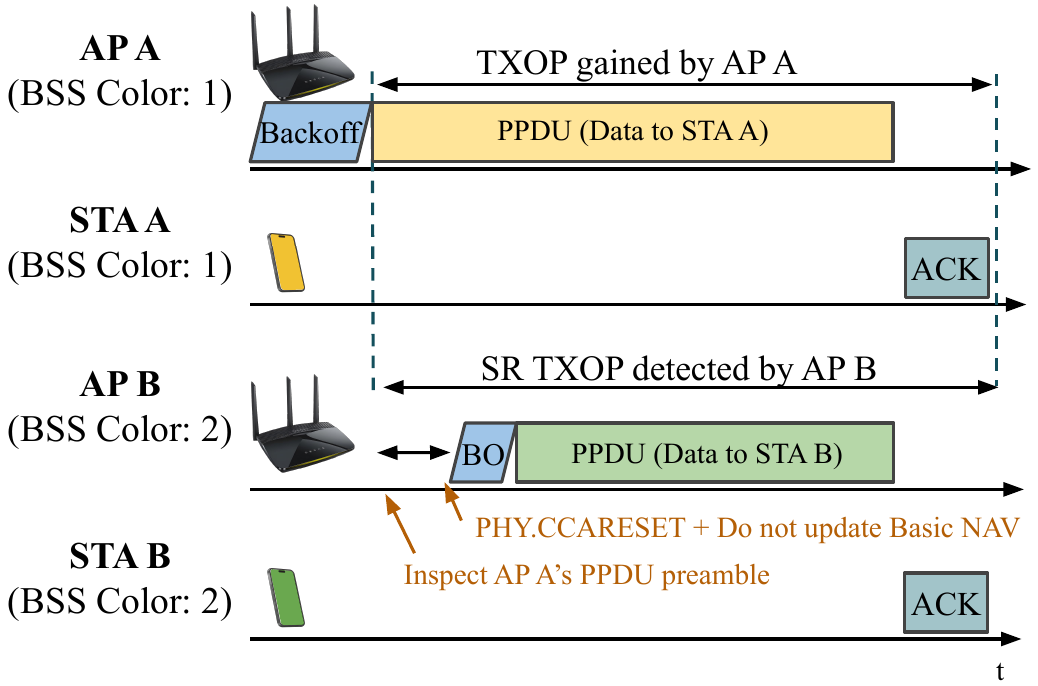}\label{fig:spatial_reuse_a}} 
\hfill
\subfloat[PSR.]
{\includegraphics[width=\columnwidth]{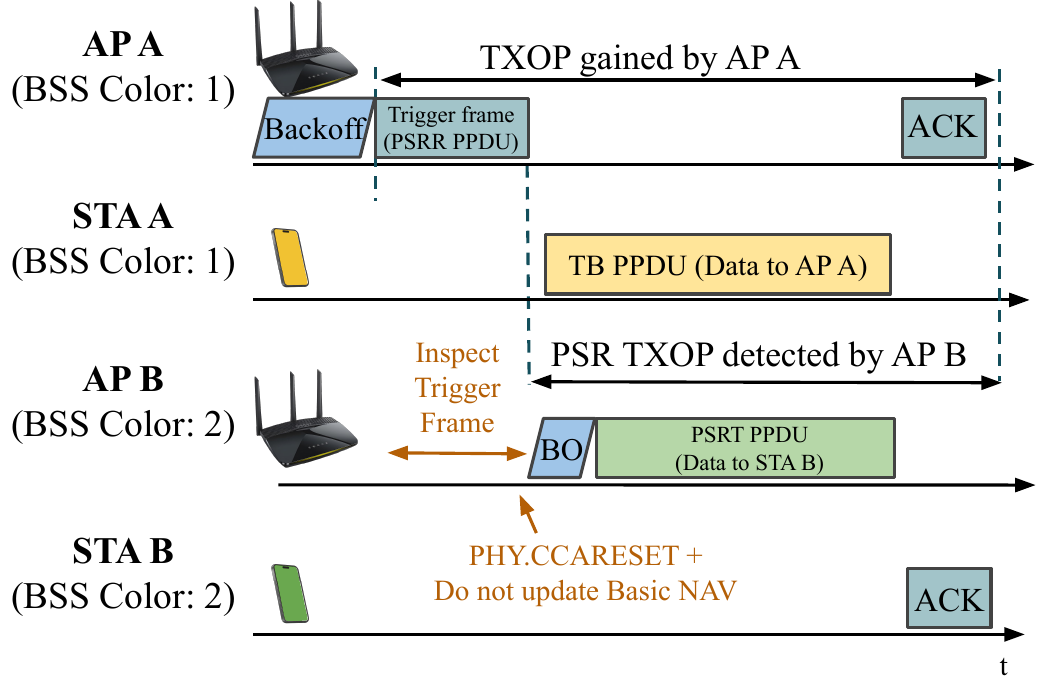}\label{fig:spatial_reuse_b}} 
\caption{Spatial reuse operation. (a) OBSS PD-based SR\blue{: \gls{ap} B detects an \gls{obss} PD-based \gls{sr} opportunity upon inspecting \gls{ap} A's transmission and initiates a subsequent transmission to \gls{sta} B.} (b) PSR\blue{: \gls{ap} A wins the \gls{txop} and sends a trigger frame to allow \gls{ap} B to transmit during the same \gls{txop}, leading to simultaneous transmissions in both \glspl{bss}.}}
\label{fig:spatial_reuse}
\end{figure}

Given the importance of efficiently using the spectrum in dense scenarios, 802.11ax introduced a mechanism called \gls{obss} \gls{pd}-based \gls{sr}~\cite{wilhelmi2021spatial}. \blue{The procedure is depicted in Fig.~\ref{fig:spatial_reuse_a} and works as follows:
\begin{enumerate}
    \item A device (\gls{ap} B) detects a \gls{ppdu} transmission (from \gls{ap}~A).
    \item It determines the origin of the transmission by reading the \gls{bss} color from the packet headers.
    \item Based on the color from \gls{ap} A (BSS Color 1), which is different from its own color (BSS Color 2), \gls{ap} B determines it is an inter-\gls{bss} transmission.
    \item An \gls{obss} \gls{pd} value is applied to check whether the detected transmission can be ignored or not. The \gls{obss} \gls{pd} threshold is a value between -82\,dBm and -62\,dBm that is defined for 20\,MHz transmissions and increased by 3\,dB as the channel width is doubled.
    \item The transmission can be ignored, so the PHY.CCARESET primitive is activated, which resets the \gls{phy} \gls{cca}.
    \item After exhausting the backoff, \gls{ap} B initiates a transmission. In exchange for relaxing the \gls{cca} requirements and using a higher \text{OBSS PD} value, a transmit power limitation is applied during the detected \gls{sr} \gls{txop}.
\end{enumerate} }


\blue{In \gls{obss} \gls{pd}-based \gls{sr},} the maximum transmit power allowed by the \gls{obss} \gls{pd}-based \gls{sr} operation depends on the selected \text{OBSS PD}, and is defined as\vspace{-0.1cm}
\begin{equation}
\resizebox{0.9\columnwidth}{!}{$\text{TX\_PWR}_{\max} = \text{TX\_PWR}_{\text{ref}} - (\text{OBSS PD} -\text{OBSS PD}_{\min})$},\vspace{-0.1cm}
\label{eq:obss_power_restriction}
\end{equation}
where $\text{OBSS PD}_{\min} = -62$ dBm and $\text{TX\_PWR}_{\text{ref}}$ is a reference power (e.g., 21\,dBm or 25\,dBm, depending on the device's capabilities).

Later, 802.11be extended the \gls{sr} operation with a second mechanism, \gls{psr}~\cite{de2020latency}, which had already been discussed during the standardization of 802.11ax. The mechanism is illustrated in Fig.~\ref{fig:spatial_reuse_b}. It is conceived as an \gls{ul} variant of \gls{sr}, in which an \gls{ap} shares with devices from other \glspl{bss} the \gls{txop} where it will receive \gls{ul} triggered transmissions from its associated \glspl{sta}. \blue{In particular:
\begin{enumerate}
    \item An \gls{ap} (\gls{ap} A) wins a \gls{txop}.
    \item Thanks to \gls{psr}, it can share the \gls{txop} with other devices from different \glspl{bss}. It does so by sending a Trigger frame with \gls{psr} information (including the maximum transmit power that can be used during the \gls{txop}).
    \item The Trigger frame is detected by other \gls{psr} devices (by \gls{ap} B in Fig.~\ref{fig:spatial_reuse_b}) as a \gls{psrr} \gls{ppdu}
    \item \gls{ap} B identifies a \gls{psr} opportunity, which is similar to the \gls{sr} opportunity defined above, and initiates a transmission.
\end{enumerate}}

\blue{In \gls{psr}, a transmit power limitation is imposed to ensure that inter-\gls{bss} interference does not disrupt the original \gls{ppdu} transmission. Specifically, the \gls{sta} that identifies the \gls{psr} opportunity (\gls{ap} B in Fig.~\ref{fig:spatial_reuse_b}) must apply the following transmit power limitation when transmitting a \gls{psrt}:\vspace{-0.05cm}
\begin{equation}
\text{TX\_PWR}_{\text{PSR}} - 10 \times \log_{10} N^{20\text{MHz}} \leq \text{PSR}_{\min} - \text{RPL}_{\text{PSRR}}^{20\text{MHz}},\vspace{-0.05cm}
\end{equation}
where $\text{TX\_PWR}_{\text{PSR}}$ is the transmit power during a \gls{psr} opportunity, $N^{20\text{MHz}}$ is the number of 20\,MHz channels used, and $\text{RPL}_{\text{PSRR}}^{20\text{MHz}}$ is the normalized received signal power on at least one of the involved 20\,MHz channels in which the \gls{psrr} and \gls{psrt} are transmitted. As for $\text{PSR}_{\min}$, it is the minimum value that the \gls{ap} holding the \gls{txop} (\gls{ap} A in Fig.~\ref{fig:spatial_reuse_b}) indicates (using the "\gls{sr}" field of a \gls{psrr} \gls{ppdu}) to ensure that other inter-\gls{bss} transmissions do not affect its own transmission. Its value is computed as
\begin{equation}
\text{PSR\_INPUT} = \text{TX\_PWR}^{(\text{AP})} + \gamma^{\text{(AP)}},
\label{eq:psr_1}
\end{equation}
where $\gamma^{\text{(AP)}}$ defines the acceptable receiver interference level at the \gls{ap}, in dBm. It is recommended to set $\gamma^{\text{(AP)}}$ as the expected signal power for the highest \gls{mcs} minus the minimum \gls{snr} value that yields at most a 10\% \gls{per}, minus an additional safety margin of up to 5\,dB.}


\section{Wi-Fi Multi-user Technologies}
\label{sec:Multiuser}

The latest 802.11 amendments have been introducing transformative multi-user technologies designed to enhance network performance, particularly in dense deployments with numerous connected \glspl{sta}. These technologies improve spectral efficiency, reduce latency, and ensure fair resource allocation among \glspl{sta}. The two cornerstone innovations in this realm are \gls{ofdma} and \gls{mu}-\gls{mimo}, which exploit the frequency and spatial dimensions, respectively, to enable simultaneous transmissions to multiple \glspl{sta}. 
These technologies are closely integrated in 802.11, with \gls{mu}-\gls{mimo} transmissions coordinated within the \gls{ofdma} framework. \blue{Specifically, multiple data streams can be spatially multiplexed within each frequency sub-channel (\gls{ofdma} \gls{ru}). This allows assigning the same \gls{ru} to multiple \glspl{sta} which then use \gls{mu}-\gls{mimo} for simultaneous transmission. The allocation is announced in the EHT/UHR-\gls{sig} of the \gls{ppdu} preamble (see Section~\ref{subsubsec:pramble}) for downlink multi-user transmissions and in the Basic Trigger frame for uplink multi-user transmissions (detailed in Section~\ref{subsubsec:downlink_OFDMA}). In this way,} the \gls{ap} can dynamically allocate radio resources to \glspl{sta} based on their traffic requirements and channel conditions, ensuring efficient spectrum utilization and improved network performance.


\subsection{Orthogonal Frequency Division Multiple Access (OFDMA)}
\label{sec:OFDMA}

\Gls{ofdma}, introduced in 802.11ax, enables efficient utilization of the frequency spectrum by dividing the \gls{rf} channel where the network is operating into smaller sub-channels called \glspl{ru}, each of which combines multiple adjacent \gls{ofdm} subcarriers~\cite{yin2006ofdma}. The \glspl{ru} can be independently assigned to different \glspl{sta}, which tune their radios to the allocated \glspl{ru}. This enables the \gls{ap} to transmit data to multiple \glspl{sta} simultaneously through different \glspl{ru} within a single transmission opportunity, based on the \glspl{sta} communication needs (\textit{downlink \gls{ofdma}}). Moreover, \textit{uplink \gls{ofdma}} enables multiple \glspl{sta} to transmit data simultaneously to the \gls{ap} over their assigned \glspl{ru}~\cite{avallone2021will}. The \gls{sta}-\gls{ru} association is decided by the \gls{ap}, which triggers this transmission mode. 

Downlink and uplink \gls{ofdma} allow increasing the efficiency in the spectrum utilization. Indeed, as introduced in Section~\ref{sec:AP_scheduled}, the \gls{edca} channel access protocol becomes increasingly inefficient as the number of \glspl{sta} grows, leading to high delays in obtaining a \gls{txop} and, in turn, reduced network performance. Uplink \gls{ofdma} addresses this challenge by aggregating multiple transmissions within a single \gls{txop}, preventing the sharp decline in capacity observed with \gls{csma}/\gls{ca} in dense environments. In~\cite{liu2023first}, the authors show that the improvement of using \gls{ofdma} is clearly visible starting from a network with four \glspl{sta}. Moreover, the high flexibility in the \glspl{ru} assignment for downlink and uplink \gls{ofdma} enables a more balanced distribution of \glspl{ru} of different bandwidths to different \glspl{sta}, which can be done based on their specific traffic requirements. This allows the network to address varying traffic demands efficiently. For instance, \glspl{sta} transmitting a high amount of data, such as video, are assigned larger \glspl{ru} to accommodate their higher bandwidth needs. On the other hand, \gls{iot} sensors sending small, infrequent packets are allocated smaller \glspl{ru} to save spectrum resources and maintain low power consumption, thus extending battery life. This way, \gls{ofdma} facilitates smaller and more frequent transmissions, which lowers latency and jitter, being particularly valuable for time-sensitive applications like voice-over-Wi-Fi and video streaming. In the following, we describe the \glspl{ru} allocation process and the way \gls{ofdma} is implemented in the downlink and uplink directions. 


\subsubsection{Resource Unit (RU) Allocation}\label{sec:ru_alloc}
\begin{figure}
\centering
\includegraphics[width=\figwidth]{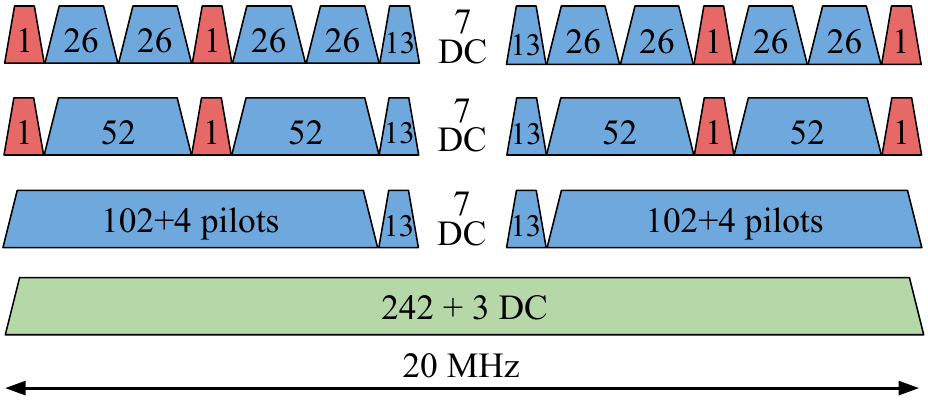}
\caption{OFDMA RU allocation in 802.11be for a 20~MHz channel.}
\label{fig:RU_allocation}
\end{figure}

Proper \gls{ru} allocation is crucial to enable full exploitation of the flexibility and benefits offered by \gls{ofdma} transmissions~\cite{magrin2023performance}. 
\commentfigtable
As shown in Fig.~\ref{fig:RU_allocation} for a 20~MHz channel, 
\endcommentfigtable
The smallest \gls{ru} in \gls{ofdma} consists of 26 adjacent subcarriers (tones), which allows up to 9 \glspl{sta} to share a 20\,MHz channel in the downlink or uplink direction. Note that the central 26-tone \gls{ru} is obtained by combining the 13 subcarriers before and after the DC subcarriers. Indeed, it is important to note that not all subcarriers in an \gls{ru} can be used for data transmission. As detailed in Section~\ref{sec:ofdm}, some subcarriers are reserved for guard bands to minimize interference, while others function as DC or pilot tones to assist with signal synchronization and demodulation, ensuring reliable communication. Moreover, some null subcarriers are located near the DC or guard tones to reduce transmit center frequency leakage, receiver DC offset, and interference from neighboring \glspl{ru}. Adjacent 26-tone \glspl{ru} can be grouped to form 52-tone, 106-tone, and 242-tone \glspl{ru} on 20\,MHz channels. Larger \glspl{ru} are available increasing the bandwidth and grouping more contiguous 26-tone \glspl{ru}. Specifically, 40\,MHz channels supports \glspl{ru} with up to 484 tones while one and two 996-tone \glspl{ru} can be formed in 80\,MHz and 160\,MHz channels~\cite{henry2024wi}. This hierarchical structure of \gls{ru} sizes enables precise resource allocation tailored to the traffic characteristics of each \gls{sta}. 

\commentfigtable
\endcommentfigtable


\subsubsection{Downlink OFDMA}

In the downlink case, the \gls{ap} simultaneously transmits data to multiple \glspl{sta} using a single transmission. This bundling mechanism minimizes contention time and significantly improves spectrum efficiency. Downlink \gls{ofdma} transmissions are set as follows:
\begin{itemize}
    \item The \gls{ap} transmits an \gls{mu-rts} trigger frame to the \glspl{sta} that are expected to receive downlink traffic. The participating \glspl{sta} simultaneously respond with a \gls{cts} frame. 
    \item The \gls{ap} bundles frames intended for each \gls{sta} and modulates them on the \gls{sta}-assigned \glspl{ru}.
    \item An MU-\gls{ppdu} (see Section~\ref{subsec:ppduformat}) is formed by using a single preamble spanning all frequencies (full-band transmission), followed by individually addressed frames within each \gls{ru}. The \gls{sta}–\gls{ru} allocation is detailed in the SIG field of the preamble. To ensure uniform transmission length, padding is used on shorter frames to match the length of the longest frame. Moreover, smaller \glspl{ru} can be allocated to shorter frames to reduce padding overhead.
    \item An \gls{mu}-\gls{bar} trigger frame may be transmitted by the \gls{ap} at the end of the multi-user transmission to request acknowledgment from multiple \glspl{sta} simultaneously, reducing overhead.
    \item A \gls{ba} is simultaneously transmitted by the \glspl{sta} to the \gls{ap} providing information about the correctness of the received data. The \gls{ba} can also be untriggered when the \gls{ul} parameters are included in the data transmission.
\end{itemize}
Note that the actual number of supported \glspl{sta} can be further increased by integrating spatial multiplexing through \gls{mu}-\gls{mimo}. In this case, sounding should be performed before actual data transmission, as discussed in Section~\ref{sec:MU-MIMO}. 

\subsubsection{Uplink OFDMA}\label{subsubsec:downlink_OFDMA}

While downlink \gls{ofdma} is relatively straightforward due to centralized control by the \gls{ap}, uplink \gls{ofdma} requires more complex synchronization mechanisms. Indeed, precise synchronization is required to ensure that all \glspl{sta}' preamble symbols arrive at the \gls{ap} simultaneously and, in turn, can be properly decoded. This synchronization is facilitated through careful scheduling performed by the \gls{ap}, which specifies the exact timing, \gls{ru} allocation, and transmission parameters for each \gls{sta}. An uplink \gls{ofdma} transmission proceeds as follows:
\begin{itemize}
    \item The \gls{ap} transmits an \gls{mu-rts} trigger frame to the \glspl{sta} that are expected to transmit uplink traffic. The \glspl{sta} simultaneously respond with a \gls{cts} frame. 
    \item The \gls{ap} transmits a \gls{bsrp} Trigger frame in broadcast to solicit the \glspl{sta} indicated in the Info List field of the Trigger frame to report the amount of buffered uplink data inside a dedicated Buffer Status Report frame, a \gls{sifs} after the trigger. This information helps the \gls{ap} to optimize \gls{ru} scheduling.
    \item \blue{Then, the \gls{ap} transmits a Basic Trigger frame} to trigger the coordinated uplink transmission, indicating the \gls{sta}-\gls{ru} association inside the User Info List field of the trigger frame. This frame provides coordination for: (i) \emph{time synchronization}, ensuring all \glspl{sta} start their transmissions simultaneously to avoid overlapping symbols (timing tolerance is 0.4 $\mu$s)~\cite{henry2024wi}, (ii) \emph{frequency alignment}, maintaining consistent frequency offsets across \glspl{sta} to prevent interference between adjacent \glspl{ru}, and (iii) \emph{amplitude calibration}, adjusting signal strength to ensure uniform reception across all \glspl{ru}. 
    \item All \glspl{sta} indexed in the User Info List field of the Trigger frame, transmit their data in the uplink direction simultaneously after an \gls{sifs} from the reception of the trigger using a \gls{tb} PPDU (see Section~\ref{subsec:ppduformat}).
    \item The \gls{ap} uses a Multi-STA \gls{ba} to inform the different \glspl{sta} about which frames have been received correctly. 
\end{itemize}

\begin{figure}
\centering
\includegraphics[width=\linewidth]{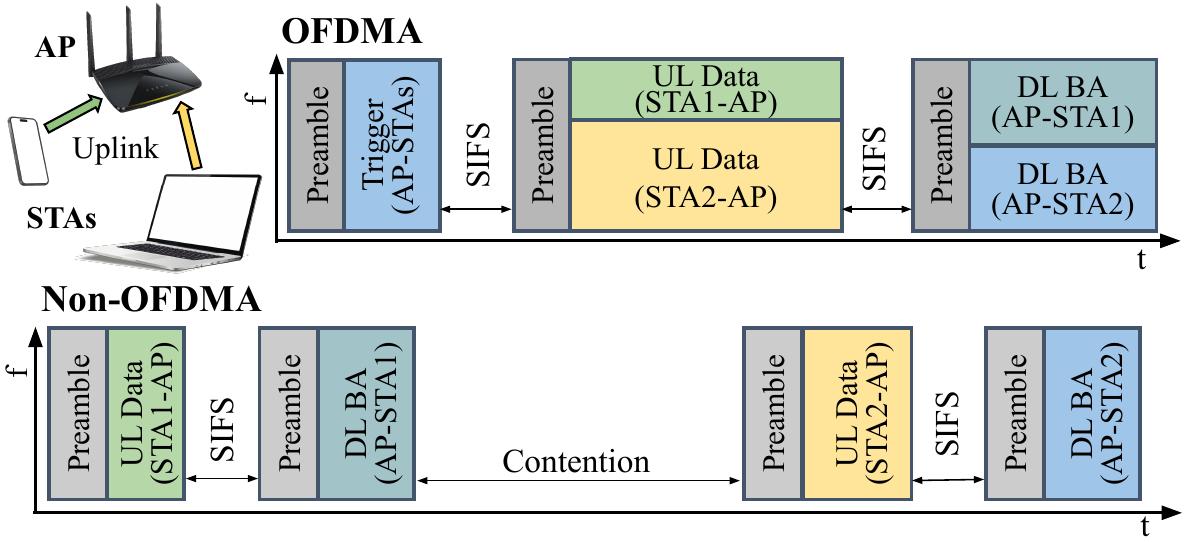}
\caption{\blue{Example of uplink OFDMA (top) and non-OFDMA (bottom). In the OFDMA case, the \gls{ap} sends a trigger frame and then receives simultaneous uplink data from two \glspl{sta} on different frequency resource units, followed by simultaneous downlink block acknowledgments (DL BAs) after SIFS intervals. In the non-OFDMA case, each \gls{sta} contends independently and transmits its uplink data and receives its DL BA in separate transmission opportunities, incurring additional contention overhead.}}
\label{fig:OFDMA}
\end{figure}

The upper part of Fig.~\ref{fig:OFDMA} summarizes \blue{the last part of this packet exchange} considering two \glspl{sta} that simultaneously transmit data to the \gls{ap} using UL-\gls{ofdma}. The bottom part of the figure shows how the same two-\gls{sta} uplink transmission would be carried out using \gls{csma}/\gls{ca} where data streams from the two \glspl{sta} are transmitted sequentially \blue{after independent channel contention}. \blue{In \gls{ofdma}, the \gls{ap} manages the transmissions in a deterministic way, significantly reducing contention overhead and consequently reducing communication latency for scheduled transmissions. Moreover, as discussed at the beginning of this section, \gls{ofdma} enables a more efficient utilization of the spectrum in case of unequal transmission requirements: the \gls{ap} can assign small \glspl{ru} to \glspl{sta} having a small amount of data to transmit (\gls{sta} 1 in Fig.~\ref{fig:OFDMA}), while reserving more space for \glspl{sta} with a higher amount of data to be transmitted (\gls{sta} 2 in Fig.~\ref{fig:OFDMA}).}

With 802.11ax, another strategy to identify the \glspl{sta} with packets to transmit has been introduced~\cite{henry2024wi}. This procedure substitutes the \gls{mu-rts} and \gls{cts} frame exchange and entails the \gls{ap} to broadcast an \gls{ndp} Feedback Report Poll, which includes information about the \glspl{sta} that are allowed to respond and their assigned \glspl{ru}. Hence, \glspl{sta} with uplink traffic respond to the \gls{ap} with an \gls{ndp} Feedback Report and wait for the \gls{bsrp}.

\subsubsection{Multi-RU Support}

One key limitation of \gls{ofdma} in 802.11ax is that each \gls{sta} can only be assigned a single \gls{ru}. This constraint can lead to underutilization of available bandwidth, particularly in scenarios with few \glspl{sta} and when preamble puncturing is employed to enable the simultaneous use of non-contiguous portions of the radio spectrum as detailed in Section~\ref{sec:wideband_op}. Considering the preamble puncturing example in the rightmost part of Fig.~\ref{fig:PreamblePuncturing}, if the \gls{ap} has data for only one \gls{sta}, it can only assign the primary 20\,MHz channel to that \gls{sta}, leaving the remaining 40\,MHz bandwidth (the majority of the spectrum) unused. While this restriction in 802.11ax helps avoid potential regulatory issues, such as protecting incumbents in the 6\,GHz band, it significantly reduces transmission efficiency and throughput in scenarios with low \gls{sta} density. 

The 802.11be standard amendment addresses this limitation by introducing \gls{mru} support, which allows the \gls{ap} to assign multiple \glspl{ru} to a single \gls{sta}, thus significantly improving spectrum utilization. \Gls{mru} support also enhances performance when leveraging it for diversity gain: by transmitting the same data across multiple \glspl{ru}, the \gls{ap} can increase the likelihood of successful packet delivery in environments with high interference or variable channel conditions. 

\begin{figure}
\centering
\subfloat[Small-size MRUs: 52+26-tone and 106+26-tone.\label{fig:multi_RU_small}]
{\includegraphics[width=0.95\linewidth]{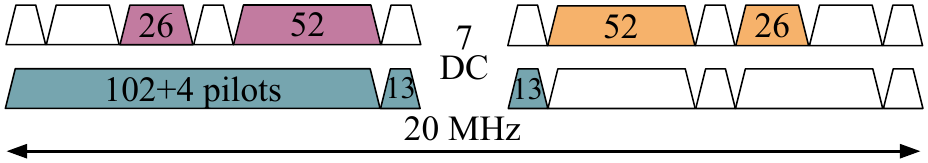}}
\hfill
\subfloat[Large-size MRUs: 484+242-tone.\label{fig:multi_RU_large}]
{\includegraphics[width=0.95\linewidth]{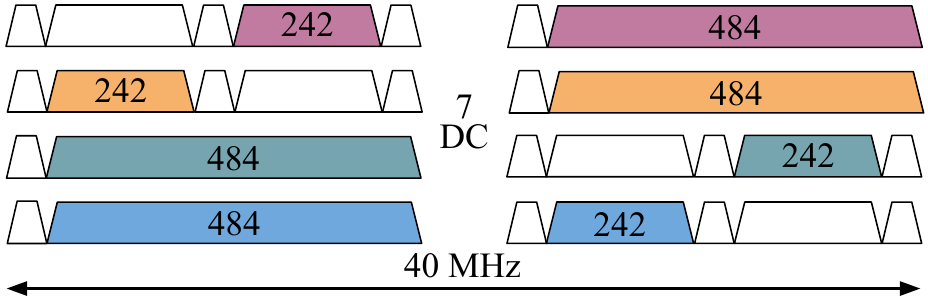}}
\caption{\blue{Examples of different MRU combinations. Top (a): small-size MRUs within a 20\,MHz channel, combining 26- and 52-tone (or 106- and 26-tone) resource units around the DC subcarrier and pilot tones. Bottom (b): large-size MRUs over 40\,MHz with 484- and 242-tone RUs in various placements across the channel bandwidth.}}
\label{fig:multi_RU}
\end{figure}

\Glspl{ru} are grouped into two categories based on their size: \textit{small-size \glspl{ru}}, including 26-tone, 52-tone, and 106-tone \glspl{ru}, and \textit{large-size \glspl{ru}}, comprising 242-tone or larger \glspl{ru}. A small-size \gls{ru} can only be combined with other contiguous small-size \glspl{ru} to form a small-size \gls{mru}, while large-size \glspl{mru} can be created by combining different large-size \glspl{ru}, which may be non-contiguous. Specific combinations of \glspl{ru} are predefined to ensure compatibility and minimize overhead. 
Small-size \glspl{mru} include 52+26-tone and 106+26-tone configurations, obtained by combining a 52-tone or 106-tone \gls{ru} with an adjacent 26-tone \gls{ru} within the same 20\,MHz channel, as depicted in Fig.~\ref{fig:multi_RU_small}.
Large-size \glspl{mru} include 484+242-tone, 996+484-tone, 2×996+484-tone, 3×996-tone, and 3×996+484-tone combinations, formed by grouping \glspl{ru} within the same 80\,MHz frequency channel. The \glspl{ru} that can be combined to create small or large-size \glspl{mru} are defined in the 802.11be amendment based on configurations that offer the highest performance gains. Large-size \glspl{mru} can also be used for non-\gls{ofdma} transmissions to multiple \glspl{sta}, in which case an additional \gls{ru} combination, i.e., 996+484+242-tone \gls{mru}, may be employed. 
Examples of 484+242-tone \glspl{mru} are shown in Fig.~\ref{fig:multi_RU_large}.

Considering the same preamble puncturing example used above, rather than restricting an \gls{sta} to the primary 20\,MHz channel, the \gls{ap} can employ \gls{mru} and  allocate to the \gls{sta} a secondary 40\,MHz channel, thus creating a 484+242-tone large size \gls{mru} and effectively tripling spectrum utilization. 

\subsubsection{Distributed RU in 802.11bn}

\Gls{ofdma} implementations up to 802.11be use \glspl{ru} composed of continuous subcarriers. \blue{This constraint is expected to be relaxed in the 802.11bn amendment, with the introduction of \glspl{dru}. A \gls{dru} consists of non-adjacent tones which are evenly distributed along the operational bandwidth (excluding the central DC and null subcarriers). \Gls{dru}-based \gls{ofdma} will addresses the limitations that continuous subcarrier placement has} in shared spectrum scenarios, such as in the 6\,GHz band, where devices should operate either in Standard Power or Low Power Indoor modes (see Section~\ref{sec:operating_rules}). Specifically, the Standard Power mode uses a central frequency coordinator that decides on which channel and at what power the \gls{ap} should operate. This requires precise localization of the \gls{ap}, which may be difficult to achieve. On the other hand, the Low Power Indoor mode sets a maximum on the \gls{eirp} per MHz, e.g., -1\,dBm/MHz. Considering the 78.125\,kHz sub-spacing used since 802.11ax, the cumulative transmit power used on the 12.8 subcarriers hosted on each MHz should meet the \gls{eirp} limit. This reduces power efficiency and leads to degradation of the \gls{ofdma} transmission. \blue{\Glspl{dru} are expected to overcome this limitation.} The main idea is to reduce the number of tones per MHz, concentrating power over fewer tones and thereby increasing the effective transmission power per tone while remaining within regulatory constraints~\cite{henry2024wi}. \blue{By evenly distributing tones along the operational bandwidth, \glspl{dru}} will support a comparable number of units per channel bandwidth as standard \gls{ofdma} without degrading spectral efficiency, thus ensuring that they can handle diverse traffic demands effectively. \blue{The regular distribution of tones} improves channel smoothing, which enhances the performance of \gls{dru}-based transmissions through a better estimation of the wireless channel. \Glspl{dru} also tend to be more robust in environments with high channel fading. The hierarchical tone mapping structure and pilot tone placement used in standard \gls{ofdma} is preserved, ensuring compatibility and ease of implementation. The combination of these features could allow \glspl{dru} to improve power and spectral efficiency without complicating implementation or sacrificing performance compared to standard \gls{ofdma} operations. Note that \glspl{dru} are currently defined for use in uplink \gls{ofdma} only. The trigger frame used to solicit multi-user uplink transmissions has been updated to include signaling for \gls{dru}. 


\subsection{Multi-user MIMO}
\label{sec:MU-MIMO}

As Wi-Fi has evolved, \glspl{ap} have incorporated more antennas and advanced spatial multiplexing techniques to support the increasing number of connected \glspl{sta}. \Gls{mu}-\gls{mimo} builds on these advancements, allowing the transmission of data streams to multiple \glspl{sta} simultaneously. This enhances spectral efficiency, reduces contention for accessing the channel, and improves overall network performance~\cite{perahia_stacey_2008}. In the following, we review the use and implementation of \gls{mu}-\gls{mimo} in the \gls{ul} and \gls{dl} directions.

\subsubsection{Downlink MU-MIMO and MU-MIMO-OFDMA}

The 802.11n amendment only supported \gls{su}-\gls{mimo}, where the \gls{ap} transmits multiple streams to a single multi-antenna \gls{sta} as detailed in Section~\ref{sec:MIMO}. Its successor, 802.11ac, integrated \gls{dl}-\gls{mu}-\gls{mimo} in the standard, allowing leveraging at full the benefits of multi-antenna systems. Indeed, \gls{su}-\gls{mimo} only allows multiplexing a few streams, as most user devices (\glspl{sta}) have limited antenna arrays (one or two antennas for smartphones and laptops). In \gls{mu}-\gls{mimo}, the \gls{ap} leverages the lack of correlation among the channels of the different \glspl{sta} to serve them simultaneously with multiple data streams. Specifically, the rank of the combined channel matrix, indicating the number of simultaneous data streams that can be transmitted, is equal to the minimum between the number of \glspl{sta}' antennas and the antennas available at the \gls{ap}. 
Moreover, as introduced in Section~\ref{sec:MIMO} for \gls{su}-\gls{mimo}, 802.11 regulates the maximum number of simultaneous transmissions. The 802.11ac amendment dictates that \glspl{ap} can transmit up to 8 spatial streams to 4 \glspl{sta} simultaneously. In 802.11ax, the number of \glspl{sta} that can be simultaneously served increased to 8 while the total number of streams remained 8. The same limits apply to 802.11be and are expected to remain unchanged in 802.11bn. 

Downlink \gls{mu}-\gls{mimo} follows a similar procedure as the one described in Section~\ref{sec:MIMO} for \gls{su}-\gls{mimo}. However, in this case, the \gls{ap} should collect the channel information from all connected \glspl{sta} to properly precode the streams and reduce their cross-interference, thus enabling accurate decoding. In turn, the channel sounding procedure is carried out for all the \glspl{sta} involved in the \gls{mu}-\gls{mimo} transmission as detailed in the sequel. Note that \gls{mu}-\gls{mimo} can be integrated with \gls{ofdma}, where multiple \glspl{sta} are served through \gls{mu}-\gls{mimo} over the same \gls{ru}. In this case, channel state information is reported only for the intended portion of the bandwidth following the indication in the \gls{ndpa} frame. 

\begin{figure}
\centering
\includegraphics[width=\figwidth]{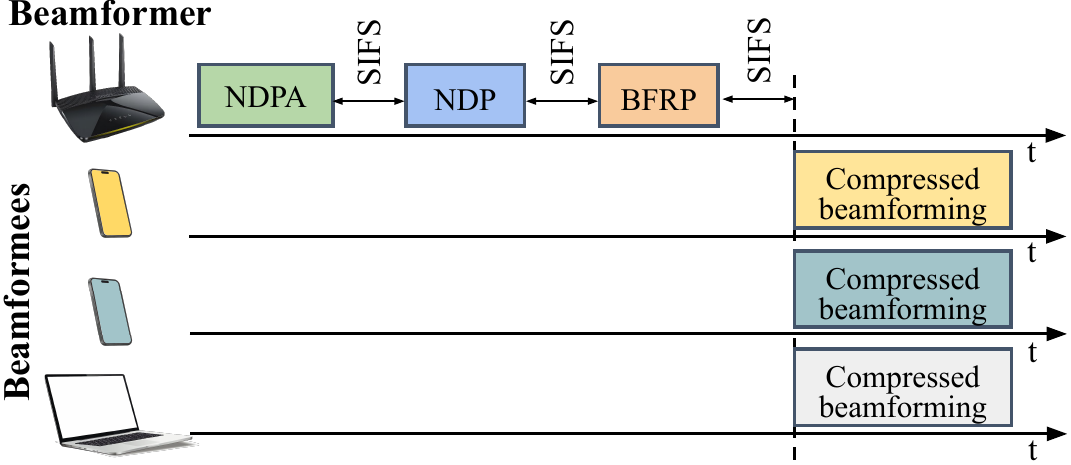}
\caption{\blue{Example of a TB channel sounding round with three \glspl{sta}. The beamformer (\gls{ap}) first transmits an NDPA frame, followed after SIFS by the corresponding NDP and a beamforming report poll (BFRP). In response, the three beamformees send their compressed beamforming feedback frames, so that the AP acquires channel state information from all STAs within a single sounding exchange.}}
\label{fig:MU-MIMO}
\end{figure}

\subsubsection{Multiuser Channel Sounding}

To obtain channel feedback for \gls{mu}-\gls{mimo} transmissions, the \gls{mu}-\gls{mimo} beamformer (\gls{ap}) shall not use the non-\gls{tb} sounding procedure described in Section~\ref{sec:MIMO}. \Gls{mu}-\gls{mimo} sounding is initiated by the \gls{ap} through the transmission of a single \textit{broadcast} \gls{ndp}. In this case, \blue{the \gls{ndpa} that comes before the \gls{ndp} includes a list of the \glspl{aid} corresponding to the involved \gls{mu}-\gls{mimo} \glspl{sta}. The \gls{aid} is an integer value in the range $[1, 2007]$ that is assigned to each \gls{sta} by the \gls{ap} as part of the association process upon reception of the Association Request frame from the \gls{sta}. The \gls{aid} is selected among the available \gls{aid} pool---with the constraint that the \gls{aid} should be unique within a \gls{bss}---and transmitted back to the \gls{sta} in the Association Response frame.} Moreover, in case of \gls{mu}-\gls{mimo}-\gls{ofdma} transmissions, the \gls{ndpa} indicates the \glspl{ru} or \glspl{mru} for which channel estimate should be fed back using the Partial BW Info subfield. This information is also included in the BW and Puncturing Channel Information fields, which are part of the U-\gls{sig} field of the sounding \gls{ndp}.


Upon \gls{ndp} reception, each \gls{sta} estimates the \gls{cfr} and compresses it following the procedure described in Section~\ref{sec:MIMO}. In the \gls{mu}-\gls{mimo} case, the beamforming feedback report also includes a \gls{mu}-\gls{mimo} Exclusive Beamforming Report field which carries delta \gls{snr} values indicating the difference between the subcarrier and the average \glspl{snr}. This information can be used by the beamformer to compute the steering matrix. 

At this point, the \glspl{sta} have to feed back their compressed channel estimate. Pre-802.11ax \glspl{ap} perform this collection of beamforming feedback frames from different \glspl{sta} sequentially: the \gls{sta} associated with the first \gls{aid} in the \gls{aid} list transmits back its feedback after a \gls{sifs} from the \gls{ndp} reception; hence, all the other \glspl{sta} are triggered to feed back their information by the \gls{ap} through the transmission of Beamforming Report Poll frames indexing each of them. The 802.11ax amendment introduced a new \gls{tb} sounding procedure which increase spectrum efficiency~\cite{zhang2023implementation} and is illustrated in Fig.~\ref{fig:MU-MIMO}. Instead of transmitting the different feedback frames separately, they are simultaneously transmitted to the \gls{ap} using an uplink multi-user strategy specified in a new frame named \Gls{bfrp} which is transmitted an \gls{sifs} after the \gls{ndp}. If the number of \glspl{sta} that should report the feedback is higher than the maximum allowed by the multi-user transmission, multiple rounds of \gls{bfrp} and feedback transmissions are allowed. This \gls{tb} sounding may also be used for \gls{su}-\gls{mimo} feedback if the beamformee supports it.

Upon reception of each compressed beamforming feedback frame, the \gls{ap} retrieves the angles and uses them to reconstruct the $\mathbf{V}_{i,k}$ matrices for all the subcarriers $k$ for each \gls{sta} $i$ (as detailed in Section~\ref{sec:MIMO} for \gls{su}-\gls{mimo}). These matrices---which are a proxy for the \gls{cfr}---are used by the \gls{ap} to decide which \glspl{sta} can be scheduled together in the \gls{mu}-\gls{mimo} transmission to maximize the \gls{qos} and spectrum efficiency. The idea is to avoid combining in the same \gls{mu}-\gls{mimo} group \glspl{sta} that will create interference to each other. A possible strategy consists of computing the inter-\gls{sta} correlation and selecting the \glspl{sta} for which this is minimum~\cite{Hou2022Muster}. However, \gls{sta} grouping is an implementation-dependent feature and, in turn, each vendor is free to decide the best strategy to adopt. \blue{Having obtained the set of \glspl{sta} in the same group,} the $\mathbf{V}_{i,k}$ matrices for subcarrier $k$ and all the \glspl{sta} are combined into a single $\mathbf{V}_{k}$ matrix over the spatial stream (column) dimension thus obtaining a $N_{\rm TX} \! \times N_{\rm SS} \! \times \! N_{\rm SD}\!+\!N_{\rm SP}$ matrix where $N_{SS}=\sum_i N_{{\rm SS},i}$ and $N_{{\rm SS},i}$ is the number of spatial streams that should be directed to \gls{sta} $i$. This matrix is used to design the precoding for subcarrier $k$. For example, using zero-forcing the beamforming steering matrix is $\mathbf{Q}_k \!=\!\! \mathbf{\Tilde{V}_k}\!\left(\mathbf{\Tilde{V}_k}^\dag\mathbf{\Tilde{V}_k}. \right)^{\!-1}$. In the ideal case, this makes the streams directed to one \gls{sta} orthogonal to the channel between the \gls{ap} and each of the other \glspl{sta}, thus reducing interference. 

Note that as the feedback is transmitted back unencrypted, security issues may arise. Indeed, a malicious user in the network may capture the legitimate feedback of other stations and properly craft a malicious feedback that leads to a wrong precoding which drastically degrades the network performance~\cite{Hou2022Muster,meneghello2024whack,Meneghello2025BREAK}, or enable eavesdropping on a victim's traffic~\cite{Tung2014Vulnerability,Wang2017eavesdropping,wang2019user,yang2020securing}.  

\subsubsection{Uplink MU-MIMO and MU-MIMO-OFDMA}
In 802.11ax, \gls{mu}-\gls{mimo} was extended to support both downlink and uplink transmissions. Uplink \gls{mu}-\gls{mimo}, allows multiple \glspl{sta} to transmit data simultaneously to the \gls{ap}. The process is orchestrated by the \gls{ap}, which sends a Basic Trigger frame specifying the \gls{ru} and timing for each \gls{sta}. The \glspl{sta} then transmit data in their assigned \gls{ru}, and the \gls{ap} decodes the individual data streams. As discussed for uplink \gls{ofdma}, uplink \gls{mu}-\gls{mimo} reduces contention, as multiple \glspl{sta} can transmit concurrently, and is particularly beneficial in environments with a high density of connected \glspl{sta}~\cite{naik2018performance}. To manage the added complexity introduced by uplink \gls{mu}-\gls{mimo} transmissions, 802.11ax introduced synchronization mechanisms for precise alignment of time, frequency, and power across \glspl{sta} to avoid interference and ensure efficient decoding at the \gls{ap}~\blue{ \cite{khorov2018tutorial}. The alignment in time is achieved by requiring the \glspl{sta} to transmit their UL data a \gls{sifs} after the Basic Trigger frame transmitted by the \gls{ap}. The \gls{ap} also indicates in the Basic Trigger frame the power that each \gls{sta} should use for transmission to ensure that the different UL transmissions are received at almost the same power at the \gls{ap}. A pre-compensation of the \gls{cfo} is also performed using the training fields in the Basic Trigger Frame~\cite{deng2017ieee}.}



\section{Wi-Fi Energy Savings}
\label{sec:energy}

Power consumption in wireless networks has become a growing concern, particularly as energy efficiency becomes a priority for both \glspl{sta} and \glspl{ap}. An 802.11 card can account for anywhere between 3\% and 10\% of a smartphone's total power consumption~\cite{mentorApPowerSave4}, which underscores the need for power-saving mechanisms. This section explores the energy-saving solutions adopted (\gls{psm} and \gls{twt}) and those under consideration for 802.11bn (\gls{dps}).


\subsection{Power Save Mode (PSM)}

\gls{psm} was introduced in the initial release of 802.11 to alleviate the power consumption of battery-powered 802.11 devices. It introduced two power states for \glspl{sta}, \emph{awake} (active) and \emph{doze} (power save mode), depending on their expected capability to transmit or receive data. The main procedures held within \gls{psm} are as follows.
\begin{itemize}
    \item When an \gls{sta} enters doze mode, it notifies the \gls{ap} by setting the Power Management bit to 1 in the Frame Control field of any frame. This allows the \gls{ap} to buffer the frames intended for such an \gls{sta} and keep them until it wakes up again.
    \item Dozed \glspl{sta} wake up to receive the Beacon frames from their \gls{ap}, in order to check if there are any \glspl{bu} available for them. The existence of buffered traffic for a particular \gls{sta} is informed by the corresponding \gls{aid} in the \gls{tim} bitmap (1 indicates the existence of unicast data).
    \item When an \gls{sta} wakes up, it notifies the \gls{ap} with a PS-Poll or \gls{uapsd} trigger frame. This action makes the \gls{ap} aware that the \gls{sta} is ready to receive the buffered data.  
    \item Finally, the \gls{ap} starts sending the buffered frames, which can be done in multiple transmissions if the More Data bit in the Frame Control is set to 1.
\end{itemize}

While effective at the time it was introduced, \gls{psm} forced \glspl{sta} to wake up frequently---often several times per second---to check the Beacon and send a Trigger frame for data retrieval, even if they did not have any data to transmit and receive. This approach resulted in significant power consumption for low-duty-cycle \glspl{sta}. Later, with the introduction of voice-over-Wi-Fi in 802.11e, a more advanced mechanism, \gls{uapsd}, was introduced. \Gls{uapsd} allowed \glspl{sta} to sleep during beacon intervals and wake up only when they needed to transmit or receive data, reducing unnecessary wake-ups. In addition, 802.11v introduced a new sleep mode, the \gls{wnm} sleep mode, whereby \glspl{sta} do not need to listen to every \gls{dtim} Beacon. However, \gls{psm} was still limited to relatively short sleep intervals and did not provide the long-term scheduling flexibility required by many modern \gls{iot} devices~\cite{mozaffariahrar2025efta}.

\begin{figure}
\centering
\includegraphics[width=1\linewidth]{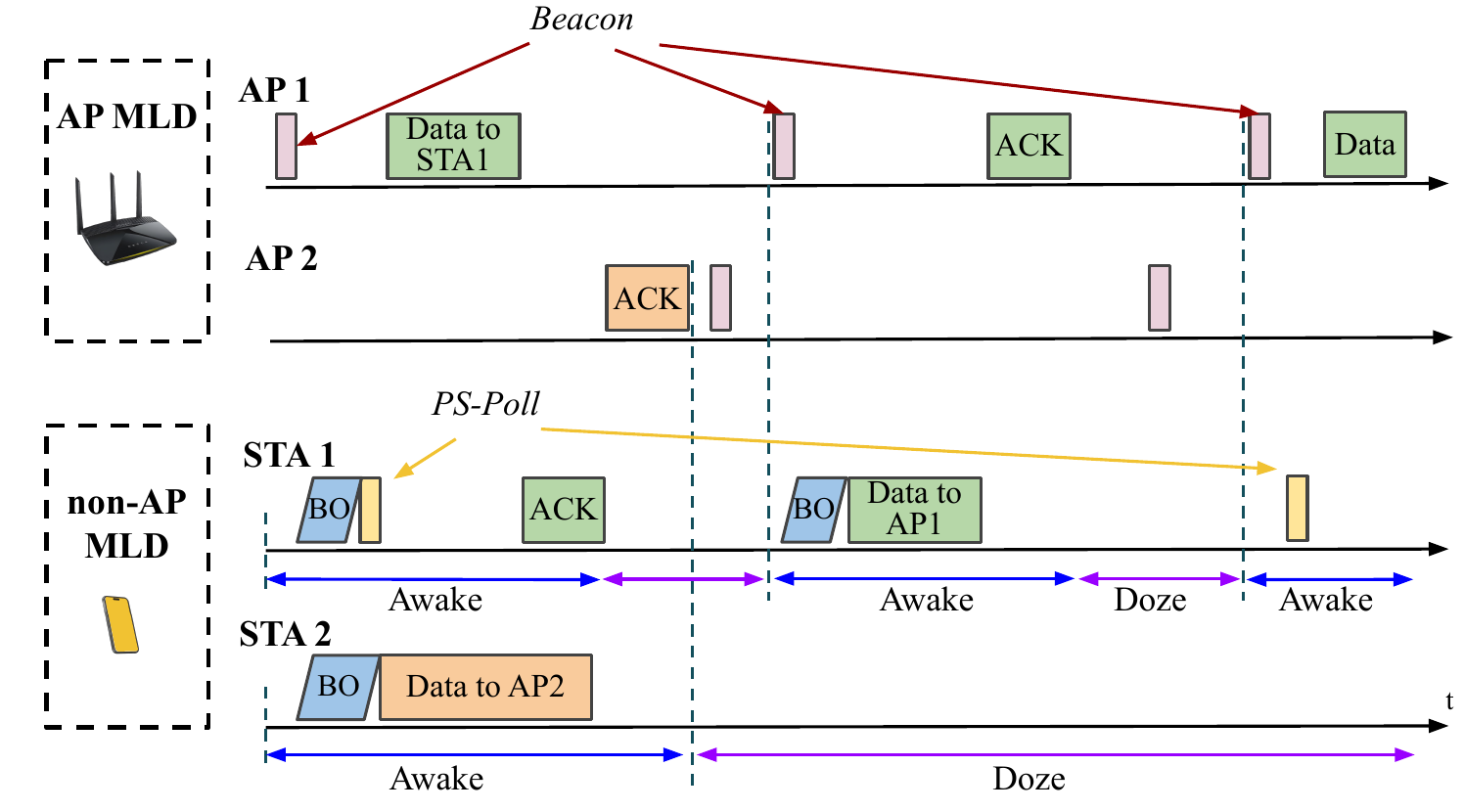}
\caption{Example of the multi-link power-save operation. \blue{The non-AP MLD activates only one link (STA 1) to listen for Beacon frames from the AP MLD. This mechanism allows the non-AP MLD to activate the other links (STA 2) only when required for data transmission, thereby saving energy.}}
\label{fig:psm}
\end{figure}

Today, \gls{psm} has turned into multi-link power-save operation in 802.11be, which extends the feature to multiple links, thus unleashing a high flexibility to perform more efficient energy saving operations thanks to dynamic link switching. As shown in Fig.~\ref{fig:psm}, one non-\gls{ap} \gls{mld} (described later in Section~\ref{sec:MLO}) can wake up to listen to Beacons from an \gls{ap} \gls{mld} on one link only, thus allowing the other link(s) to remain in doze mode when there is no need to use them.

\subsection{Target Wake Time (TWT)}
\label{sec:twt}

\blue{To address the limitations of earlier power saving mechanisms like \gls{psm}, \gls{twt} was introduced in 802.11ah (commercially known as Wi-Fi HaLow), a specialized amendment designed for low-power, long-range \gls{iot} networks in the sub-1 GHz license-exempt bands, but its adoption was limited due to the niche application of 802.11ah devices. While \gls{twt} was explicitly designed for power efficiency and extensive range requirements (at the cost of high speed), its wider adoption was initially limited due to the niche nature of 802.11ah devices.} 

\blue{To meet both power efficiency and high speed performance,} the 802.11ax amendment reintroduced and expanded \gls{twt}, making it a widely applicable power-saving feature for modern devices. \gls{twt} is especially valuable for battery-operated \glspl{sta} that communicate infrequently, such as sensors and other \gls{iot} endpoints. \gls{twt} overcomes some of the limitations of \gls{psm} and, instead of relying on frequent wake-ups, allows \gls{ap} and \glspl{sta} to negotiate a wake-up schedule tailored to the specific needs of the devices. \blue{Therefore, \gls{twt} moves away from statically defined sleeping periods and introduces a much more flexible approach, where the \gls{ap} and the \glspl{sta} define sleep schedules jointly.} The core of \gls{twt} lies in the negotiation and agreement of \glspl{sp} between the \gls{ap} and the \glspl{sta}. The \glspl{sp} define the periods when an \gls{sta} is active for communication. In what follows, we explain how \gls{twt} agreements are established and how \glspl{sp} are implemented~\cite{nurchis2019target}.

\subsubsection{TWT Agreement}

To initiate the negotiation of a \gls{twt} agreement, an \gls{sta} sends a \gls{twt} Request frame to the \gls{ap}. A request can include parameters such as Target Wake Time (offset time to start the first \gls{sp}), \gls{twt} Wake Duration (the awake time within a \gls{sp}), \gls{twt} Wake Interval (the interval between consecutive \glspl{sp}), Minimum \gls{twt} Wake Duration (the minimum period a device stays awake to exchange frames), \gls{twt} Channel (the primary channel on which communication will occur during the \gls{twt} session), \gls{twt} Protection (indication to protect \gls{twt} sessions through mechanisms like \gls{rts}/\gls{cts}), and Negotiation Type \blue{(indicating soft or strict requirements)}. Then, the \gls{ap} responds with a \gls{twt} Reply, which can simply accept the request or renegotiate it.

\Gls{twt} agreements can be \emph{individual} or \emph{broadcast} (collective). In individual \gls{twt}, an \gls{sta} negotiates a wake-up schedule that matches its traffic requirements, minimizing unnecessary wake-ups (Fig.~\ref{fig:TWT_a}). In broadcast \gls{twt}, the \gls{ap} establishes a shared wake-up schedule for a group of \glspl{sta}, optimizing spectrum use and reducing contention (Fig.~\ref{fig:TWT_b}). Broadcast \gls{twt} agreements are announced in Beacon frames, and \glspl{sta} wake up only for relevant Beacons and follow the broadcast schedule. \glspl{sta} can request membership in an existing broadcast agreement or propose new ones to the \gls{ap}. Each \gls{sta} can maintain up to eight \gls{twt} agreements with an \gls{ap}, identified by a 3-bit value. This capability allows devices to adapt to traffic patterns and changing network conditions, improving both performance and power efficiency.

\subsubsection{SP Operation}

\begin{figure}
\centering
\subfloat[Stations using trigger-enabled \gls{twt} under an individual agreement.\label{fig:TWT_a}]
{\includegraphics[width=1\linewidth]{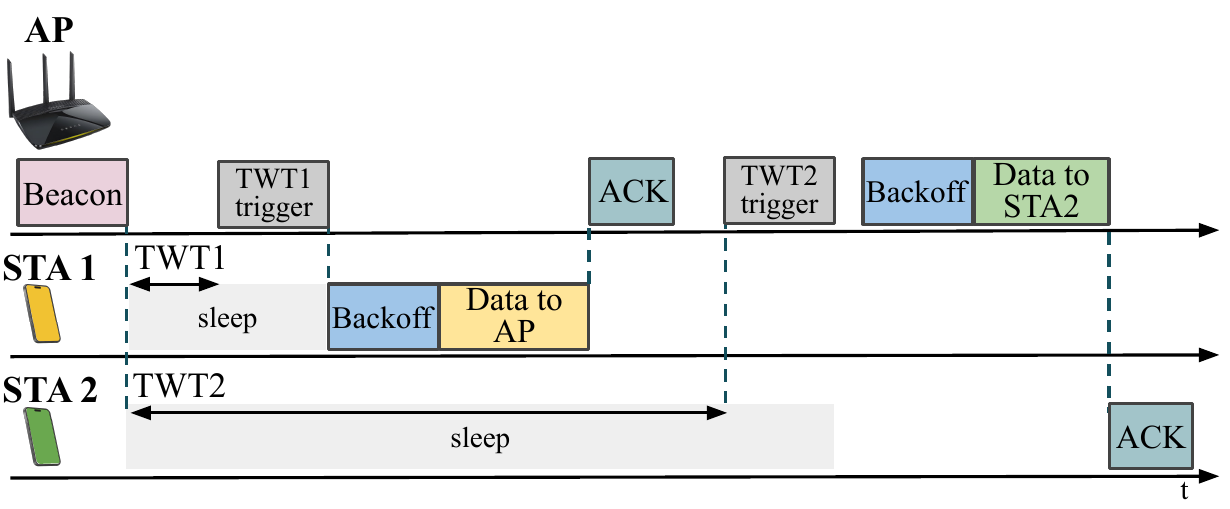}} 
\hfill
\subfloat[Stations using non-trigger-enabled \gls{twt} under a broadcast agreement.\label{fig:TWT_b}]
{\includegraphics[width=1\linewidth]{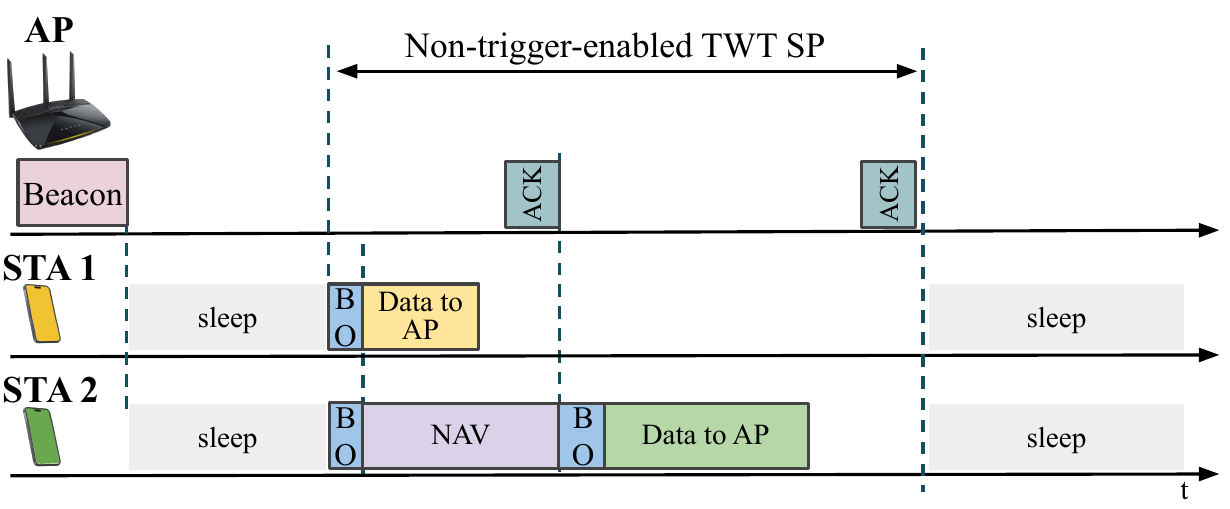}} 
\hfill
\caption{Example of a two-STA network operating using \gls{twt}. \blue{(a) Trigger-enabled TWT: the AP sends a trigger frame to announce the transmission or reception of data for specific STAs during the wake intervals. (b) Non-trigger-enabled TWT: the different STAs allocated to a given SP contend for the medium during the wake intervals.}}
\label{fig:TWT}
\end{figure}

During the agreed \gls{sp}, the \gls{sta} wakes up from its doze state and becomes active to transmit and receive data. Depending on the direction of the data, \gls{dl} or \gls{ul}, the \gls{sta} can use different operation modes. First, uplink data can be transmitted in two ways: \emph{trigger-enabled} mode, whereby the \gls{ap} uses trigger frames to explicitly schedule the transmission and reception of data during the current \gls{twt} wake interval (Fig.~\ref{fig:TWT_a}), and \emph{non-trigger-enabled} mode, where the \gls{sta} autonomously decides when to transmit or receive data during its wake interval (Fig.~\ref{fig:TWT_b}). Regarding the downlink, two more modes are supported: \emph{announced \gls{twt}}, where the \gls{sta} must send a request (e.g., a PS-Poll) to the \gls{ap} to retrieve buffered data, and \emph{unannounced \gls{twt}}, where the \gls{ap} can proactively deliver buffered data without waiting for a request, leveraging the fact that the \gls{sta} is awake at the scheduled time. 

\gls{twt} enables devices to remain in low-power sleep mode for extended intervals, often minutes, hours, or even days, depending on the application. When the agreed wake time arrives, the \gls{sta} wakes up, exchanges data with the \gls{ap}, and returns to sleep. This process ensures highly efficient communication for \glspl{sta} with predictable or periodic data requirements. Finally, with the enhancements provided by \gls{mlo} (see Section~\ref{sec:MLO}), non-\gls{ap} \glspl{mld} can negotiate multiple \gls{twt} \glspl{sp}, one per active link, by sending multiple requests through one of the links.

\subsection{Dynamic Power Save}\label{sec:dynamic_power_save}

\subsubsection*{\gls{sta} \gls{dps}} Unlike previous power-saving methods, 802.11bn defines a \gls{dps} mechanism that allows nodes to dynamically switch between \gls{hc} and \gls{lc} modes in real time,
based on the exchange of specific control frames. \gls{dps} applies to both \glspl{sta} and \glspl{ap}, though it is expected to be more commonly used by \glspl{sta}, which typically face stricter power constraints.

The \gls{hc} mode functions similarly to the traditional awake state, whereas \gls{lc} is a partially active mode, not a full doze. In \gls{lc} mode, the \gls{sta} can receive a limited subset of legacy frames, constrained by parameters such as bandwidth, spatial streams, and maximum \gls{mcs}. For example, \gls{lc} mode may restrict operation to 20\,MHz, one spatial stream, and a low \gls{mcs}~\cite{sanchez2025primer}. This enables the \gls{sta} to wake instantly upon receiving an \gls{icf}, transmitted by its associated device.

Wake-up introduces latency due to the need to initialize \glspl{pll} and activate RF and digital circuits. During a preliminary configuration phase, the \gls{sta} specifies its required delay, which the associated device accommodates by inserting padding into the \gls{icf}. An \gls{ifcs} is included before the padding to allow decoding in advance, ensuring the \gls{sta} is ready by the end of the \gls{icf} transmission. Upon meeting certain conditions (e.g., traffic reduction), an \gls{sta} in \gls{hc} mode may return to \gls{lc} mode.

\subsubsection*{\gls{ap} \gls{dps}} Historically, Wi-Fi power-saving mechanisms have focused on \glspl{sta}, given their reliance on batteries, while \glspl{ap} are typically mains-powered. However, \glspl{ap} still consume significant energy while idle, due to continuous medium sensing. With increasing environmental concerns, energy costs, and \gls{ap} complexity, reducing their power consumption is a growing priority~\cite{sanchez2025primer, guerin2023overview}.

To address this, 802.11bn introduces a \gls{dps} mode for mobile \glspl{ap}, with possible extension to non-mobile \glspl{ap} under consideration~\cite{mentorApPowerSave}. 
Like \gls{sta} \gls{dps}, the mechanism defines \gls{lc} and \gls{hc} states that alternate based on communication needs. AP \gls{lc} mode may include strategies such as maintaining one active link, operating on 20\,MHz, reducing spatial streams, or capping \gls{mcs}.

A key challenge for \gls{ap} power saving, unlike for \glspl{sta}, is maintaining compatibility with legacy (pre-802.11bn) \glspl{sta}. Any power-saving scheme must preserve uninterrupted communication with the associated \glspl{sta}. At the time of writing, the final specification for \gls{ap} power save remains open. Several proposals exist: one suggests that the \gls{ap} remains in \gls{lc} mode for listening, switching to \gls{hc} only when transmitting or receiving~\cite{mentorApPowerSave4}. Another proposes that a mobile \gls{ap} stays in \gls{lc} mode by default unless explicitly triggered by an \gls{sta}, e.g., via an \gls{icf}~\cite{mentorApPowerSave2}. Additionally, periodic \gls{ap} power-saving schedules may be coordinated using broadcast \gls{twt}~\cite{mentorApPowerSave3}.


\section{Wi-Fi Multi-link Operation}
\label{sec:MLO}

\Gls{mlo} is widely regarded as the standout innovation introduced in 802.11be~\cite{IEEE80211beD70_2024}, enabling devices to operate dynamically on multiple channels. 
\Gls{mlo} opens up great opportunities, including increased throughput via link aggregation, reduced latency by improving channel access likelihood, enhanced reliability through redundancy across links, and the ability to separate traffic types---such as control and data planes---across different frequency bands~\cite{UnderstandMLO_Carrascosa2023,PerfCoexMLO_Carrascosa2023}. 
Existing Wi-Fi devices already use multiple links in the 2.4\,GHz, 5\,GHz, and 6\,GHz bands, but these links operate through independent associations. In contrast, \gls{mlo} integrates and manages multiple links through a single association, leveraging them in a coordinated and highly dynamic manner for maximum efficiency and performance. \gls{mlo} is also regarded as a future-proof framework for integrating and coordinating additional links, such as those operating at higher frequency carriers. This presents an opportunity to reconsider existing 802.11 amendments for mmWave technology (see Appendix~\ref{subsubsec: 80211bq}).


\subsection{Multi-link Architecture and Key Procedures}

\Gls{mlo}-capable devices adopt a new architecture to support multi-link operations. In turn, \gls{mlo} comes with new namings for \gls{ap} and \gls{sta} devices, which are now referred to as \gls{ap} \gls{mld} and \gls{sta} (non-\gls{ap}) \gls{mld}. Unlike previous definitions of \glspl{ap} and \glspl{sta}, \glspl{mld} have multiple \gls{phy} interfaces, each capable of operating on different frequency bands or channels. Therefore, an \gls{ap}/\gls{sta} \gls{mld} comprises two or more legacy \glspl{ap}/\glspl{sta} that support multi-link operations. Links are established between pairs of legacy \glspl{ap} and \glspl{sta} within the \glspl{mld}. The architecture divides responsibilities across two sublayers: the upper \gls{mac} sublayer handles \gls{mld}-level shared functions, the lower \gls{mac} sublayer manages per-link operations. 

\commentfigtable
\begin{figure}
\centering
\includegraphics[width=\figwidth]{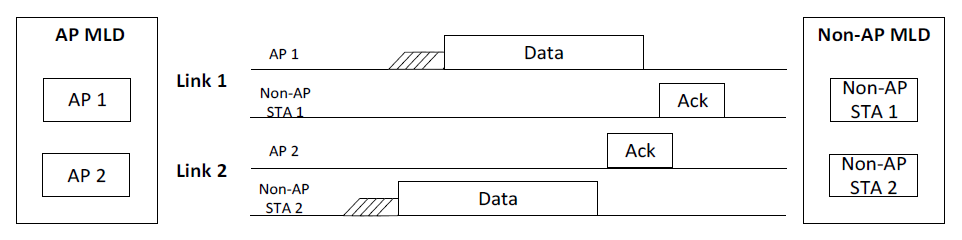}
\caption{MLD and link level architecture \cite{IEEE80211beD70_2024}.\hl{To be redone with the final template.}}
\label{fig:MLD_architecture}
\end{figure}
\endcommentfigtable

\blue{In the following,} we delve into the key aspects of \gls{mlo}, beginning with its discovery and setup procedures and extending to the practical implementations and control mechanisms that enable multi-link functionality.


\subsubsection{Multi-link Discovery}

This phase involves either passive or active scanning procedures. In passive scanning, an \gls{mld} listens on each link and waits to detect Beacon or unsolicited Probe Response frames. In contrast, active scanning involves the \gls{mld} sending Probe Request frames on each of its available links, soliciting faster responses. 
The discovery process works as follows: an \gls{sta} identifies an \gls{ap} and requests information to determine whether to associate with it. Multi-link discovery extends this process to \glspl{mld}, where an \gls{sta} \gls{mld} gathers information about multiple \glspl{ap} across multiple links by allowing a \gls{mld} to advertise information about all its available links through a single link. \blue{As a result, an \gls{sta} \gls{mld} does not have to perform separate scanning on each band/link, reducing 
the need for redundant scanning and accelerates link setup.} 
During multi-link discovery, the \gls{mld} can share either basic or complete information:
\begin{itemize}
    \item 
    \emph{Basic information} includes essential details about a reported \gls{ap} on a specific link, such as operating channel, \gls{bssid}, short \gls{ssid}, and \gls{bss} parameters. To facilitate this, each legacy \gls{ap} in an \gls{ap} \gls{mld} includes a \gls{rnr} element in its Beacon or Probe Response frames. This \gls{rnr} element provides basic information about all other \glspl{ap} within the same \gls{ap} \gls{mld}.
    \item 
    \emph{Complete information} comprises all parameters related to a reported \gls{ap} on a specific link, equivalent to the data included in a Beacon or Probe Response for that link. For this purpose, a new frame type, the Multi-Link Probe Request, allows an \gls{sta} \gls{mld} to request complete information for all \glspl{ap} in an \gls{ap} \gls{mld} in a single transmission on one link. Upon receiving the request, the \gls{ap} \gls{mld} responds with a Multi-Link Probe Response frame, containing complete information for all affiliated \glspl{ap}.
\end{itemize}


\subsubsection{Multi-link Setup and Traffic-to-link Mapping}

Multi-link setup is a key procedure introduced to streamline the association and configuration of multiple links within an \gls{mld}. The association request and response can be performed on a single link, carrying the necessary information to set up all the links in the \gls{mld} and eliminating the need for separate associations for each link. 

Traditionally, the association process allowed only one link to be set up per association, meaning three links would require three separate association executions. Multi-link setup addresses this limitation by enabling the exchange of capabilities and setup procedures for all links in a single execution. This is achieved by reusing existing (Re)Association Request and (Re)Association Response frames, enhanced with a new Multi-link Element. 
The Multi-link Element includes a Common Info field that carries \gls{mld}-level information shared across all \glspl{sta}, plus one or more \gls{sta} Profile sub-elements that provide complete information for each \gls{sta} operating on a specific link.
While legacy \gls{ba} agreements are link-specific, \gls{mld} \gls{ba} agreements apply across all links. Additionally, the receive status of \gls{qos} Data frames may be signaled on other links, ensuring seamless multi-link operation.

A critical procedure in multi-link operation is mapping traffic across different links. This includes modifying traffic indication maps to reflect the entire \gls{mld} and associating different \gls{tid} with individual links to optimize traffic separation and prioritization. The \gls{tim} element indicates buffered data for the entire \gls{mld}, providing a unified view across links. In Default Mode, all \glspl{tid} are mapped to all links. In Optional Mode, specific subsets of \glspl{tid} can be mapped to particular links, enabling traffic separation and prioritization. A link is considered enabled if at least one \gls{tid} is mapped to it. 
By combining the multi-link setup and efficient traffic-to-link mapping, 802.11 networks can achieve enhanced performance, better traffic management, and improved resource utilization across multiple links.


\subsection{Multi-link Implementations}

\Gls{mlo} can be implemented in various forms, each offering distinct capabilities and trade-offs. These implementations include \gls{emlsr} and \gls{emlmr}, illustrated in Fig.~\ref{fig:MLO} and summarized below.

\subsubsection*{Enhanced Multi-link Single-radio (EMLSR)}  
An \gls{emlsr} device listens to two or more links simultaneously, but only for \gls{cca} and limited control frame reception. This approach supports opportunistic spectrum access at a reduced cost, as it requires only one fully functional 802.11be radio, supplemented by low-capability radios that can decode control frame preambles. Upon receiving an initial control frame on one of the links, the device switches to that link and operates using all antennas. 

\subsubsection*{Enhanced Multi-link Multi-radio (EMLMR)}  
\begin{figure}
\centering
\includegraphics[width=\linewidth]{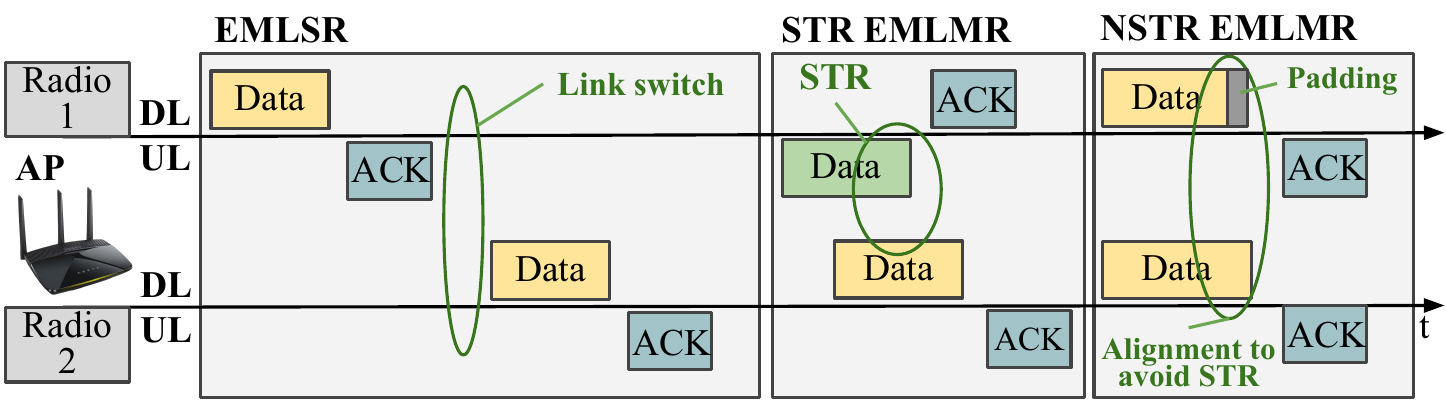}
\caption{\blue{Illustration of different MLO implementations. Left: EMLSR operation, where traffic is switched between two links (radios) over time. Middle: STR EMLMR enables simultaneous transmissions on both links with overlapping data and ACK frames. Right: NSTR EMLMR uses padding and temporal alignment so that transmissions on the two links do not overlap, thereby avoiding STR operation.}}
\label{fig:MLO}
\end{figure}

In \gls{emlmr}, all radios are 802.11be-compliant, enabling concurrent operation across multiple links. \Gls{emlmr} is further classified into:
\begin{itemize}
    \item \emph{\Gls{str}:} Allows simultaneous uplink and downlink transmissions over a pair of links, maximizing flexibility and throughput.
    \item \emph{\Gls{nstr}:} Prevents simultaneous transmissions to avoid self-interference. For example, data and \gls{ack} transmissions over multiple links must start and end at the same time, as illustrated in Fig.~\ref{fig:MLO}.
\end{itemize}
The flexibility of \gls{emlmr} lies in its ability to allocate variable \gls{rf} chains across links. For example, with four RF chains, three could be assigned to a 5\,GHz link and one to a 6\,GHz link, allowing the device to optimize the number of spatial streams on each link.

\subsubsection*{Considerations on MLO implementations}
Studies indicate that under low contention, \gls{str} \gls{emlmr}---the most flexible implementation---can support significantly higher traffic loads and throughput compared to single-link configurations, given the same delay requirements~\cite{Lacalle_LatencyAnalysis_MLO_2021,DelayAnalysis_Bellalta2023,UnderstandMLO_Carrascosa2023,PerfCoexMLO_Carrascosa2023,PerfMLOXR_Carrascosa2024,AnalyticalMLO_Korolev2022}. However, in high-load and high-contention scenarios, \gls{str} \gls{emlmr} devices frequently access multiple links, potentially blocking contending \glspl{bss}. This can occasionally result in larger worst-case delays compared to legacy single-link systems \cite{CarGerKni2024, BelCarGal2023}.


\subsection{Seamless Roaming}\label{subsec:seamless_roaming}

Seamless roaming in 802.11bn~\cite{80211bnD1.1} builds upon the \gls{mlo} framework and is designed to enhance the mobility of \gls{sta} \glspl{mld} by allowing them to transition smoothly between different \gls{ap} \glspl{mld} without losing connectivity. This capability is crucial for maintaining continuous service in environments with multiple overlapping wireless networks, such as large enterprises or public areas. 
The seamless roaming process consists of two main procedures: preparation and execution, as detailed next.

\subsubsection{Roaming Preparation Procedure}
This involves the transfer or renegotiation of the context related to the \gls{sta} \gls{mld} from the current \gls{ap} \gls{mld} to the target \gls{ap} \gls{mld}. The context includes essential information required for maintaining the connection and ensuring a seamless transition.
The preparation also includes setting up the necessary link(s) with the target \gls{ap} \gls{mld}. This step ensures that the \gls{sta} \gls{mld} is ready to switch to the new \gls{ap} \gls{mld} when the execution phase begins, while avoiding any interruption of the data stream thanks to the possibility of maintaining connectivity with the current \gls{ap} \gls{mld}. 
The specifics of what context can be transferred or renegotiated are yet to be determined.

\subsubsection{Roaming Execution Procedure}
The \gls{sta} \gls{mld} initiates the execution phase by sending a Request frame to the current \gls{ap} \gls{mld}. This frame signals the intention to transition to a new \gls{ap} \gls{mld}.
Upon receiving the request, the current \gls{ap} \gls{mld} may continue to send individually addressed downlink data frames to the \gls{sta} \gls{mld} for a predetermined period. This ensures that the \gls{sta} \gls{mld} continues to receive data during the transition.
The current \gls{ap} \gls{mld} is responsible for transferring the necessary context to enable operations with the target \gls{ap} \gls{mld}. This context transfer is crucial for service continuity.
After the context transfer or renegotiation is completed, the current \gls{ap} \gls{mld} sends a response frame to the \gls{sta} \gls{mld}, confirming the completion of the transition process.

Overall, seamless roaming in 802.11bn is focused on minimizing the time during which connectivity is lost, ensuring that applications requiring real-time data transmission, such as \gls{voip} and video streaming, are unaffected by the transitions between \gls{ap} \glspl{mld}.

\section{Wi-Fi Multi-AP Coordination}
\label{sec:MAPC}

Due to their distributed nature, one of the primary limitations of 802.11 networks arises from inter-\gls{bss} interactions, which can lead to collisions and unpredictable worst-case delays. This issue is especially pronounced in dense and heavily loaded environments, 
where contention and interference effects are magnified. Although recent 802.11 amendments have introduced mechanisms to improve channel access within a \gls{bss},
such as \gls{bss} coloring (see Section~\ref{sec:BSScolor}),
the absence of explicit coordination among \glspl{ap} belonging to different \glspl{bss} remains a major obstacle to achieving high reliability in Wi-Fi. \blue{In the past, standardization bodies like \gls{wfa} or \gls{ietf} have developed protocols and functionalities that provide inter-\glspl{ap} backhauls (e.g., Wi-Fi EasyMesh, CAPWAP), enabling the development of wireless controllers and unlocking functionalities such as roaming, mesh networking, or optimized radio resource management. However, these solutions often lack interoperability and are incapable of providing high-grained performance optimization (because they operate at layers above the \gls{phy} and \gls{mac}) and are instead better suited for longer-term network management (e.g., channel allocation).}

The 802.11bn amendment aims to overcome the limitations of spectrum sharing in unlicensed bands and advance the reliability of future Wi-Fi. 
\Gls{mapc} is expected to play a central role in this by \blue{enabling multiple \glspl{ap} to coordinate over-the-air and perform more efficient operations,}
thereby reducing channel contention and interference, and improving resource utilization. 
This section presents the framework and coordination modes currently under consideration in 802.11bn~\cite{P80211bnPAR}: \gls{ctdma}, \gls{crtwt}, \gls{csr}, and \gls{cbf}.

\commentsection
The potential of \gls{mapc} for Wi-Fi 8 goes beyond the existing management Wi-Fi solutions based on, for instance, \gls{wfa}'s \textit{Wi-Fi EasyMesh}~\cite{wfa_easymesh} or \gls{wba}'s \textit{Operator Managed Wi-Fi}~\cite{wba_omwi_reference}. Proprietary mesh Wi-Fi systems, while popular, mainly focus on the management of a limited set of \glspl{ap} that are controlled by a single network manager (e.g., roaming, load balancing, channel selection). Moreover, they lack finer-grained optimizations at the \gls{phy} and \gls{mac} layers needed to significantly enhance multi-\gls{ap} system performance in an \gls{obss}. Wi-Fi 8's \gls{mapc}, in contrast, operates at the \gls{mac} layer, allowing real-time cooperation between the \glspl{ap} at the \gls{txop} level, thus allowing for fine-grained optimization. 
\endcommentsection


\begin{figure*}[t]
\centering
    \subfloat[Co-TDMA.]
    {\includegraphics[width=0.4\columnwidth]{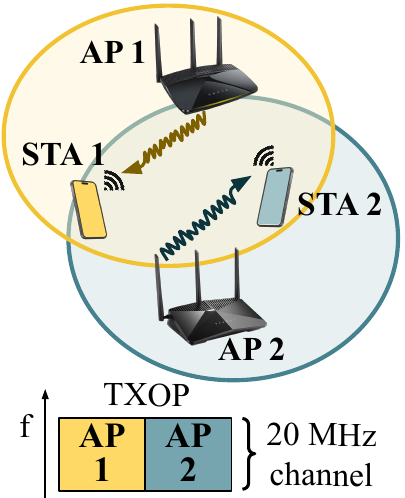}\label{fig:MAPC_a}} 
    \hfill
    \subfloat[Co-R-TWT.]
    {\includegraphics[width=0.4\columnwidth]{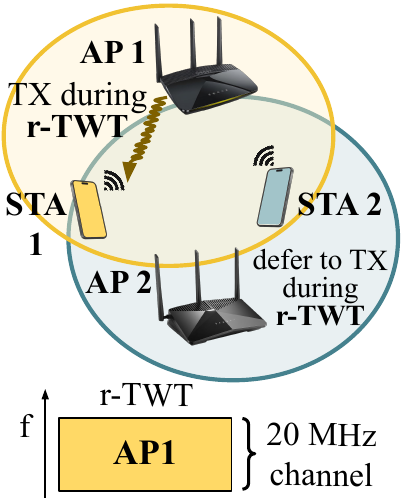}\label{fig:MAPC_b}} 
    \hfill
    \subfloat[Co-SR.]
    {\includegraphics[width=0.4\columnwidth]{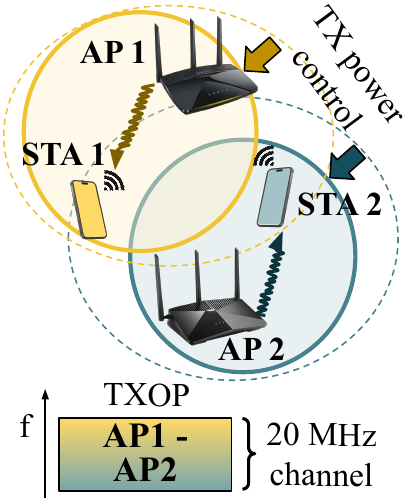}\label{fig:MAPC_c}} 
    \hfill
    \subfloat[Co-BF.]
    {\includegraphics[width=0.4\columnwidth]{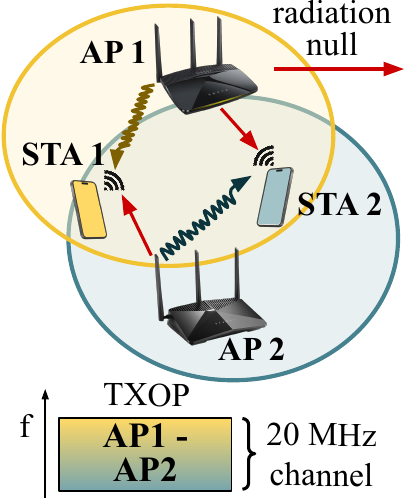}\label{fig:MAPC_d}} 
    \hfill    
\caption{\blue{Illustration of four different MAPC mechanisms for 802.11bn in a scenario with two overlapping APs and two STAs. In (a), Co-TDMA enables time-division sharing of the same 20\,MHz channel between AP1 and AP2 within a TXOP. In (b), Co-R-TWT reserves an rTWT service period for AP1 while AP2 defers its transmissions. In (c), Co-SR uses coordinated transmit-power control so both APs can transmit simultaneously with limited inter-BSS interference. In (d), Co-BF employs joint beamforming to create radiation nulls toward non-intended receivers, enabling fully overlapping concurrent transmissions on the same channel.}}
\label{fig:MAPC}
\end{figure*}

\subsection{MAPC Framework}

Each \gls{mapc} mechanism targets specific challenges in resource allocation, interference mitigation, and performance enhancement, offering the flexibility needed to adapt to diverse deployment scenarios. 
As a result, the implementation complexity and signaling requirements of each \gls{mapc} approach may vary, depending on the volume, type, and periodicity of data exchanged between \glspl{ap}.

802.11bn defines a common framework for \gls{mapc}, designed to support multiple coordination schemes. This framework includes the following key operations:
\begin{itemize}
    \item \emph{\gls{mapc} discovery:} 
    To initiate coordination, 
    \glspl{ap} must advertise their coordination capabilities, e.g., support for responding in a \gls{tb} \gls{ppdu}.
    These capabilities are typically broadcast through management frames, such as Beacons or Probe Request/Response frames, allowing neighboring \glspl{ap} to discover and evaluate potential coordination partners.    
    \item \emph{\gls{mapc} negotiation:}
    Once neighboring \glspl{ap} discover one another and agree to coordinate, 
    they establish a formal agreement defining the conditions for coordinated transmissions. This includes negotiating roles (e.g., coordinator vs. responder), maximum \gls{txop} durations for simultaneous transmissions, and transmit power limits. Negotiation occurs via management frames,
    such as Public Action frames or new individually-addressed Action frames.
    While 802.11bn limits each \gls{mapc} agreement to two \glspl{ap},
    it permits multiple concurrent agreements.
    
    \item \emph{\gls{mapc} session:}
    With an agreement in place, 
    \glspl{ap} can initiate coordinated transmissions.
    Although each \gls{mapc} scheme has distinct requirements,
    a unified mechanism is defined to trigger coordination. 
    Specifically, the \emph{sharing \gls{ap}}, i.e., the one holding the \gls{txop}, sends a Trigger frame to notify the \emph{shared \gls{ap}} of the start of a coordinated transmission.
    
    \item \emph{\gls{mapc} teardown:} 
    A coordinated session can be terminated using new teardown messages,
    which may be initiated by any participating \gls{ap}.
\end{itemize}

All the aforementioned procedures require a significant architectural transformation from 802.11be,
which currently lacks the necessary messages and mechanisms to support inter-\gls{ap} communication.
As noted earlier,
new fields, formats, and even message types,
such as \gls{mapc} Discovery, \gls{mapc} Negotiation Request, and \gls{mapc} Negotiation Response,
will need to be defined. \blue{The 802.11bn amendment considers over-the-air communication among \glspl{ap}, which, unlike alternative wired backhauls, allows a sub-millisecond synchronization that is necessary to perform joint transmission operations in real time.} 

To ensure the \gls{mapc} framework is generalizable across coordination schemes, it is essential to introduce flexible signaling structures, such as a dynamic \gls{usig} field, where \gls{mapc}-specific information can be adapted as needed. Additionally, new \gls{mlme} primitives will be required to support core operations, e.g., to trigger the transmission of a \gls{mapc} Negotiation Response upon receipt of a corresponding request.

From a management perspective, 
identifiers such as \gls{ap} IDs and session IDs must be maintained to track eligible and coordinated \glspl{ap}, as well as active and past coordination sessions. 
Finally, security mechanisms must be extended to enable the secure exchange and management of cryptographic keys between coordinating \glspl{ap}, 
potentially leveraging mechanisms like \gls{pasn}.

\commentsection
The MAPC framework begins with defining an AP candidate set, which includes APs eligible to initiate or participate in MAPC, forming a trusted coordination region. Only APs within the same candidate set can engage in MAPC with one another. The process for forming this candidate set is implementation-specific and depends on the network configuration. MAPC transmissions involve a sharing AP, which obtains a TXOP and coordinates transmissions with one or more shared APs. During the MAPC transmission, the Sharing AP manages resource allocation for the Shared APs. APs need to obtain CSI from OBSS to estimate channel conditions for non-associated neighboring devices. This information supports frequency resource management, transmit power adjustments, and beamforming to minimize OBSS interference. Efficient CSI acquisition will be crucial to balance performance gains against overhead.
\endcommentsection


\subsection{MAPC Mechanisms}
\label{sec:mapc_features}

This section provides an overview of the four candidate \gls{mapc} mechanisms proposed for 802.11bn, as illustrated in Fig.~\ref{fig:MAPC}.

\subsubsection{Coordinated TDMA (Co-TDMA)}

\Gls{ctdma} is one of the foundational approaches to \gls{mapc}, 
leveraging coordination in the time domain~\cite{val2025wi}. 
In \gls{ctdma}, the sharing \gls{ap} divides its \gls{txop} into multiple time slots,
distributing each slot among itself and the shared \glspl{ap}.
This method aims to reduce channel contention and improve resource allocation efficiency.
As illustrated in Fig.~\ref{fig:MAPC_a}, 
two \glspl{ap} (\gls{ap}~1 and \gls{ap}~2) can split the duration of a \gls{txop} to transmit to their respective \glspl{sta}, 
showcasing the fundamental principle of time-based coordination in \gls{ctdma}.


\gls{ctdma} operates in three phases~\cite{val2025wi}: 
the sharing \gls{ap} first polls shared \glspl{ap} with an \gls{icf}; then allocates \gls{txop} slots based on responses and access category priorities; and finally, allows early \gls{txop} return if a shared \gls{ap} finishes before its slot ends.
The detailed scheduling logic is implementation-specific and may vary across vendors and system objectives.
To prevent interference during coordination, \gls{ctdma} uses enhanced NAV protection mechanisms, including short NAV settings during polling, dynamic updates during \gls{txop} return, and backup strategies like \gls{edca} timers or \gls{rts} gating~\cite{val2025wi}.

A key limitation of \gls{ctdma} is its inability to achieve spatial reuse.
The time slots allocated to shared \glspl{ap} are carved directly from the sharing \gls{ap}'s own \gls{txop},
resulting in no net increase in overall airtime utilization. 
Furthermore, performance may degrade in the presence of mismatched traffic loads or access categories. 
For instance, in unbalanced  \gls{obss} scenarios~\cite{val2025wi},
an \gls{ap} serving mostly low-latency traffic may find limited benefit in sharing with an  \gls{ap} focused on throughput-heavy traffic, 
leading to poor slot utilization, wasted return time, or coordination inefficiencies.

Despite these limitations, \gls{ctdma} provides a simple and deterministic framework for multi-\gls{ap} coordination. By avoiding contention and introducing structured access, it improves latency predictability, particularly in dense or delay-sensitive deployments.


\subsubsection{Coordinated R-TWT}

\Gls{crtwt} is a mechanism that enhances reliability by allowing \glspl{ap} to share and coordinate their \gls{rtwt} schedules
(see Section~\ref{sec:rtwt})~\cite{haxhibeqiri2024coordinated}. 
The ability to secure unlicensed spectrum resources across multiple \glspl{bss} represents a major breakthrough for Wi-Fi.
Fig.~\ref{fig:MAPC_b} illustrates the \gls{crtwt} principle, 
where \gls{ap}~1 successfully protects its transmission to \gls{sta}~1. 
This is achieved through a coordination agreement with another \gls{bss} (\gls{ap}~2 and \gls{sta}~2),
which remain silent during \gls{ap}~1's scheduled \gls{sp}.
Such lightweight coordination makes \gls{crtwt} particularly suitable for reliability-sensitive applications in \gls{obss} scenarios.\!

To enable \gls{crtwt}, the two \glspl{ap} must negotiate an \gls{rtwt} schedule by exchanging frames, either existing or newly defined management frames (to this date, the specific negotiation and advertisement of \gls{rtwt} schedules between \glspl{ap} has not been defined).
These frames convey key information such as the \gls{twt} Interval, the Broadcast \gls{twt} ID, or the Nominal Minimum \gls{twt} Wake Duration 
(notice that other parameters might also be included later, such as the parameters needed to time synchronize the \glspl{ap}).
Multiple \gls{rtwt} agreements can be negotiated in parallel,
each of which is included as a \gls{crtwt} Parameter Set field.
Once these agreements are in place,
each \gls{ap} must ensure that neither its own transmissions nor those from its associated \glspl{sta} overlap with the scheduled \glspl{sp} of the other \gls{ap}.
For that, each involved \gls{ap} must advertise,
e.g., by extending Beacon frames,
the \gls{obss} \gls{rtwt} schedules to its \glspl{sta}.

Similarly to \gls{ctdma}, 
\gls{crtwt} focuses on a better utilization of time resources, so no net gain is expected. However, \gls{crtwt} offers good potential for achieving reliability in \gls{obss} scenarios.


\subsubsection{Coordinated Spatial Reuse (Co-SR)}

\gls{csr} enables \glspl{ap} to cooperatively manage their transmit power, 
facilitating concurrent transmissions and improving spectral efficiency in dense environments~\cite{wilhelmi2023throughput}. 
This approach enhances the moderately adopted spatial reuse operation
(see Section~\ref{Sec:CoexistenceAndSpatialReuse}), 
where one \gls{ap} transmits at full power,
while others reduce their power,
often resulting in suboptimal \gls{sinr} for some \glspl{sta}. Fig.~\ref{fig:MAPC_c} illustrates the principle of \gls{csr}, 
where \gls{ap}~1 and \gls{ap}~2 adjust their transmit power to enable concurrent transmissions within a shared \gls{txop}, maximizing spectral reuse without compromising link quality.

To enable \gls{csr}, two \glspl{ap} exchange the necessary information to derive transmit power levels that ensure an acceptable \gls{sinr} at all receiving \glspl{sta} during simultaneous transmissions.
A practical method involves measuring and sharing the \gls{rssi} of interfering links,
e.g., between the sharing \gls{ap} and the shared \gls{ap}'s \gls{sta}. 
Existing mechanisms support frequent \gls{rssi} measurements and centralized reporting at the \gls{ap}, making \gls{csr} viable with minimal coordination overhead.

Once transmit power values are negotiated, the coordinated transmission is triggered by the shared \gls{ap} through the transmission of a Trigger frame (which one remains to be decided), indicating the shared \gls{txop} duration. From the shared \gls{ap} side, the transmit power employed cannot exceed the one previously negotiated. 802.11bn defines two \gls{csr} transmission modes, differentiated by the indication of different information in the \gls{usig}: \textit{Mode~1}, whereby the different combinations of \gls{uhr} and \gls{eht} \glspl{sta} are considered, and \textit{Mode~2}, which defines transmissions involving \gls{uhr} devices only. In either case, and for the sake of ensuring reliable decoding, 
both the sharing and shared \gls{ap} must begin and complete their \gls{ppdu} transmissions simultaneously. In the \gls{uhr}-only case (Mode~2), the coordinated transmission starts a \gls{sifs} after the trigger.

\gls{csr} is a feature that only accounts for \gls{su} \gls{dl} transmissions and which limits the maximum number of spatial streams to four. However, its potential for improving the delay is substantial thanks to the added capability of enabling concurrent transmissions from multiple \glspl{bss}, provided those \glspl{bss} have data ready to send at the same \gls{txop}.


\begin{figure*}[t]
    \centering
    \includegraphics[width=0.82\linewidth]{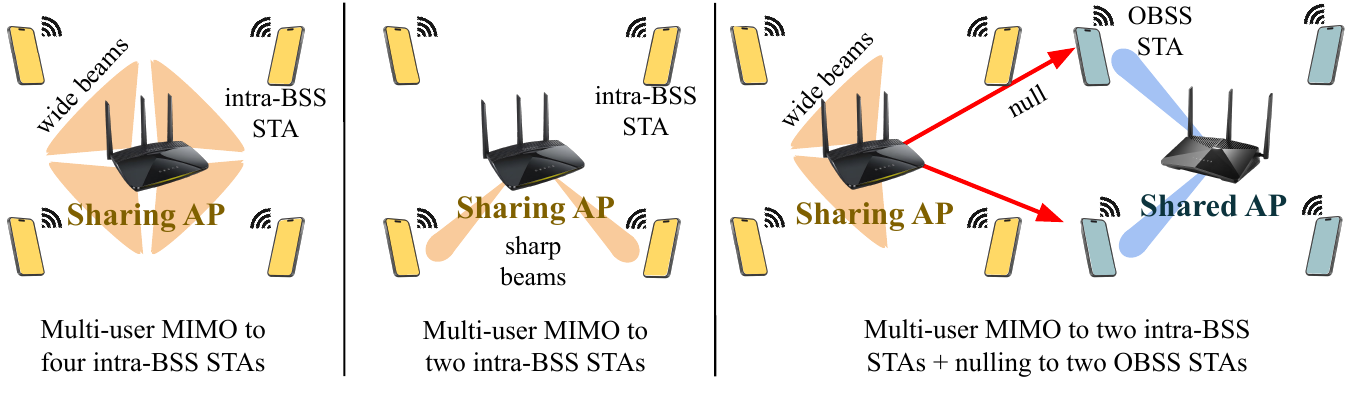}
    \setlength\abovecaptionskip{-0.05cm}
    \caption{Three examples of employing four spatial \gls{dof} in \gls{cbf}. Left: all four DoF are used to spatially multiplex four intra-\gls{bss} \glspl{sta}, resulting in limited beamforming gain. Middle: two \gls{dof} are used to spatially multiplex two intra-\gls{bss} \glspl{sta}, while the remaining two \gls{dof} provide additional beamforming gain. Right: two \gls{dof} are used to spatially multiplex two intra-\gls{bss} \glspl{sta}, and the remaining two \gls{dof} place nulls at two \gls{obss} \glspl{sta}, enabling spatial reuse.\vspace{-0.5cm}}
    \label{fig:CBF_tradeoffs}
\end{figure*}

\subsubsection{Coordinated Beamforming (Co-BF)}

\gls{cbf} is a spatial-domain coordination mechanism in which multi-antenna \glspl{ap} suppress interference to and from neighboring \glspl{sta} belonging to other \glspl{bss}~\cite{shen2020coordinated}.
By leveraging spatial degrees of freedom, 
each \gls{ap} can serve its own associated \glspl{sta}, 
while simultaneously placing radiation nulls toward non-associated \glspl{sta}~\cite{GerGarLop2017, GarGerGio2018}. 
This suppression of spatial interference reduces contention and enables concurrent transmissions, improving both spectral reuse and worst-case latency in dense deployments.
Fig.~\ref{fig:MAPC_d} illustrates a concurrent transmission scenario enabled by \gls{cbf}, 
where \gls{ap}~1 and \gls{ap}~2 steer radiation nulls towards \gls{sta}~2 and \gls{sta}~1, respectively, to suppress mutual interference.

In \gls{cbf}, the sounding phase is critical, and therefore, it is being carefully designed. In particular, two sounding procedures are envisioned in 802.11bn: (i) \emph{Sequential sounding}, whereby \gls{ndpa} and \gls{ndp} exchanges and \gls{cfr} feedback reporting occur in one \gls{bss} at a time (the same \gls{eht} Compressed Beamforming/CQI report and procedures detailed in Section~\ref{sec:MIMO} are adopted), and (ii) \emph{Joint sounding}, where a new protocol allows for \blue{multiple \glspl{ap} to simultaneously perform sounding with} a given \gls{sta}. The type and conditions associated with sounding, e.g., the number of \gls{obss} sounding reports that can be stored at a time, are agreed upon during the negotiation phase of \gls{cbf}. Furthermore, before starting each sounding phase, the \glspl{ap} involved in \gls{cbf} are expected to exchange certain frames (which are to be decided) to indicate availability for participating in the process and inform about the capabilities of the \glspl{sta} to be sounded. Finally, to address \gls{cfo} correction, it is worth noting that the sharing and shared \glspl{ap} take the role of \emph{sync-reference} and \emph{sync-follower} \glspl{ap}, respectively.

Once beamforming information is available at both \gls{ap} sides, the coordinated transmission can take place. Before initiating the transmission of the \gls{cbf} \glspl{ppdu}, the sharing \gls{ap} first sends a newly defined \gls{cbf} Invite frame, which is used to indicate parameters such as the minimum and maximum number of supported \gls{ofdm} symbols, the selected bandwidth, information regarding the punctured channel, the total number of spatial streams allowed for the sharing \gls{ap}, or the number of \gls{cbf} \glspl{sta} and corresponding spatial streams in the sharing \gls{bss}. Such a message is replied to by the shared \gls{ap} with a \gls{cbf} Response frame, which indicates the acceptance or not of undergoing a \gls{cbf} transmission, plus the suggested number of \gls{ofdm} symbols (not smaller than the initially proposed) and information about the number of \glspl{sta} and their assigned \gls{mcs} and spatial streams in the shared \gls{bss}. The setup is concluded by a \gls{cbf} Sync frame, which the sharing \gls{ap} sends to confirm the final set of parameters.

\blue{Unlike joint transmission schemes where all \glspl{ap} cooperate to jointly serve all \glspl{sta}~\cite{GerGarLop_WCNC2018}, 
\gls{cbf} does not require joint data processing or tight backhaul integration.} 
Each \gls{sta} is served by a single \gls{ap},
and only the channel information is required to coordinate null placement. 
This design makes \gls{cbf} more practical for decentralized deployments, 
albeit still dependent on reliable \gls{cfr} acquisition.

However, the performance of \gls{cbf} hinges on several critical factors. First, spatial \gls{dof}---limited by each \gls{ap}'s antenna array---must be split among competing goals: (i) spatial multiplexing of multiple intra-\gls{bss} \glspl{sta}, (ii) enhancing \gls{sinr} via beamforming gain, and (iii) placing nulls to suppress interference to \gls{obss} \glspl{sta}. Fig.~\ref{fig:CBF_tradeoffs} shows three representative configurations using four \gls{dof}, highlighting the trade-offs among these objectives.

Second, the accuracy of available \gls{cfr} is constrained by the channel coherence time, 
which depends on \gls{sta} mobility and environmental dynamics. As the channel ages, outdated \gls{cfr} may result in imperfect nulls, degrading \gls{sinr}, and reducing throughput. While allocating more resources to channel estimation can reduce these effects, it introduces overhead that must be carefully managed.

Lastly, each \gls{ap} must intelligently select which intra-\gls{bss} \glspl{sta} to serve and which \gls{obss} \glspl{sta} to null. These decisions are influenced by the quality of \gls{cfr}, the \gls{snr} of each candidate \gls{sta}, and the spatial correlation between them. Together, these challenges underscore the importance of efficient channel information acquisition, robust scheduling, and dynamic user selection to achieve the full potential of \gls{cbf}.



\commentsection

\subsection{Other Discussed MAPC Approaches}



\subsubsection{Coordinated OFDMA (Co-OFDMA)}

\subsubsection{Joint Transmission}

JTX, also referred to as distributed MIMO, is an advanced multi-AP coordination technique that leverages the spatial domain by enabling non-co-located APs to jointly transmit and receive data to and from multiple STAs. This approach transforms neighboring APs from potential sources of interference into cooperative data transmitters, significantly enhancing network performance. 
JT offers the potential for high throughput and low latency by suppressing interference without reducing the number of spatial streams. However, its implementation is complex, requiring: (i) a distributed CSMA/CA protocol to coordinate transmissions; (ii) tight synchronization across cooperating APs in time, frequency, and phase; and (iii) data sharing among participating APs, necessitating a reliable out-of-band backhaul, most likely a wired connection. 
Despite these challenges, JTX holds significant promise for optimizing spectral efficiency and achieving superior network performance in dense Wi-Fi deployments.
\endcommentsection


\commentsection

\subsection{Performance and Trade-offs in MAPC \blue{[Lorenzo, Boris]}}

Multi-AP coordination (MAPC) involves several trade-offs, many of which can be particularly well-illustrated using CBF as an example. 

\begin{figure}
\centering
\includegraphics[width=0.9\linewidth]{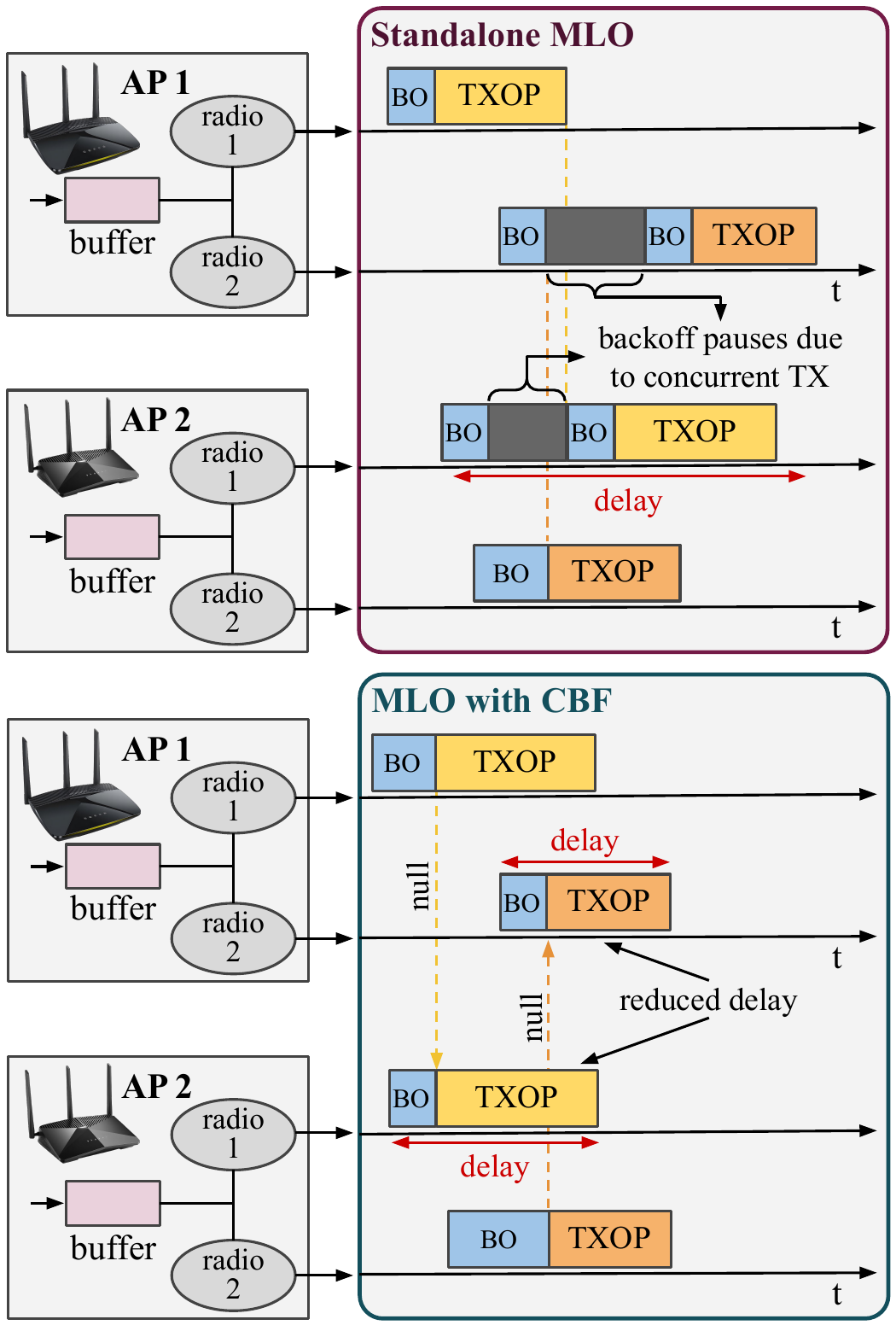}
\caption{Illustration of standalone MLO (top) and MLO with CBF
(bottom). When CBF is enabled, imperfect nulling may lead to a
lower MCS and increased transmission delays. However, this may be
more than compensated for by the increased spatial reuse. \hl{BB: Is there any reason to limit the channel widht to 20 MHz? I would just remove the 20 MHz channel text from the figure.}}
\label{fig:CBF_MLO_diagram}
\end{figure}

This section presents a preliminary evaluation of the performance trade-offs when CBF is combined with MLO~\cite{GalGerCar2024}, as envisioned for 802.11bn. The study considers a Wi-Fi 7 network with two overlapping BSSs.
Each BSS comprises a single AP with four antennas and one associated STA with two antennas. Both BSSs support MLO (specifically EMLMR) and operate on two 160\,MHz links while implementing CBF. Each AP transmits two spatial streams to its associated STA, using the remaining two spatial DoF to create nulls toward the other BSS when CBF is enabled. 
As illustrated in Fig.~\ref{fig:CBF_MLO_diagram}, we compared standalone MLO (top) with MLO combined with CBF (bottom). Combining MLO with CBF creates additional reuse opportunities, highlighted in yellow, and reduces delay by enabling concurrent transmissions across overlapping BSSs. However, the performance of CBF depends critically on the accuracy of null placement. Imperfect nulling leaves residual interference, which may necessitate using a lower MCS, impacting overall throughput. 

\begin{figure}
\centering
\includegraphics[width=\figwidth]{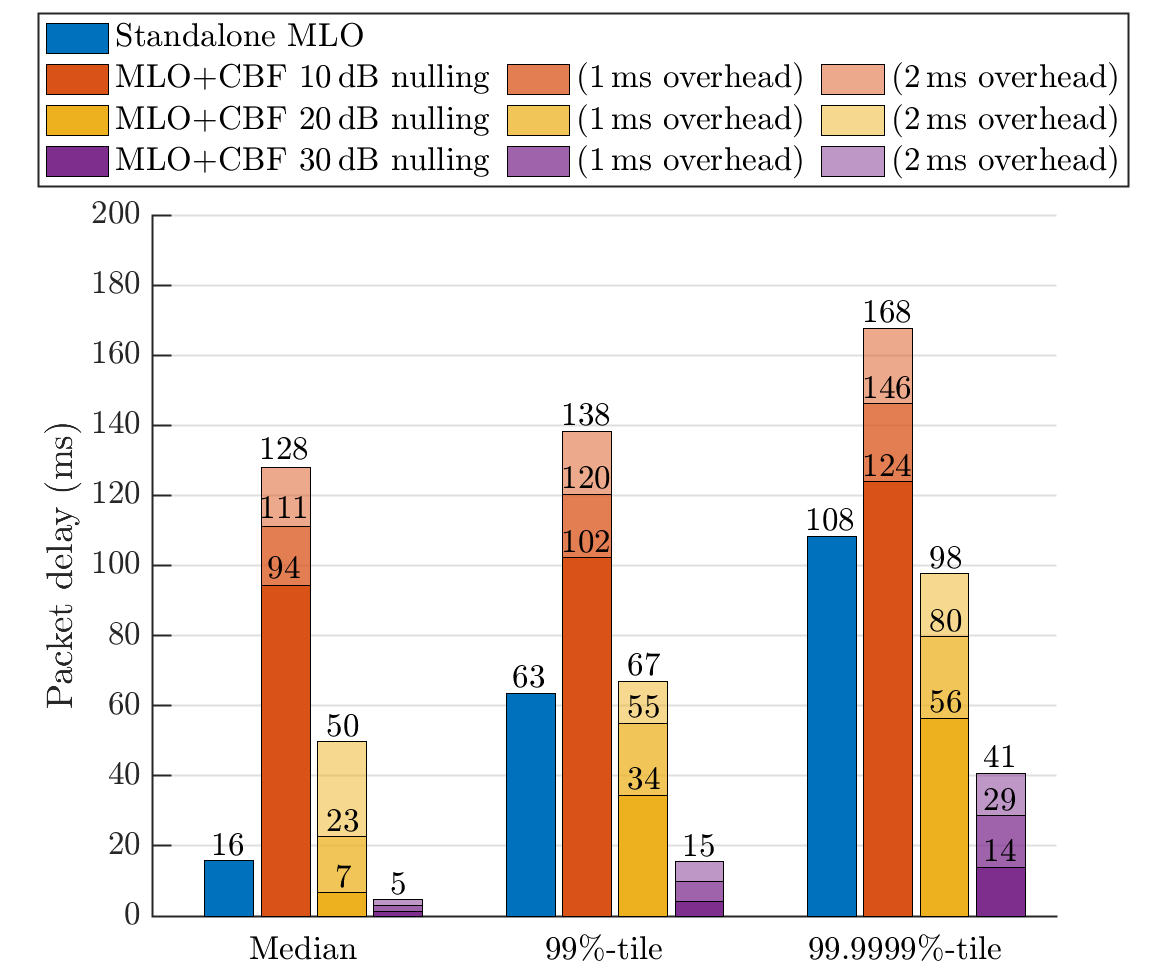}
\caption{Delay incurred by standalone MLO and when combining MLO
with CBF under a variable nulling accuracy and overhead.}
\label{fig:CBF_performance}
\end{figure}

The study evaluates the performance of CBF by modeling the effects of CSI acquisition signaling and its aging. Fig.~\ref{fig:CBF_performance} presents median, 99th-percentile, and six-nines reliability delay values for MLO combined with CBF, as interference suppression accuracy increases from 10 to 30\,dB. Suppression accuracy is represented by colors ranging from red (10\,dB) to purple (30\,dB), while CSI acquisition overheads, ranging from 0 to 2\,ms, are shown with varying opacities. For comparison, standalone MLO performance is depicted in blue bars. 
Key observations include:
\begin{itemize}
    \item Under standalone MLO, six-nines delay exceeds 100\,ms due to medium contention.
    \item With CBF and a null accuracy of 10\,dB, performance initially worsens as residual interference degrades the MCS.
    \item As null accuracy improves to 20\,dB or higher, the reduction in contention outweighs the impact of MCS degradation. At 30\,dB accuracy, the highest MCS is achieved, leading to a nearly tenfold reduction in the worst-case delay.
    \item High CSI acquisition overheads can offset the advantages of CBF, emphasizing the need for efficient and accurate CSI acquisition to fully realize the benefits of MAPC.
\end{itemize}

\endcommentsection


\section{Conclusions}\label{sec:conclusion}
With Wi-Fi over 25 years old, it is perhaps not surprising that there are decades of Wi-Fi tutorials published already, including some by the authors of this paper. Notwithstanding, this paper is the first tutorial that spans Wi-Fi’s full evolution from 802.11b to the emerging 802.11bn (Wi-Fi~8). 
We describe how use cases have evolved from basic web browsing to high-resolution video requiring high throughput, to real-time critical applications requiring low latency and high reliability. To keep up with, and indeed drive such application demand, the underlying technologies of Wi-Fi have radically transformed over these generations. Major early milestones include introduction of OFDM and MIMO. The next generations brought features such as multi-user technologies including uplink and downlink OFDMA and multi-user MIMO. Energy saving methods such as Power Saving Mode and Target Wake Time are increasingly important as Wi-Fi overwhelmingly dominates data usage in mobile phones. Recently, \blue{Multi-Link Operation and Multi-AP Coordination} are yielding new capabilities for providing not only higher throughput, but also radically improving reliability and reducing latency. This article aims to be a comprehensive resource for researchers and practitioners to describe \mbox{Wi-Fi}'s technical journey, current status and the road ahead.

\appendices



\blue{Building on the advancements envisioned for 802.11bn, the Appendix explores parallel efforts shaping the future of \mbox{Wi-Fi}. Key areas of focus include mmWave operations, sensing, security and privacy, and the integration of AI/ML techniques into Wi-Fi.}


\section{\blue{Task Group 802.11bq: Integrated mmWave Operations}}
\label{sec:11bq}

The primary fuel for innovation in wireless technology has always been the availability of spectrum. The regulatory experiment in the 2.4\,GHz band with approximately 80\,MHz allocated for unlicensed operation gave rise to technologies such as Wi-Fi and Bluetooth in late 1990s. Realizing the success of these technologies, regulators around the world made additional unlicensed spectrum available in the 5\,GHz band in early 2000s and, more recently, in the 6\,GHz band. However, wireless spectrum is a finite resource with a demand much greater than the available supply. Therefore, there is a growing fear that finding a large contiguous swath of unlicensed spectrum below 10\,GHz will be extremely difficult, if not impossible. Among other things, this means that increasing the channel bandwidths of present Wi-Fi above 320\,MHz or adding new channels will be challenging, which then limits Wi-Fi's future ability to meet the stringent demand of emerging applications in highly dense environments, such as use of VR/AR in classrooms and wireless docking stations in enterprises.

To address this challenge, proposals have emerged to leverage the large swaths of unlicensed spectrum available in the 60\,GHz mmWave band (e.g., 57\,GHz to 71\,GHz in the USA and EU) to enable an extension to Wi-Fi that can deliver, among other things, channels wider than 320\,MHz for high throughput operation, ultra-low latency wireless connectivity, and networks that can operate in highly dense environments given the large number of channels that can be created. Needless to say, operation in mmWave bands is not without its challenges. \blue{Besides a higher energy consumption,} operating in these bands entails facing rapid signal attenuation and susceptibility to blockages. One approach to mitigate these losses is through beamforming gain, where most spatial degrees of freedom are allocated for beamforming rather than for transmitting multiple spatial streams. This enables improved link reliability and performance, even under challenging propagation conditions.  

In early 2025, the 802.11bq Task Group, also known as Integrated mmWave (IMMW), was formed to explore these opportunities further. The goal is to develop a new amendment for dynamically operating mmWave links using the \gls{phy}/\gls{mac} functionalities derived from 802.11ac, 802.11be, and 802.11bn. Future high-end devices are expected to operate not only in the traditional sub-7\,GHz bands, but also in the 60\,GHz band. 


\subsection{Background: 802.11ad and 802.11ay (WiGig)}

801.11bq is not the first project in 802.11 dealing with operation in mmWave. Between 2010 and 2020, two other amendments were developed that defined operation in mmWave: 802.11ad and 802.11ay, collectively referred to as WiGig. 802.11ad operates on a single channel with a bandwidth of 2.16\,GHz, while 802.11ay supports channel bonding (up to four 2.16\,GHz channels) and \gls{mu}-\gls{mimo} (up to 4 spatial streams), significantly enhancing throughput and capacity. Despite these improvements, the market adoption of 802.11ad and 802.11ay has been limited and mainly in specific segments such as fixed wireless access. The main reasons for this limited adoption are the costs and complexity associated with defining a completely new mmWave modem that does not reuse any component of the existing 2.4/5/6\,GHz modem. These have a ripple impact on the design, development, silicon area, power consumption, and validation of such solutions.


\subsection{Overview of 802.11bq}\label{subsubsec: 80211bq}

802.11bq is the current project that promises to overcome the shortcomings faced with 802.11ad and 802.11ay. 802.11bq has a stated goal of building on the existing 802.11ac/be/bn \gls{phy} and \gls{mac} features and channelization schemes. This means that the basic modulation scheme of 802.11bq will be based on \gls{ofdm}, upclocking the modem used in the 2.4/5/6\,GHz bands. As for channelization, 802.11bq could adopt channel sizes that are multiple of 160\,MHz or 320\,MHz as defined in the lower bands. 

\begin{figure}
\centering
\includegraphics[width=\figwidth]{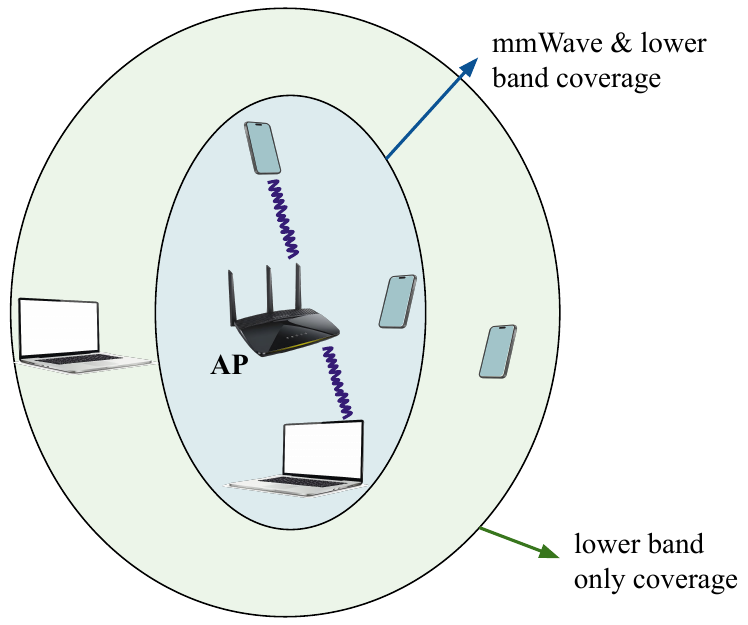}
\caption{MLO can dynamically switch operation between lower and mmWave bands.}
\label{fig:mmWave}
\end{figure}

In the \gls{mac}, a key innovation will be the reuse of the \gls{mlo} framework introduced in 802.11be (see Section~\ref{sec:MLO}), which could be extended to manage operations across the lower and mmWave bands, dynamically activating links as needed to adapt to varying conditions. 
\commentfigtable(see Fig.~\ref{fig:mmWave})\endcommentfigtable
With \gls{mlo}, the \gls{ap} would be able to dynamically (in the order of microseconds) switch between lower and mmWave bands to exchange packets with \glspl{sta}. Moreover, given that links in the lower bands will be more robust than those in mmWave, using \gls{mlo} will make it possible to move all control and management traffic to the lower bands while dedicating the use of the high capacity and ultra-low-latency links in mmWave bands for data exchanges only. For example, the lower bands could be used for beacon frame transmissions, authentication and key establishment, association, and similar management exchanges, which would also encompass operation in mmWave. Then when data transmission is needed with an \gls{sta} that finds itself in mmWave coverage, the AP can choose whether to use the lower, mmWave, or even both bands to exchange data with the \gls{sta}.

\commentfigtable
\endcommentfigtable

\section{\blue{Task Group 802.11bf: Wi-Fi Sensing}}
\label{sec:11bf}

Sensing is a transformative application of Wi-Fi, which will allow devices to ``see'' the surrounding environment, providing new services to the users, and improving the network management procedures~\cite{adib2013see}. The increasing interest in this feature led to the instantiation of a new \gls{tg} in 2020 to design an amendment, identified as 802.11bf, that will support sensing in the sub-7\,GHz and mmWave portions of the radio spectrum.
The key idea is that wireless devices are inherently able to sense the environment thanks to the channel estimation procedure that is continuously performed for data decoding and precoding as thoroughly detailed in Section~\ref{sec:Multiuser}. The \gls{cfr} provides information about how transmitted signals are modified on their way toward the receiver due to the multipath propagation phenomenon. As the multiple paths are generated by objects and people in the environment, the \gls{cfr} can be effectively used as a proxy of the physical environment to passively sense it. Importantly, the sensing targets do not need to carry a Wi-Fi device: their presence is detected as they act as reflectors, diffractors, or scatterers for the signal propagation. By analyzing this information through over-the-top signal processing algorithms and machine learning tools, researchers have been able to design and implement several applications such as human activity and gesture recognition~\cite{meneghello2022sharp}, people tracking~\cite{pegoraro2023rapid}, vital sign monitoring~\cite{wang2020csi}, and many others~\cite{ma2019wifi,Liu2022Integrated}. 

\subsection{Motivation}

A question that always arises when talking about Wi-Fi sensing is about the motivations behind and the benefits of performing sensing through radio waves, and in particular through a wireless network. Indeed, several other technologies are already available in the market to perform sensing, such as cameras and lidars. However, these technologies face problems in bad illumination conditions and when obstacles are present in the environment. Moreover, the use of camera systems comes with privacy issues when they operate in the presence of people. Radio waves allow overcoming all these issues as they can provide sensing information also in the dark and in the presence of smoke, and they do not capture the images of people around. For these reasons, radar devices have been extensively used to perform sensing. The use of Wi-Fi signals for this comes with additional benefits linked with the widespread adoption of this technology that, in turn, enables reusing already deployed infrastructures, extending their utility beyond traditional connectivity. However, this requires designing new procedures to integrate the two functionalities, which have quite different requirements~\cite{meneghello2023toward}. While communication waveforms are generated only when there is data to transmit, sensing requires a continuous transmission of signals to trigger the channel estimation at the monitoring devices. Moreover, communication devices are usually characterized by a smaller bandwidth with respect to radars, which impacts the sensing resolution. Indeed, the resolution is inversely proportional to the bandwidth $B$ as $c / (2 \!\times \! B)$ or $c /B$ for mono-static and bi-static sensing, respectively, where $c$ is the speed of light. Operating in the 2.4 and 5\,GHz bands, where bandwidth is limited, provides meter-level resolution, while larger bandwidths available in the mmWave portion of the spectrum enable centimeter-level resolution. However, sensing accuracy improves with multiple receivers and a larger number of \mbox{Wi-Fi} transmitters, making Wi-Fi sensing particularly promising given the proliferation of Wi-Fi~\cite{haque2025beamsense}.

Current sensing algorithms mostly rely on custom firmware modifications which are crafted for some specific devices, such as Nexmon CSI~\cite{nexmoncsi2019}, AX CSI~\cite{axcsi2021}, and PicoScenes~\cite{picoscenes2021}. Another approach is to rely on the beamforming feedback matrices introduced in Section~\ref{sec:MIMO} as described in~\cite{wi-bfi_haque}. This last strategy is hardware-agnostic as all Wi-Fi devices operating in \gls{su}-\gls{mimo} or \gls{mu}-\gls{mimo} mode have to feed back such information. However, all these approaches require triggering the transmission of data for generating the \gls{cfr} or the beamforming feedback matrices. Instead, the goal of the 802.11bf \gls{tg} has been to define modifications to the 802.11 \gls{phy} and \gls{mac} to enhance sensing capabilities in 802.11-compliant devices. Most sensing operations are based on existing beamforming protocols and are time-separated from data communications. Note that while the measurement report formats have been standardized through the 802.11bf amendment, specific sensing algorithms and application layers remain open for vendor-specific implementation~\cite{meneghello2023toward}.


\subsection{Sensing Procedure in 802.11bf}

\begin{figure}
\centering
\includegraphics[width=\columnwidth]{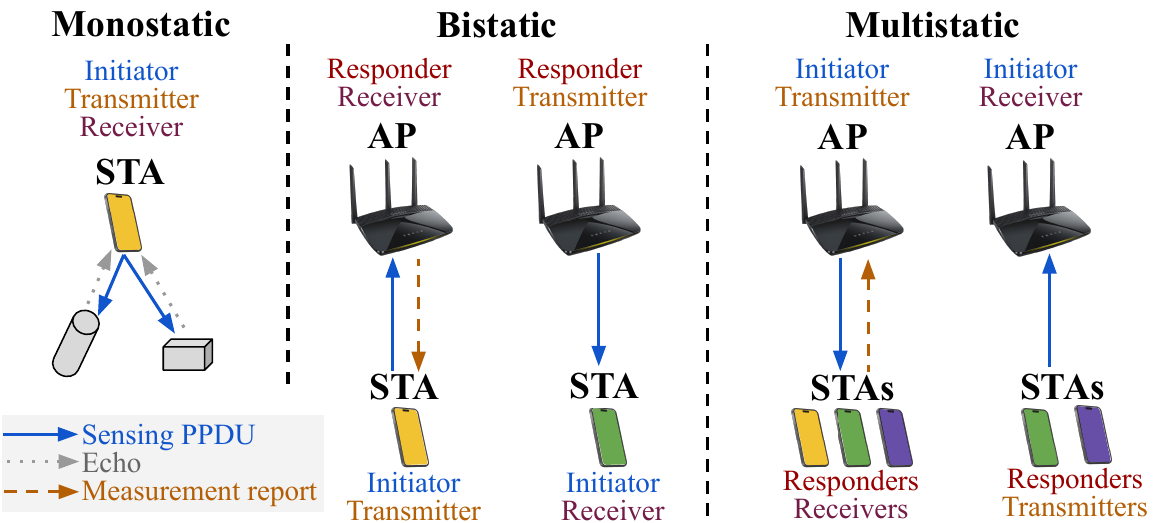}
\caption{\blue{Examples of monostatic, bistatic and multistatic sensing configurations. In the monostatic setting, the sensing is entirely executed by a single device that initiates the procedure by transmitting the sensing PPDU and collects the echoes from the environment. In bistatic sensing, the transmitter and receiver are distinct devices, e.g., an AP and a STA, and each of them can act as initiator. In multistatic, more sensing transmitter or receiver are involved.}}
\label{fig:sensing_architecture}
\end{figure}

802.11bf defines a unified procedure for obtaining channel measurements between two or more devices (bistatic and multistatic sensing), or between the transmitter and receiver antennas of a single device (monostatic sensing)~\cite{chen2022wi}. The devices involved can be either \glspl{ap} or \glspl{sta}, and their roles can be identified as follows:
\begin{itemize}
\item \emph{Sensing transmitter}: Transmits \glspl{ppdu} for sensing measurements.
\item \emph{Sensing receiver}: Receives \glspl{ppdu} from the sensing transmitter and obtains sensing measurements. In monostatic sensing, this device is the same as the sensing transmitter, while in bistatic sensing, they are distinct entities. 
\item \emph{Sensing initiator}: Initiates the sensing procedure and is where the sensing application is hosted. It may participate in sensing as a sensing transmitter, or sensing receiver, or both a transmitter and a receiver, or neither a transmitter nor a receiver. 
\item \emph{Sensing responder}: Participates in the sensing procedure. It may participate in sensing as a sensing transmitter, or sensing receiver, or both a transmitter and a receiver.
\end{itemize}
In addition to monostatic and bistatic sensing, multistatic sensing can be performed by involving multiple transmitters and/or receivers. Some examples of sensing configurations in the three different scenarios are depicted in Fig.~\ref{fig:sensing_architecture}. The IEEE 802.11bf amendment also introduces a ``sensing by proxy'' mode to enable an \gls{sta} to enhance sensing accuracy through diversity, with assistance from an \gls{ap}. While an \gls{sta} can only perform sensing measurements on the link between itself and the \gls{ap}, the latter has access to and can perform measurements of the channel of all other \gls{sta}~\cite{chen2022wi}.

A typical sensing procedure consists of four phases:
\begin{itemize}
\item \emph{Sensing capabilities exchange}: Devices capable of sensing discover each other and establish security. For associated \glspl{sta}, this is handled automatically through regular association.
\item \emph{Sensing measurement session}: The sensing initiator and responder(s) negotiate operational parameters specific to the application, such as role assignments, \gls{phy} parameters, measurement report types, and scheduling details.
\item \emph{Sensing measurement exchange}: The actual sensing measurements are performed as detailed in Section~\ref{subsec:sensing_sub7} and Section~\ref{subsec:sensing_submmwave}.
\item \emph{Sensing measurement session termination}: The procedure concludes after the channel information collection is completed.
\end{itemize}
Note that for monostatic sensing, the \emph{sensing session setup} and \emph{sensing measurement setup} phases are not required.



\subsection{Sensing in sub-7\,GHz Bands}\label{subsec:sensing_sub7}
\begin{figure}
\centering
\includegraphics[width=\columnwidth]{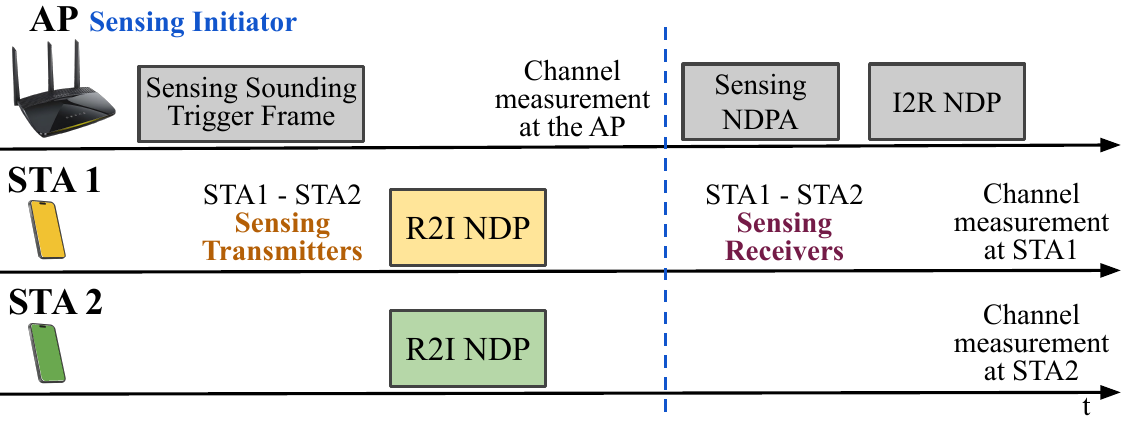}
\caption{\blue{Illustration of two AP-initiated Sub-7 GHz sensing approaches. On the left, the STAs transmit NDP for channel estimation in response to the AP transmitted trigger frame; the channel is then measured at the AP. On the right, the AP transmits a NDP for channel measurement at the STAs, preceded by a NDPA announcement frame. `R2I' indicates a frame from the responder to the initiator, while `I2R' indicates a frame from the initiator to the responder.}}
\label{fig:sensing_pipeline}
\end{figure}

Only bistatic and multistatic sensing are supported at sub-7\,GHz, while monostatic sensing is not allowed. The process may be initiated by the \gls{ap} or an \gls{sta}. \Gls{ap}-initiated sensing follows a \gls{tb} mechanism, as depicted in Fig.~\ref{fig:sensing_pipeline}. The \gls{ap} can transmit a Sensing Sounding Trigger Frame to stimulate \glspl{sta} to transmit an \gls{ndp} for channel measurement at the \gls{ap} (left part of Fig.~\ref{fig:sensing_pipeline}). The \gls{ap} can also transmit a Sensing \gls{ndpa} frame followed by a Sensing \gls{ndp} to collect measurements performed by the \gls{sta} through the latter frame (downlink sensing, right part of Fig.~\ref{fig:sensing_pipeline}). The \gls{sta}-initiated sensing is a non-\gls{tb} mechanism where the \gls{sta} initiates the sensing by transmitting a Sensing \gls{ndpa} frame to the \gls{ap} (single responder) followed by a Sensing \gls{ndp} frame for uplink sensing. The \gls{ap} subsequently transmits a Sensing \gls{ndp} frame for downlink sensing.

\subsection{Sensing in the 60\,GHz Band}\label{subsec:sensing_submmwave}
Sensing in the 60\,GHz band is built on the 802.11ad and 802.11ay amendments~\cite{chen2022wi}. In monostatic sensing, the same \gls{sta} transmits an (E)DMG \gls{ppdu}, receives the echo, performs channel measurements leveraging the training fields, and reports the results to the initiator. Bistatic sensing is instead based on the \gls{brp}: the sensing transmitter sends a \gls{brp} frame with appended training fields to the sensing receiver for measurements. The receiver then responds with a \gls{brp} frame containing the channel measurements. Multistatic sensing in the 60\,GHz band requires more coordination among devices: the initiator sends a DMG Sensing Request frame to each responder, each of which replies with a DMG Sensing Response and aligns its receiver antennas toward the initiator. To accommodate varying antenna configurations, sensing \glspl{ppdu} are appended with unique synchronization fields for each \gls{sta}.


\section{\blue{Task Groups 802.11bh and 802.11bi: Security and Privacy Extensions}}
\label{sec:bh_bi}

Security and privacy have been a major focus of every new generation of Wi-Fi~\cite{wifiprivacy}. In March 2021, two new task groups were formed to improve the security and privacy of 802.11 networks: 802.11bh and 802.11bi.

\subsection{The 802.11bh Standard Amendment}
802.11bh examined the specific implications of random and changing MAC addresses. \Gls{mac} address randomization can improve user privacy, but at the same time its use can have a significant implication in how existing Wi-Fi networks and services on these networks operate. For example, many existing networks and applications running on such networks rely on \gls{mac} addresses to recognize devices, e.g., captive portals, \gls{dhcp}, and troubleshooting. If these addresses change often, this can disrupt the seamless onboarding of devices into the network, device operation within a network and the delivery of network service to devices. As a result, 802.11bh defines two opt-in methods to identify a device to a network. Both methods are based on sharing information securely over an initial connection that can then be used when the devices (\gls{sta} and any \gls{ap} in the network) next encounter each other:
\begin{itemize}
\item \Gls{irm}: an \gls{sta} tells the \gls{ap} what \gls{mac} address it will use the next time it engages with the \gls{ap}. 
\item Device \gls{id}: an \gls{ap} gives the \gls{sta} a secret identifier to be provided to the network the next time the two engage.
\end{itemize}

An \gls{sta} participates in this identification only when it opts-in. The mechanisms are designed to prevent third-party tracking of the \gls{sta} by its identification. The intention is to only allow the \gls{ap} to track the \gls{sta} (i.e., recognize it on future interactions) when there is a shared security context, which prevents network spoofing. 802.11bh also introduces mechanisms that provide the ability to support a device identification pre-association through \gls{pasn} extension and a Beacon Measurement extension.

\subsection{The 802.11bi Standard Amendment}
802.11bi was formed with a focus on improving personal and device privacy, since MAC address randomization (as defined in 802.11bh) has been shown to be necessary but insufficient to ensure user privacy~\cite{11bipaper}. Users and regulatory agencies are concerned about protecting personal information such as location, movement, contacts, and activities of \glspl{sta} and their users. 802.11 devices are ubiquitous and protecting users from tracking and profiling attacks has become a major market need. While Wi-Fi is already being capable of encrypting all frame exchanges between an AP and a station after the association process, prior to association frame exchanges are still transmitted in the open. This can lead to privacy attacks and fingerprinting. To address these challanges, 802.11bi defines key derivation in Authentication frames and subsequent encryption of (Re)Association Request and Response frames. By encrypting (Re)Association Request and Response frames, all configuration-specific information that today is carried in the open within these frames and that could otherwise be used for fingerprinting becomes protected, thereby ensuring a very high level of privacy for an \gls{sta} and its user.


\section{\blue{AI/ML Topic Interest Group / Standing Committee}}
\label{sec:AIML}

The integration of \gls{ai} into 802.11 is another promising direction for the future of Wi-Fi~\cite{wilhelmi2024machine}. The adoption of \gls{ml} in Wi-Fi, however, requires a major transformation of the 802.11 standard, which lacks the necessary elements and procedures to support specialized \gls{ml} operations such as training/inference data collection, model training, or model output application. In this regard, the \gls{ieee} 802.11 \gls{wg} formed the \gls{ai}\gls{ml} \gls{tig} in 2022 (turned into \gls{sc} in 2024), which was entrusted to investigate the potential and feasibility of integrating \gls{ml} into future 802.11 amendments. The group’s primary focus consisted in identifying requirements for specific \gls{ml}-based use cases for Wi-Fi, which were collected in a technical report~\cite{aiml_tig_use_cases}. The included use cases are as follows:
\begin{itemize}
    \item \emph{AIML-based CSI feedback compression:} The sounding procedures in Wi-Fi lead to large overheads (see Section~\ref{sec:MU-MIMO}), especially when multiple users, spatial streams, and even multiple \glspl{bss} are involved. Henceforth, a promising and much needed use case for \gls{ai} is to reduce \gls{cfr} overheads, which can be achieved in multiple ways, e.g., leveraging similar \gls{sta} channels using unsupervised learning~\cite{deshmukh2022intelligent} or quantizing \gls{cfr} matrices efficiently using auto-encoders~\cite{sangdeh2020lb}. 
    \item \emph{Deep-learning based distributed channel access:} While valid for many years, Wi-Fi's channel access procedures based on \gls{csma}/\gls{ca} are inefficient and unreliable. For that reason, \gls{ml} is identified as a potential game changer in this use case. In particular, it is proposed to perform centralized \gls{cw} optimization using the \gls{dl}~\cite{wydmanski2021contention} or even relying on \gls{drl} to drive the access to the channel~\cite{guo2022multi}.
    \item \emph{AIML enhanced roaming:} Denser deployments make roaming very challenging since a large number of candidate \glspl{ap} need to be handled, thus high roaming times can be experienced. In this use case, \gls{ai}\gls{ml} is envisioned to provide meaningful information about the different candidate \glspl{ap} for roaming, indicating the most likely ones on a per-\gls{sta} basis, e.g., based on learned \gls{sta} patterns~\cite{wilhelmi2020flexible}. \gls{ai}\gls{ml} enhanced roaming is proposed be integrated into the 802.11k neighbor reporting mechanism.
    \item \emph{AIML based multi-AP transmission:} The added complexity introduced by \gls{mapc} (see Section~\ref{sec:MAPC}) introduces new challenges, such as deciding the set of \glspl{ap} that should collaborate. It has been suggested that \gls{ai}\gls{ml} could help in identifying devices and networks that can benefit from specific coordination features, e.g., \gls{cbf}.
    \item \emph{Efficient AIML model sharing:} This use case, which was the only one categorized as a use case that \emph{enables} \gls{ai}\gls{ml} in \glspl{wlan}, aims to address the fact that \gls{ml} applications require large information exchanges, including data, metadata, and models. Regardless of whether \gls{ml} applications that exchange data over Wi-Fi are used to improve Wi-Fi or not, it is clear that enhancing \gls{ml} data distribution would be beneficial to preserve \gls{qos}. In this regard, potential changes include the possibility of exchanging models at the \gls{mac} layer rather than at the application layer. With this, applications based on paradigms such as \gls{fl} could benefit from fast model exchanges.    
\end{itemize}

So far, it remains unclear what the role of \gls{ai} in 802.11 is going to be and what \gls{ml} features could be enabled. The fact that 802.11 focuses on the \gls{mac} and \gls{phy} layers limits the range of \gls{ai} functionalities and operations that could be defined by the standard, especially when compared to frameworks like \gls{3gpp}, where both core and \gls{ran} are defined. Furthermore, the inherent nature of Wi-Fi and its principles of decentralized channel access and backward compatibility make the adoption of \gls{ai}\gls{ml} a unique challenge for this technology.

\color{black}


\vspace{-0.1cm}
\bibliographystyle{IEEEtran}
\bibliography{bibliography}

\immediate\closeout\usedacronyms

\end{document}